%% file: main.tex
\documentclass{article}

\usepackage{microtype}
\usepackage{graphicx}
\usepackage{booktabs} % for professional tables

\usepackage{hyperref}

\usepackage{xspace}

\usepackage[accepted]{icml2024}

\usepackage{amsmath}
\usepackage{amssymb}
\usepackage{mathtools}
\usepackage{amsthm}

\usepackage[capitalize,noabbrev]{cleveref}

\theoremstyle{plain}

\theoremstyle{definition}

\theoremstyle{remark}

\usepackage{tikz}

\usepackage{filecontents}
\usepackage{graphicx}
\usepackage[caption=false]{subfig}
\usepackage{makecell}
\usepackage{listings}
\usepackage{svg}
\usepackage{amsmath}

\definecolor{codegreen}{rgb}{0,0.6,0}
\definecolor{codegray}{rgb}{0.5,0.5,0.5}
\definecolor{codepurple}{rgb}{0.58,0,0.82}
\definecolor{backcolour}{rgb}{0.95,0.95,0.92}

\definecolor{brewerpurple}{HTML}{AF4EA3}
\definecolor{brewerblue}{HTML}{377EB8}
\definecolor{NavyBlue}{HTML}{006EB8}
\definecolor{BrickRed}{HTML}{B6321C}
\definecolor{ForestGreen}{HTML}{009B55}

\lstdefinestyle{customc2}{
    emph={update\_metrics, prune\_completed\_jobs, exec\_jobs, accept, schedule, place, AcceptAll, Fifo, Consolidated, pop\_wait\_queue, update\_cluster},
    emphstyle=\bfseries\color{NavyBlue},
    commentstyle=\color{ForestGreen}\itshape\ttfamily,
    morekeywords={yield},
    stringstyle=\color{red}\ttfamily,
    emph={[2]admission\_policy, scheduling\_policy, placement\_policy},emphstyle={[2]\bfseries\color{BrickRed}},
    language=Python,
    keywordstyle=\bfseries\color{green!40!black}
}

\lstdefinestyle{mystyle}{
    backgroundcolor=\color{backcolour},   
    commentstyle=\color{codegreen},
    keywordstyle=\color{magenta},
    numberstyle=\tiny\color{codegray},
    stringstyle=\color{codepurple},
    basicstyle=\footnotesize,
    breakatwhitespace=false,         
    breaklines=true,                 
    captionpos=b,                    
    keepspaces=true,                 
    numbers=right,                    
    numbersep=5pt,                  
    showspaces=false,                
    showstringspaces=false,
    showtabs=false,                  
    tabsize=1
}

\newcommand{\dejavu}{{DéjàVu}\xspace}
\newcommand{\dejavulib}{{DéjàVuLib}\xspace}
\icmltitlerunning{DejaVu: KV-cache Streaming for Fast, Fault-tolerant Generative LLM Serving}

\begin{document}
%-------------------------------------------------------------------------------

%don't want date printed
    
%\icmltitle{Deja vu: KV-cache Streaming for Improved Performance \\ and Fault tolerance in Generative LLM Serving}
\twocolumn[
\icmltitle{DéjàVu: KV-cache Streaming for Fast, Fault-tolerant Generative LLM Serving}
%\maketitle

\begin{icmlauthorlist}
\icmlauthor{Foteini Strati}{x,ethz}
\icmlauthor{Sara Mcallister}{x,cmu}
\icmlauthor{Amar Phanishayee}{msr}
\icmlauthor{Jakub Tarnawski}{msr}
\icmlauthor{Ana Klimovic}{ethz}
\end{icmlauthorlist}

\icmlaffiliation{x}{MSR Project Fiddle Intern}
\icmlaffiliation{ethz}{ETH Zurich}
\icmlaffiliation{cmu}{Carnegie Mellon University}
\icmlaffiliation{msr}{Microsoft Research}

\icmlcorrespondingauthor{}{}
\vskip 0.3in
]
\printAffiliationsAndNotice{} % otherwise use the standard text.

\input{content/abstract}
\input{content/intro}
\input{content/motivation}

\input{content/dejavu}
\input{content/design}
\input{content/evaluation}
\input{content/relatedwork}
\input{content/conclusion}
%\input{content/impact}

%\clearpage

\bibliography{bibliography}
\bibliographystyle{plainnat}

%\newpage
\input{content/appendix}
%%%%%%%%%%%%%%%%%%%%%%%%%%%%%%%%%%%%%%%%%%%%%%%%%%%%%%%%%%%%%%%%%%%%%%%%%%%%%%%%
\end{document}

%% file: content/abstract.tex
\begin{abstract}

Distributed LLM serving is costly and often underutilizes hardware accelerators due to three key challenges: bubbles in pipeline-parallel deployments caused by the bimodal latency of prompt and token processing, GPU memory overprovisioning, and long recovery times in case of failures.
In this paper, we propose \dejavu, 
a system to address all these challenges using a versatile and efficient KV cache streaming library (\dejavulib).
Using \dejavulib, we propose and implement efficient prompt-token disaggregation to reduce pipeline bubbles, microbatch swapping for efficient GPU memory management, and state replication for fault-tolerance.
We highlight the efficacy of these solutions on a range of large models across cloud deployments.
    
\end{abstract}

%% file: content/intro.tex
\section{Introduction}

\begin{figure}[t]
\centering
    \includegraphics[width=\linewidth]{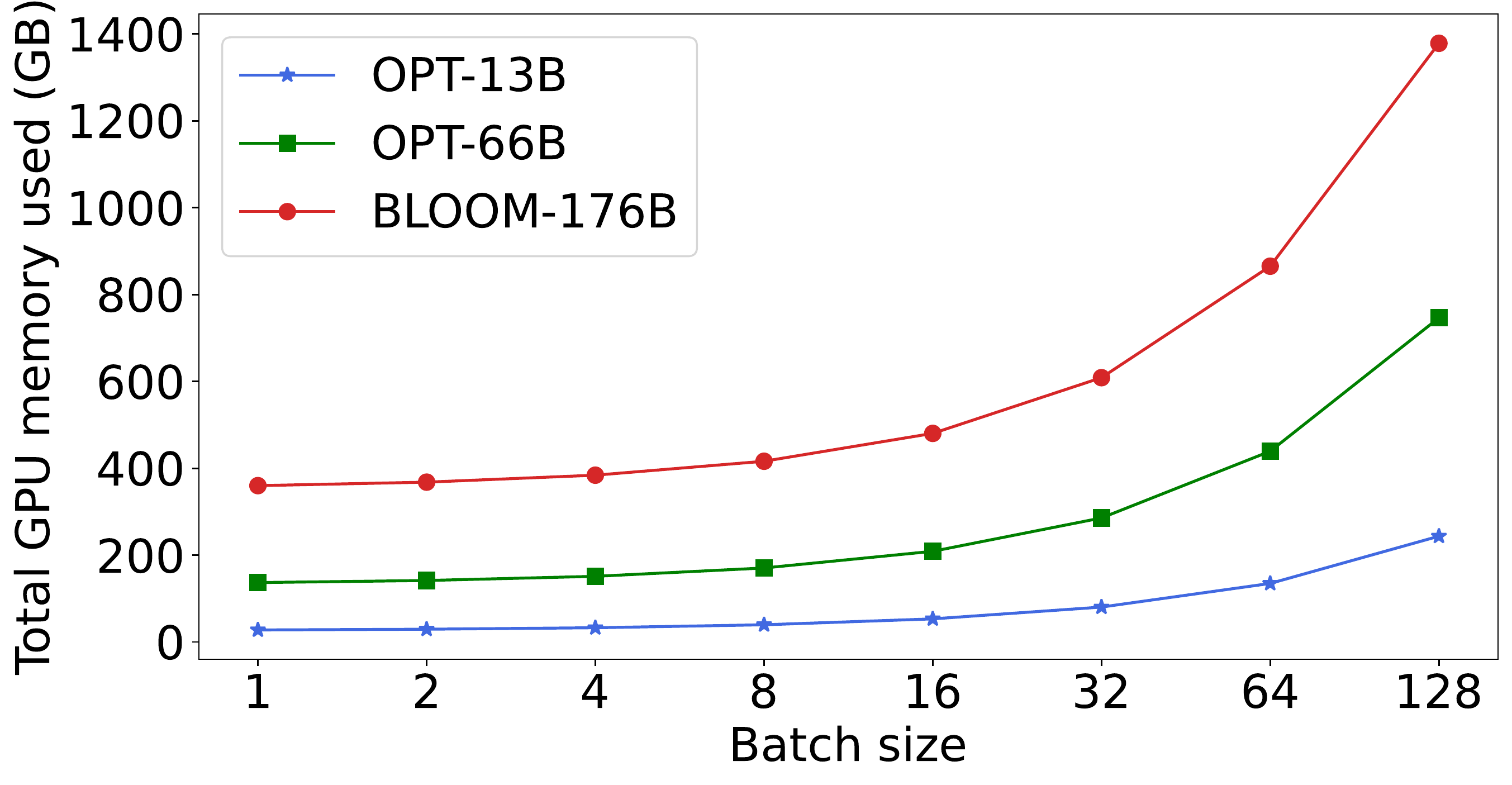}
    \caption{Memory footprint of serving various LLMs with 2K sequence length (input + generated tokens) and half precision (fp16).}\label{fig:optmem}
\end{figure}

Large Language Models (LLMs) like GPT-3~\cite{Brown2020GPT}, OPT~\cite{zhang2022opt} and BLOOM~\cite{workshop2023bloom} are widely used in chatbots~\cite{openai}, code generation, and text summarization~\cite{Github-copilot}. Two key trends in generative LLM inference have changed the landscape of ML model serving. First, large model sizes, input sequence lengths, and consequently large intermediate inference state lead to high memory footprint for LLM inference. Figure~\ref{fig:optmem} shows the GPU memory required to serve various generative LLMs with a 2K sequence length; their memory footprint greatly exceeds the capacity of a single GPU, mandating parallelization across many high-end GPUs (including tensor-model and pipeline parallel execution). Second, for low latency serving, these LLMs use a \textit{Key-Value Cache} to store prior computations as individual tokens are generated for each request~\cite{Kwon2023VLLM}. While ML inference is traditionally stateless, the use of KV cache makes generative LLM inference \textit{stateful}.  

Given these trends, we identify three key challenges in stateful, distributed large-scale LLM serving. First, we observe a substantial latency discrepancy (up to 2 orders of magnitude) between two phases of LLM serving, which leads to expensive GPU underutilization. Prompt processing depends on the input size and is compute bound. Meanwhile, token generation is memory bandwidth-bound and the time to generate a token is nearly constant when using the KV cache. Processing both prompts and tokens in the same pipeline introduces \textit{pipeline bubbles}, where GPUs idle. 

Second, state-of-the-art LLM serving systems like FasterTransformer vastly \textit{overprovision the KV cache} in pipeline parallel setups by allocating GPU memory for all microbatches upfront~\cite{NVIDIA21FasterTransformer}. Since the KV cache is only used by one microbatch at a time, there is an opportunity to allocate GPU memory more efficiently.

Third, existing LLM serving systems do not efficiently \textit{handle failures} or preemptions, which often occur in large-scale GPU deployments~\cite{Eisenman2022Checknrun, Myeongjae2019Philly}. Upon a failure, the LLM serving system crashes and stalls all in-flight requests. When KV cache state is lost, current systems process requests from scratch. These redundant computations severely increase end-to-end request latency.

To address the above challenges for pipeline-parallel distributed inference, we propose \dejavu, an efficient and fault-tolerant LLM serving system, based on KV cache streaming. First, \dejavu \textit{disaggregates} prompt processing from token generation and optimizes the number of machines for each stage to satisfy GPU memory capacity constraints and avoid GPU idle times. Second, to effectively use GPU memory capacity, \dejavu \textit{swaps} KV cache state per-microbatch between the GPU and CPU, maximizing GPU memory allocation for each microbatch being processed. Third, \dejavu \textit{replicates} KV cache state to avoid losing state and employs fast recovery mechanism to minimize lost work on failures.

The core component in \dejavu that enables all these optimizations is an efficient and versatile KV cache streaming library, \dejavulib.
We build \dejavulib as a modular set of primitives that enable fast streaming for diverse configurations, such as streaming between local or remote machines and for a variety of different KV cache structures. 

We evaluate \dejavu under different use cases. In pipeline parallel setups without failures, \dejavu improves LLM serving throughput by up to 2$\times$ compared to FasterTransformer. We show that \dejavu microbatch swapping can improve throughput by up to 1.8$\times$ by accommodating larger batch size for models that already fit in the given deployment. This enables serving even larger models that might not fit using existing state-of-the-art LLM systems. In the presence of system failures, \dejavu reduces microbatch latency by 1.54$\times$ compared to non-fault-tolerant systems.

%% file: content/motivation.tex
\section{Background and Motivation}
\subsection{Generative LLM inference}\label{sec:background}

Generative LLM inference involves two phases: \textit{prompt processing} and \textit{autoregessive token generation}. In the prompt processing phase, the model processes a user-defined sentence (i.e., prompt) provided as input and generates a new token. During autoregressive token generation, which spans multiple steps, the model generates new tokens one by one, using the token generated at step $i$ as input for step $i+1$. This continues until a user-specified number of tokens is generated or until the special EOS token is generated.

A crucial component of an autoregressive LLM is the \textit{attention} mechanism. Upon each step of token generation, each attention layer applies transformations to the input, to extract the \textit{query, key}, and \textit{value} vectors. At each generation step $i$, the attention mechanism computes the \textit{attention score} and token probability using the query vector at position $i$, and the key and value vectors at positions $[0,i-1]$. Thus, the computations and output at each step depend on the keys and values of the generated tokens at the previous steps. To avoid recomputing the key and value vectors of all processed tokens at each step, LLM inference frameworks store the vectors in the \textit{KV cache}~\cite{Ott2019fairseq}. 
During prompt processing, the key and value vectors of all tokens in the prompt are generated, populating the KV cache. Since all prompt tokens are known, computations during prompt processing use matrix-matrix multiplications and tend to be compute-bound. At each subsequent token generation step, the KV vectors for the newly generated token are appended in the KV cache. This phase is memory-bandwidth-bound~\cite{jin2023s3, Kwon2023VLLM}.

The KV cache size depends on model characteristics, such as the number of layers, hidden units per layer, floating point precision, batch size, and sequence length tokens~\cite{sheng2023flexgen}. Larger models, batch sizes, or longer generated sequences lead to a larger KV cache memory footprint.  Most LLM inference frameworks preallocate GPU memory for the KV cache for performance and often overprovision for the model's maximum supported sequence length~\cite{NVIDIA21FasterTransformer}. vLLM~\cite{Kwon2023VLLM} proposes \textit{PagedAttention} to dynamically allocate GPU memory for the KV cache. As LLM serving requires 100s of GB of GPU memory (see Figure~\ref{fig:optmem}), LLM inference is distributed across multiple GPUs, with pipeline and tensor parallelism~\cite{GyeongIn2022Orca, jiang2024hexgen}. Tensor parallelism requires very fast interconnects limiting it to single-node boundaries~\cite{Narayanan2021Megatron, jiang2024hexgen}; pipeline parallelism is additionally required for cross-node scaling\footnote{In this paper, we always use a combination of parallelization schemes: tensor-model parallel within a stage (multiple GPUs on a single server) and pipeline parallel across stages (servers)}. With pipeline parallelism, model layers are split across stages, with adjacent stages exchanging activations, and multiple micro-batches used to keep all stages busy.

Next, we highlight 3 important challenges posed by distributed LLM inference.

\subsection{Challenges of LLM serving}\label{sec:challenges}

\subsubsection{Bimodal prompt vs. token-gen  latency}\label{sec:disaggregation_challenge}

\begin{figure}[t]
\centering
    \includegraphics[width=\linewidth]{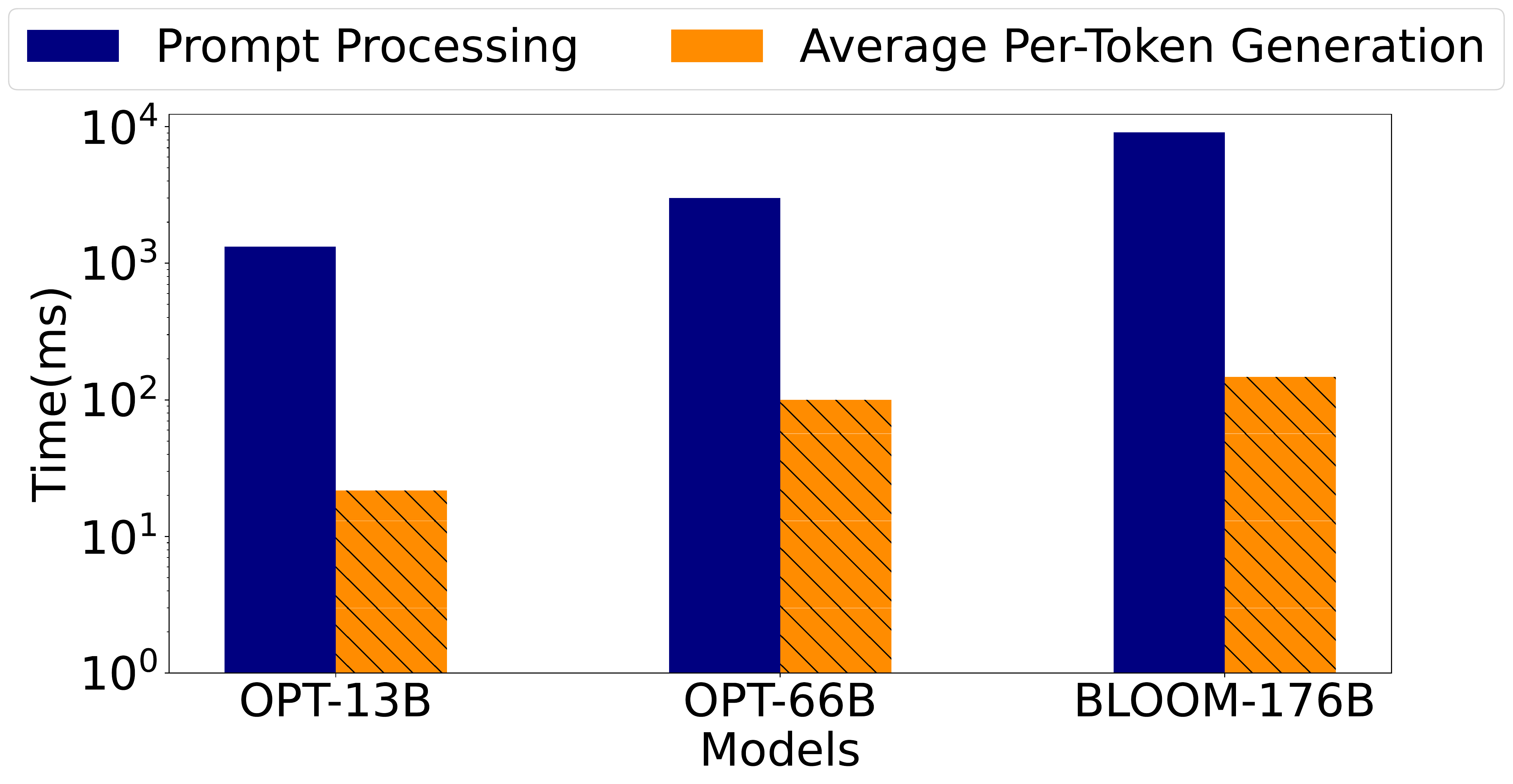}
    \caption{Prompt processing and average per-token generation time on A100 GPUs, using FasterTransformer (with batch size 8 and prompt size 1000). Y-axis is in log scale.}\label{fig:prompt-token}
\end{figure}

The first challenge in LLM serving comes from the disparity between prompt processing and token generation. Since the number of tokens processed during prompt processing is as large as the input sequence length, the prompt processing phase usually takes longer than the subsequent token generation phase. Figure \ref{fig:prompt-token} shows prompt processing and per-token generation times for various LLMs. Prompt processing time can be more than an order of magnitude higher than per-token generation time, depending on the model, batch size, and prompt length. In our study, prompt processing latency is 1.4$\times$ to 106$\times$ higher than per-token generation (see Appendix \ref{app_llm_profiling} for details).

\begin{figure*}[htp!]

\subfloat[Baseline]{
  \includegraphics[width=\linewidth]{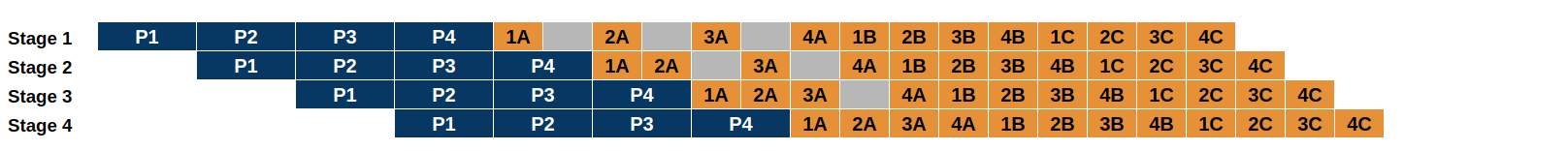}\label{fig:baseline_pipeline}%
}

\subfloat[Baseline, with request 3 stopping earlier than the rest (at $3A$)]{
  \includegraphics[width=\linewidth]{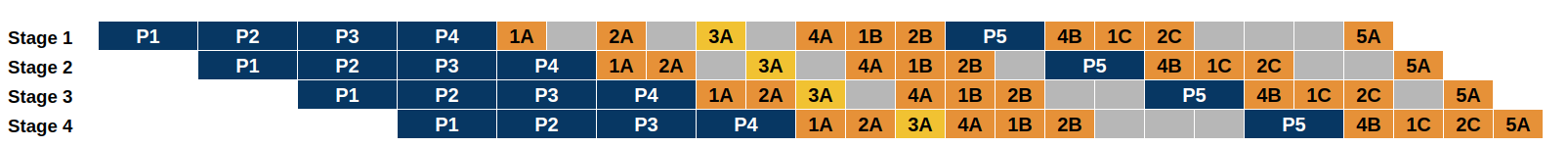}\label{fig:baseline_pipeline_es}%
}

\caption{LLM serving with 4-stage pipeline. A stage is a machine with $n$ GPUs running a set of layers with tensor model parallelism. $Px$ shows prompt processing of microbatch $x$. $X_y$ shows token generation for token $y$, microbatch $X$. For simplicity, in this figure, we assume prompt processing time takes 2$\times$ per-token processing time. In reality, the prompt-token difference can be up to 106$\times$ (see Appendix \ref{app_llm_profiling}). Grey areas are bubbles due to prompt processing vs. token generation latency discrepancy. 
} 
\label{fig:pipeline_bubbles}
\end{figure*}

With pipeline parallelism, this difference in execution time between the two stages causes \textit{pipeline bubbles}, leaving some stages idle while waiting for others to finish. For example, Figure \ref{fig:baseline_pipeline} shows a 4-stage pipeline, with 4 microbatches. Each microbatch consists of the prompt processing ($P$) step, and multiple token generation ($T$) steps. We observe bubbles in the pipeline, e.g. for token generation step \textcolor{orange}{$3A$} to start at Stage 1 for microbatch 3, Stage 4 must have completed prompt processing (generating the first token) for this microbatch  (\textcolor{blue}{$P3$}). Because prompt processing is slower than token generation, Stage 1 stalls. 

The problem of pipeline bubbles becomes even more pronounced with \textit{early stopping} of microbatches, where certain requests complete earlier than others. To keep the pipeline full, existing frameworks like vLLM~\cite{Kwon2023VLLM} will introduce a new microbatch, which will go through the prompt processing phase, disturbing the token generation of unfinished microbatches. Figure \ref{fig:baseline_pipeline} shows an example of early stopping, where microbatch 3 finishes earlier than the others (at step  \textcolor{orange}{$3A$}), and is replaced by a new microbatch. The difference in processing time between prompt and token steps introduces bubbles in the pipeline.

\subsubsection{Inefficient use of GPU memory}\label{sec:swapping_challenge}
In pipeline parallel settings, multiple microbatches should be processed concurrently by the different stages, to keep all stages busy~\cite{Narayanan2019Pipedream}. For example, in Figure \ref{fig:baseline_pipeline_es}, 4 microbatches are in-flight at each stage. While the prompt processing phase happens only once for each microbatch, each microbatch goes through multiple token generation steps.  Due to data dependencies between the different stages, the microbatches are processed in a \textit{round-robin} fashion. Each microbatch has its own KV cache.

 To increase performance, existing frameworks~\cite{NVIDIA21FasterTransformer} preallocate the KV caches of all microbatches in GPU memory. 
 However, since microbatches are processed sequentially at each stage, only the KV cache memory for a single microbatch is used at a time. Hence, memory is overprovisioned.

\subsubsection{Statefulness and failure handling}\label{sec:ft_challenge}

Due to its large memory footprint (see Figure \ref{fig:optmem}), LLM inference typically spans multiple GPUs across multiple nodes. In a distributed setup, failures are inevitable. Industry studies emphasize the prevalence of failures in ML training jobs. Meta reports that 50\% of the jobs encounter a failure within less than 16 minutes of execution~\cite{Eisenman2022Checknrun}, while Microsoft notes that training jobs often suffer from hardware or software failures~\cite{Myeongjae2019Philly}. Although these studies focus on iterative training jobs, the causes of software and hardware failures can also affect inference. Due to data dependencies between stages in a pipeline parallel inference setup, a failure in one stage leads all remaining stages to idle, or even results in timeouts and cascading failures, downgrading the throughput of LLM serving.

\begin{figure}[htp!]
\centering
    \includegraphics[width=\linewidth]{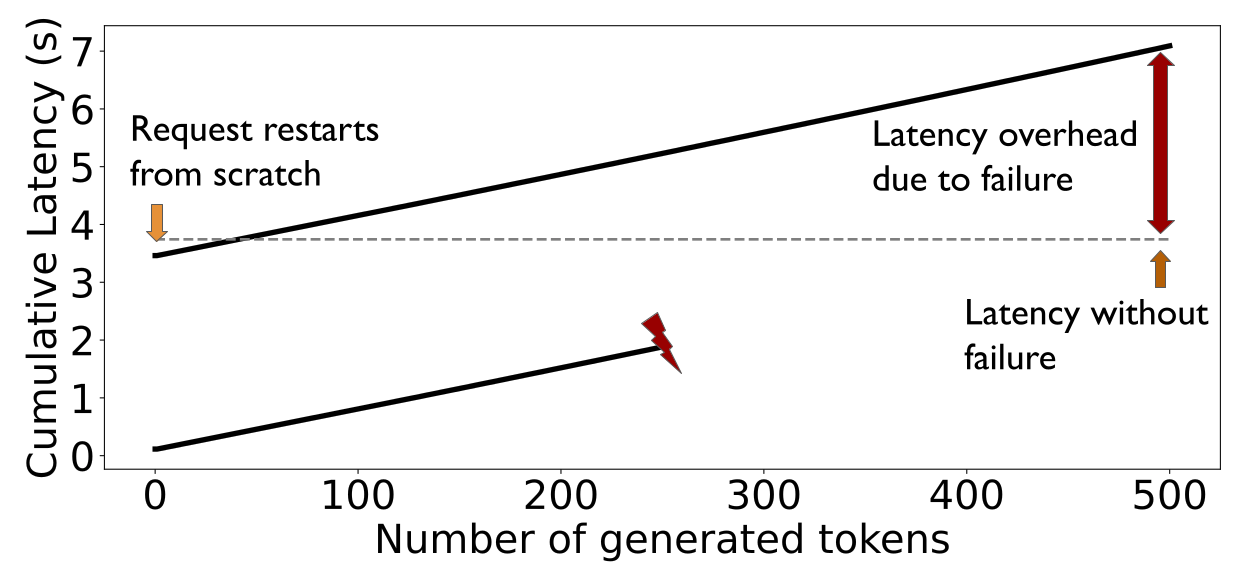}
    \caption{Effect on cumulative latency of an inference request when a failure occurs in today's systems, on a GPT2-1.5B model}\label{fig:failure}
\end{figure}

The impact of these failures in LLM inference is exacerbated by its stateful nature, due to the use of the KV cache. 
Since the KV cache is typically stored in GPU memory for fast accesses, an accelerator failure would result in loss of cached data for an inference request which in turn would require that all work for that request is redone. Figure \ref{fig:failure} shows a toy example of a GPT2-1.5B model serving a request with a prompt size of 500 tokens, and generating 500 new tokens. After the first 250 tokens are generated, a failure occurs. Existing LLM serving systems lack a fault tolerance mechanism, resorting to restarting the entire request. This involves repopulating the KV cache, thus reprocessing the prompt and regenerating tokens up to the point of failure. In our illustrative example, this approach results in a 1.89$\times$ increase in the end-to-end latency of the request. 
This issue is magnified with pipeline parallelism and multiple requests grouped in microbatches, where a failure in one stage will cause the whole pipeline and the requests across in-flight microbatches to restart from scratch.

%% file: content/dejavu.tex
\section{Proposed Solutions}\label{sec:solutions}

We now present our proposed solutions to address the challenges described in the previous section.

First, to mitigate pipeline bubbles, we propose \textit{disaggregating} prompt processing from token generation, by allocating separate machines for each task. 
By avoiding mixing prompt and token tasks, disaggregation helps reduce pipeline bubbles and leads to higher throughput. However, disaggregation's effectiveness relies on the swift transfer of the prompt KV cache, which can be a bottleneck especially since user-submitted prompts grow in size~\cite{ding2023longnet}, underscoring the need for an efficient KV cache streaming mechanism. Additionally, a key challenge is how to partition the available resources into prompt processing and token generation, to optimize system throughput. While disaggregation has also been recently proposed by concurrent related work~\cite{patel2023splitwise, zhong2024distserve}, we employ it to mitigate bubbles in pipeline-parallel settings (mandated by large generative models) and our resource allocation planner uses a principled approach to optimally partition the available resources to maximize system throughput.

Second, to optimize GPU memory usage, we propose \textit{swapping} the KV cache between GPU and CPU at the microbatch level \footnote{In contrast, vLLM\cite{Kwon2023VLLM} employs swapping of the KV cache of individual requests, instead of microbatches.}. The KV caches for all in-flight microbatches are stored in the CPU, and transferred to the GPU only when the respective microbatch is processed. 
This dramatically reduces GPU memory requirements, enabling larger batch sizes, and facilitating LLM serving under limited hardware~\cite{sheng2023flexgen}. However, CPU-GPU transfers through limited-bandwidth PCIe, can be a bottleneck, potentially leaving GPUs idle. Thus, we need an efficient mechanism to swap the KV cache in and out of the GPU.

Third, for fault tolerance, we propose \textit{replicating} the KV cache in persistent storage or remote CPU memory. In the event of a failure, \dejavu restores the most recent computed values to the failed GPUs, allowing inference to resume from the last generated token, and decreasing the recovery time compared to other LLM serving systems. To use such a system in practice, we need to minimize the overheads of KV cache streaming to storage or remote memory. We also need to make sure that failures can be detected and mitigated quickly to minimize recovery time.

These solutions require a fast and versatile KV cache streaming mechanism. Next, we describe our KV cache streaming library, \dejavulib (\S\ref{sec:dejavulib}), and how our system, \dejavu, implements the proposed solutions using \dejavulib.

%% file: content/design.tex
\section{The \dejavu LLM serving system}\label{sec:design}

\begin{figure}[tp!]
\centering
    \includegraphics[scale=0.23]{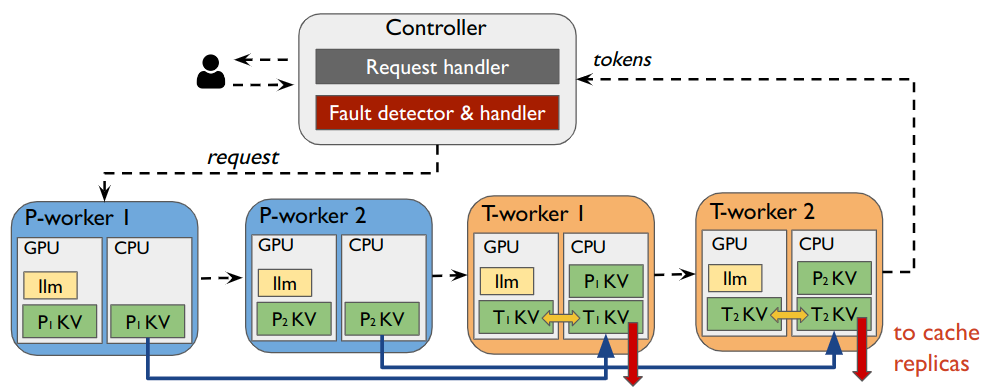}
    \caption{Full \dejavu system diagram. When disaggregation is enabled, the workers do either only prompt processing (\textit{P-worker}) or token generation (\textit{T-worker}). The \textcolor{blue}{blue} arrows stand for prompt-token cache exchange, the  \textcolor{red}{red} arrows for cache replication, and the \textcolor{orange}{orange} arrows for cache swapping.}\label{fig:system_diagram}
\end{figure}

Figure \ref{fig:system_diagram} illustrates the \dejavu system. A centralized controller is used to coordinate inference. Workers are registered with the controller to serve requests. Clients connect to the controller to submit requests, which are sent to the workers. The workers send the tokens to the controller as they are generated. Each \dejavu Worker has a \textit{cache manager} that handles KV cache streaming. The cache manager is aware of the pipeline configuration (pipeline depths, prompt or token processing, batch sizes, etc). When a Worker needs to stream the KV cache out or into the GPU, the cache manager calls the appropriate \dejavulib primitives (\ref{sec:dejavulib_primitives}). 

\subsection{\dejavulib: A KV cache streaming library}\label{sec:dejavulib}

\subsubsection{Low-level implementation of \dejavulib}

\begin{figure*}
\subfloat[Before inference starts.]{%
  \includegraphics[width=0.25\textwidth]{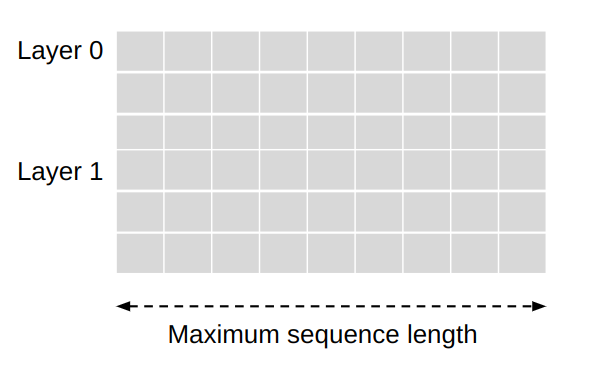}\label{fig:cache}%
}
\subfloat[After prompt (4 words)]{%
 \includegraphics[width=0.25\textwidth]{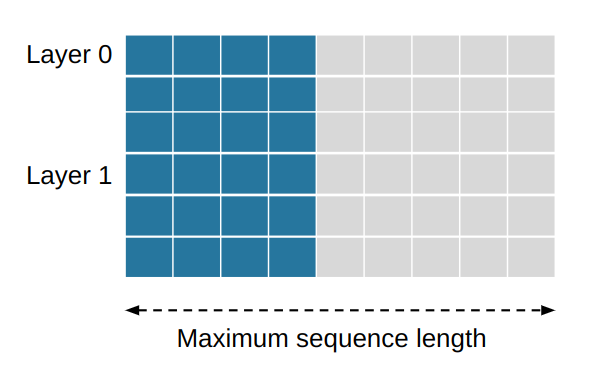}\label{fig:cache_prompt}%
}
\subfloat[After second token generation]{%
 \includegraphics[width=0.25\textwidth]{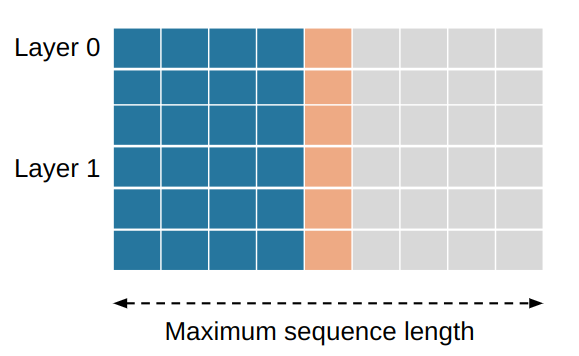}\label{fig:cache_token1}%
}
\subfloat[After third token generation]{%
 \includegraphics[width=0.25\textwidth]{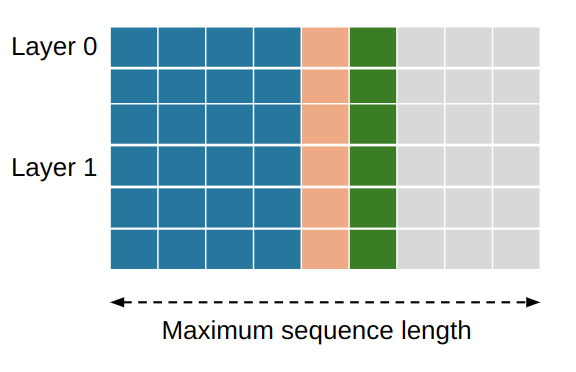}\label{fig:cache_token2}%
}
\caption{Simplified, 2D version of the key cache and the updated parts during prompt processing and token generation}
\label{fig:cache_ft}
\end{figure*}

\begin{figure*}[ht]
\centering
\subfloat[Buffered Copies: We aggregate small updates in a contiguous buffer in the GPU and then copy this buffer out.]{%
 \includegraphics[scale=0.23]{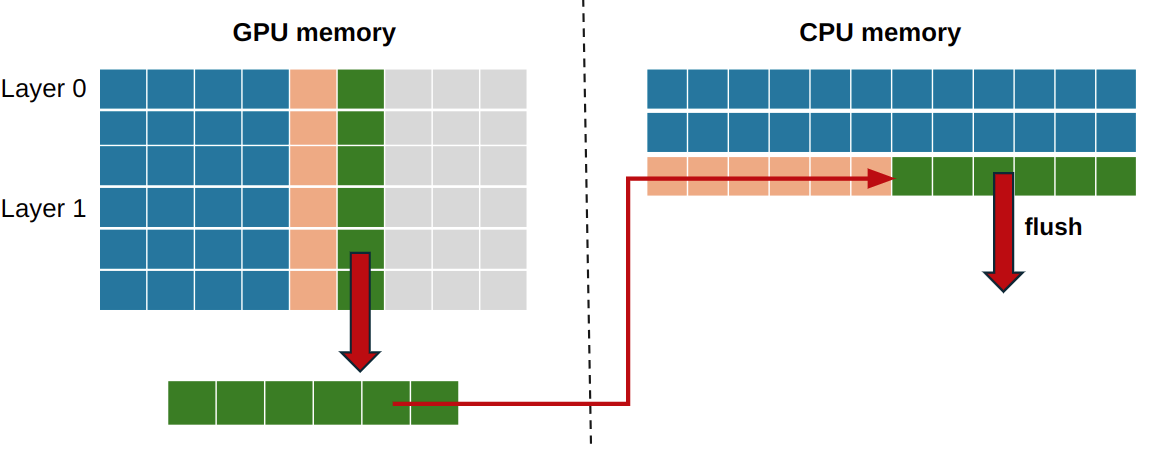}\label{fig:batch-opt}%
}

\subfloat[Layer-by-layer pipelining of prompt KV cache streaming with computation. \\ We also pipeline the streaming of microbatch $i$ with prompt processing of microbatch $i+1$]{%
  \includegraphics[scale=0.23]{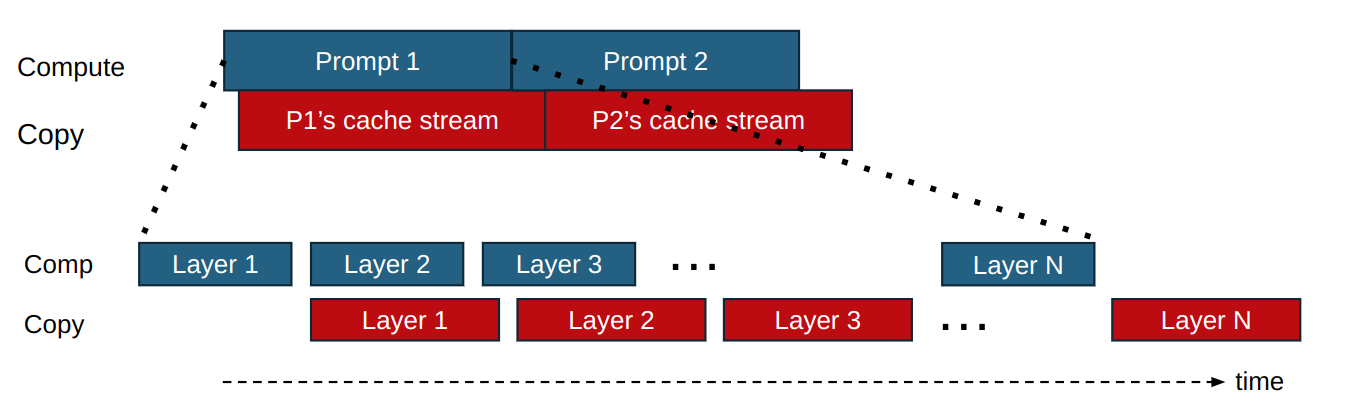}\label{fig:prompt-stream-opt}%
}
\subfloat[Pipelining of token streaming with computation]{%
 \includegraphics[scale=0.23]{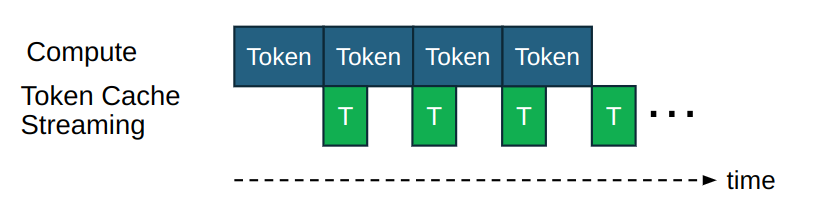}\label{fig:token-opt}%
}
\caption{\dejavulib KV cache streaming optimizations}
\label{fig:all_opt}
\end{figure*}

\dejavu is built on top of FasterTransformer\footnote{We chose FasterTransformer~\cite{NVIDIA21FasterTransformer} since it is the state-of-the-art framework with support for tensor-model and pipeline parallel inference. Other frameworks such as vLLM do not support pipeline parallelism at the moment~\cite{Kwon2023VLLM}}, supporting both tensor and pipeline parallelism. FasterTransformer preallocates GPU memory for the KV cache based on a maximum sequence length (either the maximum length supported by the model or user-defined). Figure \ref{fig:cache_ft} provides a simplified 2D representation of the key cache \footnote{The value cache follows a similar structure.}, including only the layer and the sequence length dimension.\footnote{In reality, the key cache is a 6D tensor, and the value cache is a 5D tensor. The extra dimensions include the number of attention heads and batch size.} Figure \ref{fig:cache_prompt} shows what happens to the key cache after processing a prompt of 4 words. Prompt processing occurs \textit{layer-by-layer}, populating the respective portion of the cache at each layer. After the prompt has been processed, tokens are generated one by one. Figures \ref{fig:cache_token1} and \ref{fig:cache_token2} depict key cache contents after 2 subsequent tokens have been generated. 
After the generation of each token, only a small, non-contiguous part of the Key cache is updated.  The need to copy numerous non-contiguous small memory regions results in significant overhead, necessitating optimizations that we state below:

\textbf{(1) Buffered copies (Figure \ref{fig:batch-opt}):}
Individual token generation leads to multiple non-contiguous small updates in the KV caches (Figure \ref{fig:cache_ft}). Employing multiple \texttt{cudaMemcpy} calls for copying these chunks, results in substantial overhead. Instead, we leverage the high bandwidth of GPU DRAM, and aggregate all updates in a temporary buffer within GPU memory. 
Once the temporary buffer has been populated, we copy it to the appropriate destination.  Since these buffers are reused, the overhead in GPU memory capacity is negligible.

\textbf{(2) Layer-by-layer prompt cache streaming (Figure \ref{fig:prompt-stream-opt}):}
Since prompt processing occurs in a layer-by-layer fashion, we also stream the prompt cache layer-by-layer. This is similar to \textit{wait-free backpropagation} which overlaps backward pass computations with gradient exchange in distributed ML training~\cite{Zhang17Poseidon}. In a pipeline parallel setup, we further parallelize streaming the prompt of microbatch $i$ with computation of microbatch $i+1$.

\textbf{(3) Token computation and streaming parallelization (Figure \ref{fig:token-opt}):}
Unlike prompt processing, token generation for a single request involves multiple steps. In a single-machine setup, we stream the KV cache for step $i$, while step $i+1$ is in progress. In a pipeline parallel setup, we parallelize the cache streaming of microbatch $i$, step $j$, with computation of microbatch $i+1$, step $j$. Token computation, like prompt processing, occurs layer-by-layer. However, token streaming time can be fully masked behind subsequent token computation, so we do not use layer-by-layer streaming in this case. 

We use a background CPU thread that is responsible for cache streaming, and CUDA streams to parallelize KV cache streaming with computation on the GPU~\cite{NVIDIA2015Streams}. 
In section \ref{sec:microbenchmarks} we evaluate our streaming implementation, and its overheads during inference. 

\subsubsection{\dejavulib primitives}\label{sec:dejavulib_primitives}

\begin{table*}[ht]
%\tiny
\centering
\caption{\dejavulib primitives. The \texttt{stream\_out},\texttt{stream\_in} call \texttt{scatter},\texttt{gather}, which call \texttt{flush} and \texttt{fetch} respectively. }
\vskip 0.15in
\begin{center}
\begin{small}
\begin{tabular}{c|p{12cm}}
\toprule
\textbf{Primitives} & \textbf{Functionality} \\
\midrule
\texttt{stream\_out}, \texttt{stream\_in} & Given a source (or destination) worker, the KV cache, and the inference setup (number of workers, pipeline depths, batch sizes), find the proper destinations (or sources) for the different chunks of KV cache. This might involve splitting the cache at the source or merging cache chunks at the destination.\\
\hline
\texttt{scatter}, \texttt{gather} & Given a non-contiguous region of KV cache, and a local or remote destination (or source),  chunk the region to contiguous transfers and orchestrate movement. \\
\hline
\texttt{flush},  \texttt{fetch}  & Copy a contiguous chunk of KV cache, on the same or remote host. Local copies with CUDA, and remote copies with NCCL~\cite{NVIDIA2023NCCL}, MPI~\cite{OpenMPI2023MPI}, or Boost~\cite{Boost2021Asio} are supported. \\
\bottomrule
\end{tabular}
\end{small}
\end{center}
\vskip -0.1in
\label{table:primitives}
\end{table*}

We built \dejavulib as a versatile library that handles different configurations, addressing the challenges we met when developing our solutions described in \ref{sec:solutions}. The source, destination, data volume, and transferring method of KV cache streaming depend on the pipeline setup, and network topology. For instance, when disaggregating prompt processing from token generation, the prompt KV cache is transferred from the prompt processing to the token generation machines. This can occur through various mechanisms, such as GPU-GPU or CPU-CPU copies. Moreover, the prompt and token pipelines might have different pipeline depths and batch sizes, requiring splitting or merging the KV cache at the source and destination respectively. \dejavulib aims to account for the diverse set of configurations that require KV cache streaming, abstracting away the implementation details from the high-level handling of the KV cache while offering efficient solutions depending on the type of streaming. We achieve this by offering primitives with different levels of abstraction, as shown in table~\ref{table:primitives}.

\subsection{Detailed description of the proposed solutions}

\subsubsection{Prompt-Token disaggregation}\label{sec:disaggregation_impl}

Workers are categorized into 2 groups: prompt processing, and token generation (Figure \ref{fig:system_diagram}). The Controller assigns incoming requests to the prompt 
workers, which generate the first token, populating the KV
cache, which is then transferred to the token generation machines. With disaggregation, we need to ensure that: 1) we optimally allocate resources for each phase, and 2) we transfer the KV cache from prompt to token machines with minimal overheads.

\textbf{1. Principled allocation of resources:}

\begin{figure*}[htp!]
\subfloat{%
  \includegraphics[scale=0.3]{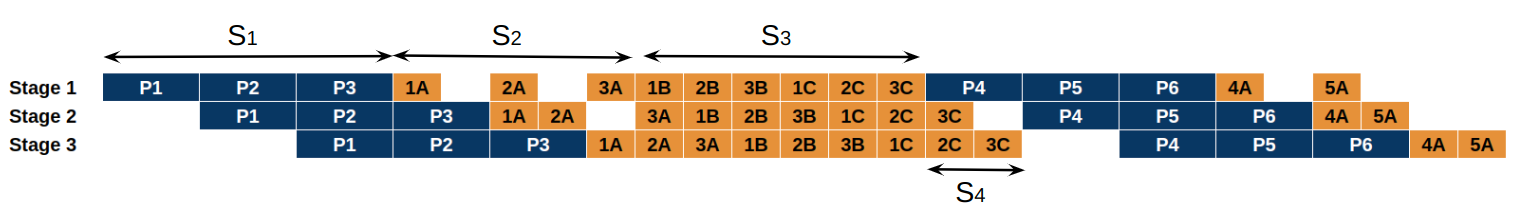}%
}
\caption{Illustration of the different phases of a 3-stage pipeline with prompt processing and token generation}
\label{fig:pipeline_thr}
\end{figure*}

Given a fixed set of machines, we want to partition them into prompt and token processing to satisfy the following requirements: 1) the aggregate memory footprint (model parameters and KV cache), for the active microbatches should fit into the aggregate GPU memory capacity for each pipeline, and 2) the throughput of the disaggregated system should be maximized, and ideally be higher than the throughput of the non-disaggregated system. We developed a resource allocation planner to address the above requirements.

Assume we are given $D$ machines, each with aggregate GPU memory capacity of $M$ GB. Assume a model has $L$ layers. For simplification, we consider only the memory requirements of the attention layers $W_i$. We also assume each layer's prompt KV cache footprint is $C_i$. The requirements that we need to satisfy are: 1) the aggregate memory footprint (model parameters and KV cache), for the active microbatches should fit into the aggregate GPU memory capacity for each pipeline, and 2) the throughput of the disaggregated system should be maximized, and ideally be higher than the throughput of the non-disaggregated system.

We first aim to satisfy requirement (1) for the prompt processing pipeline, i.e. find the prompt pipeline depth $D_p$. $P_n$ is the number of attention layers per stage. Assuming each machine corresponds to a pipeline stage, the following inequality should hold:
$$ M \geq P_n \cdot (C_0 + W_0) \Longrightarrow P_n \leq \lfloor \frac{M}{C_0 + W_0} \rfloor $$
Since $D_p = \lceil \frac{L}{P_n} \rceil $:
\begin{equation}
\label{formula:mem1}
 D_p \geq \lceil \frac{L\cdot (C_0 + W_0)}{M} \rceil
\end{equation}

Similarly, we need to satisfy requirement (1) for the token generation pipeline, i.e. find the token generation depth $D_t$. Each layer's token KV cache footprint is $K_i$. $T_n$ is the number of attention layers per stage. Since token generation involves multiple steps, and at least $D_t$ microbatches need to be on-the-fly at any given step, we have:
$$ M \geq T_n \cdot W_0 + D_t \cdot (C_i + K_i)  $$
thus:
$$ M \geq T_n \cdot (W_0 + (C_0 + K_0) \cdot D_t) $$

Since $D_t = \lceil \frac{L}{T_n} \rceil $:

$$ M \geq T_n \cdot (W_0 + (C_0 + K_0) \cdot \lceil \frac{L}{T_n} \rceil) $$

For simplicity, we assume $T_n$ divides $L$:
$$ M \geq T_n \cdot W_0 + L \cdot (C_0 + K_0) $$
thus:
$$M \geq \frac{L}{D_t} \cdot W_0 + L \cdot (C_0 + K_0) $$
\begin{equation}
\label{formula:mem2}
D_t \geq \frac{L \cdot W_0}{M - L \cdot (C_0 + K_0)}
\end{equation}

For requirement (2), we need to compute the throughput of the non-disaggragated and the disaggregated setups. For simplicity, in the following formulas, we work with the \textit{inverse throughput} of the pipelines. Thus, we would like the disaggregated case to have \textit{lower} inverse throughput than the non-disaggregated one.
Assume, for simplicity, that prompt processing of each microbatch with $D$ machines lasts $Y$ ms, and each token generation step for a single microbatch takes $t$ ms. Since we dedicate $D_p$ machines to prompt processing and $D_t$ machines to token generation, each machine will host a larger number of layers. Thus, $Y_{dis} = \frac{D}{D_p} \cdot Y$, and $t_{dis} = \frac{D}{D_t} \cdot t$. Assume, also, $N$ new tokens are generated per microbatch. 

First, we compute the inverse throughput ($I$) of the baseline. Figure \ref{fig:pipeline_thr} illustrates a toy example of a pipeline with 3 stages. In the general case of a pipeline with $D$ stages, and $D$ active microbatches at each point in time, the inverse throughput is given by:

$$ I_c = \frac{S_1 + S_2 + S_3 - S_4}{D}$$

where: 

$$S_1 =  D \cdot Y$$
$$S_2 = (D-1) \cdot Y$$
$$S_3 = N \cdot D \cdot t$$
$$S_4 = (D-1) \cdot t $$

Thus:

$$ I_c = \frac{D \cdot Y + (D-1) \cdot Y +  N \cdot D \cdot t - (D-1) \cdot t}{D}$$
$$ I_c = Y + N \cdot t + \frac{(D-1)(Y-t)}{D}$$

\begin{equation}
\label{formula:basethr}
I_c = \frac{(D-1)(Y-t)}{D} + Y + N \cdot t
\end{equation}

In steady case, the token generation pipeline with $D_t$ machines will have inverse throughput:

$$I_t = \frac{N \cdot D_t \cdot t_{dis}}{D_t} = N \cdot t_{dis} = \frac{N \cdot D \cdot t}{D_t}$$

The prompt generation pipeline with $D_p$ stages will have inverse throughput:

$$I_p = \frac{m \cdot Y_{dis} \cdot D_p}{D_p} = m \cdot Y_{dis} = \frac{m \cdot D \cdot Y}{D_p}$$

where m is the additional overhead due to cache streaming (i.e. $m \geq 1$).

The performance of the disaggregated system ($I_{dis}$) depends on the performance of the prompt processing and token generation pipelines, i.e. $I_{dis} = max(I_t, I_p)$.
Since we have $D$ machines, and we partition them into the 2 phases, allocating more machines to prompt processing, i.e. decreasing its inverse throughput, would lead to fewer machines for token generation, i.e. increasing its inverse throughput. Since we are minimizing a $\max$ function, the ideal case will be when $I_t = I_p$. Thus, we want:

$$I_{dis} = I_t = I_p < I_c$$

From $I_t = I_p$ (and the fact that $D_t + D_p = D$), we get that:

$$ \frac{N \cdot D \cdot t}{D_t} = \frac{m \cdot D \cdot Y}{D_p} \Longrightarrow D_t = \frac{D \cdot N \cdot t}{m \cdot Y + N \cdot t} $$

Given this $D_t$, the throughput of the disaggregated system will be higher than the throughput of the non-disaggragated system if 

\begin{equation}
\label{formula:perf1}
I_{dis} = I_t < I_c \Longrightarrow \frac{Y}{t} > \frac{D-1}{D \cdot (2-m) - 1}
\end{equation}

Eq. \ref{formula:perf1} holds if $m \in [1,2)$. If $m \in [1,2)$, we have:

\begin{equation}
\label{formula:token_formula}
D_t = \frac{D \cdot N \cdot t}{m \cdot Y + N \cdot t}
\end{equation}

and 

\begin{equation}
\label{formula:prompt_formula}
D_p = D - D_t = \frac{D \cdot m \cdot Y}{m \cdot Y + N \cdot t}
\end{equation}

Formulas \ref{formula:perf1}, \ref{formula:token_formula} and \ref{formula:prompt_formula} lead to a couple of observations. First, as expected, given $D$, $Y$, and $t$, with $Y>t$, the benefits of disaggregation depend on the overheads of the prompt KV cache streaming. If the streaming overhead is too high (i.e. $m \geq 2$), there will be no benefits from disaggregation.  \dejavulib employs multiple optimizations to ensure that the prompt KV cache streaming overhead is minimized. Second, the larger $N$ is, i.e. a lot of new tokens are generated, $D_t$ is increasing, i.e. we need to dedicate more machines to token generation. In contrast, when $\frac{Y}{t}$ increases, $D_p$ is increasing, thus more machines need to be dedicated for prompt processing. Moreover, as $\frac{Y}{t}$ increases, i.e. with larger prompts, the disaggregated setup becomes more beneficial, as can be seen from inequality  \ref{formula:perf1}.

\textbf{2. Fast prompt KV cache transfers: } We use optimizations (1) and (2) from \ref{sec:dejavulib} to pipeline layer-by-layer prompt KV cache streaming with prompt processing. To avoid overloading GPU memory, we transfer the KV cache to local CPU memory, and then to CPU memory of the token machine. Prompt and token pipelines might have different pipeline depths, or different batch sizes. This may require splitting the KV cache to multiple token machines or merging from different prompt machines. The cache manager invokes the  \texttt{stream\_out} primitive which calls the lower-level primitives (Table \ref{table:primitives}) based on pipeline depth and batch size. Whenever a prompt is needed, the token machines check if there are any prompt KV caches in their local CPU memory.  When prompt KV cache becomes available, it is loaded into GPU memory and token generation starts.

\subsubsection{Microbatch swapping}\label{sec:swapping_impl}

To facilitate microbatch swapping with minimal overhead we leverage all three optimizations presented in \ref{sec:dejavulib}. For a pipeline of depth $D$, where $D$ microbatches are active at a time, and each microbatch requires $M$ GB, we allocate $D \cdot M$ GB in CPU memory, and $2 \cdot M$ GB in GPU memory \footnote{or $M$ GB in GPU memory if $D==2$}. Before token generation step $t$ for microbatch $x$ starts, \dejavu prefetches the KV cache for this microbatch from CPU to GPU (\textit{swap in}). After step $t$ has finished, the \dejavu cache manager transfers the updated part of  microbatch $x$'s KV cache, corresponding to step $t$, back to the CPU (\textit{swap out}). 

\begin{figure*}[htp!]
\centering
\subfloat{%
  \includegraphics[scale=0.4]{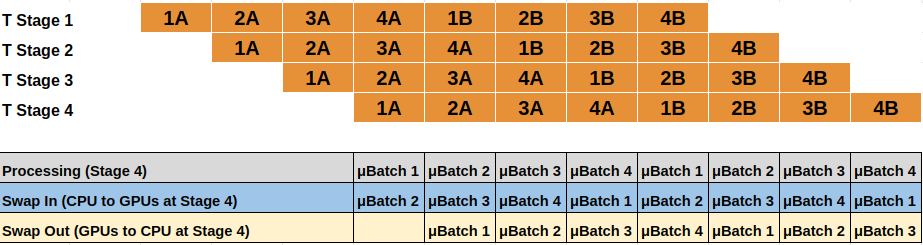}%
}
\caption{Microbatch KV cache swapping over time for a 4-stage token generation pipeline. We show microbatch swapping for Stage 4, but all other stages follow similar pattern.}
\label{fig:swapping}
\end{figure*}

Figure \ref{fig:swapping} illustrates microbatch KV cache swapping, for a pipeline of 4 stages, focusing on Stage 4. Whenever a microbatch is processed, we make sure its KV cache resides in GPU memory, while, in parallel, swapping other microbatches in and out of GPU. When Stage 4 generates a token for microbatch 1 (e.g. step $T1_1$), it swaps in the KV cache for the next microbatch to be processed, i.e. microbatch 2. When the processing of microbatch 1 has finished, the newly added contents to the KV cache of microbatch 1 are swapped out of the GPU. The same procedure follows for all microbatches: assuming a pipeline with $N$ stages, when microbatch $x$ is processed, microbatch $(x+1)\%N$ is swapped in, and microbatch $(x-1)\%N$ is swapped out.

\subsubsection{Failure handling}\label{sec:ft_impl}

\begin{figure*}[htp!]
\subfloat[A failure occurs ($F$)]{%
  \includegraphics[clip,width=\linewidth]{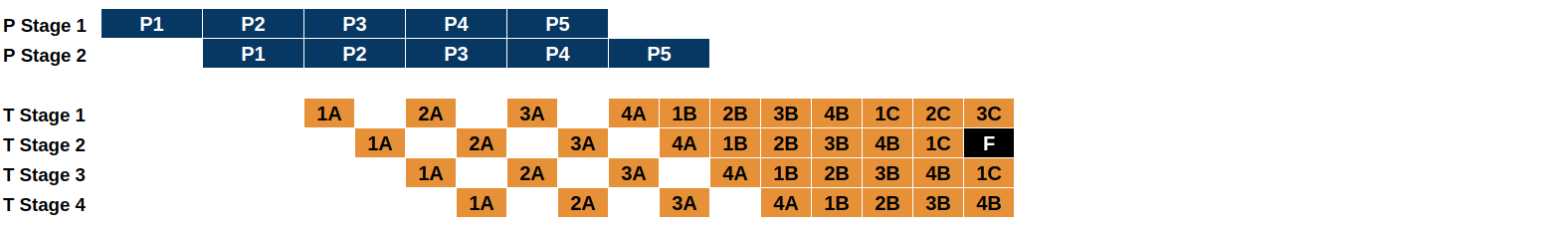}\label{fig:ft_failure}%
}

\subfloat[The failure is detected and repaired]{%
  \includegraphics[clip,width=\linewidth]{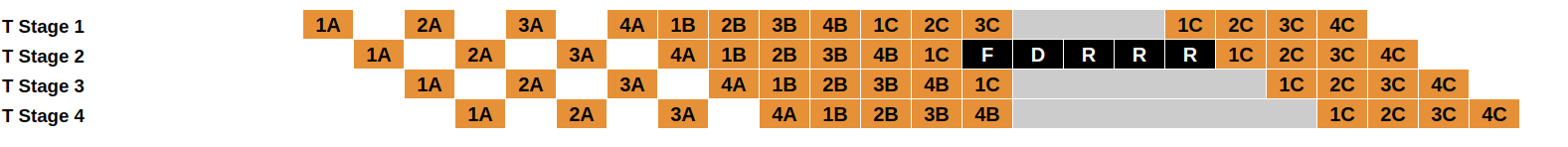}\label{fig:ft_repair}%
}
\caption{Example of a pipeline failure and recovery. Token stage 2 fails ($F$), and is detected by the \dejavu Controller ($D$). The pipeline is repaired ($R$), which includes copying the KV caches around appropriately. After repair is done, inference continues.}
\label{fig:ft_protocol}
\end{figure*}

Figure \ref{fig:ft_protocol} provides a toy example with 4 token generation workers. In practice, we need to ensure that 1) KV cache replication has minimal overheads, and 2) failures are detected and mitigated quickly to minimize recovery time. 

The \dejavu Controller is responsible for detecting and mitigating failures. The workers send heartbeats to the controller periodically. If the controller has not received a heartbeat from a worker within a specified timeframe, it identifies the worker as failed, and notifies the rest workers to stop serving requests. To recover from the failure, we need to 1) restore the lost KV caches, and 2) determine the step and microbatch from which the inference should resume. 

Each worker $x$ streams its KV cache to worker $(x+1)\%N$ (assuming an N-stage pipeline). For example, in Figure \ref{fig:ft_protocol}, the worker at Stage 1 streams its cache to Stage 2, Stage 2 to Stage 3, etc. The cache is streamed incrementally, as each token is generated, and takes place asynchronously (in parallel) to computation (see \ref{sec:dejavulib}). We define a background thread at each worker that is responsible for receiving the KV cache from its peer. When a worker $x$ fails, both its own KV cache and the replica KV cache of worker $(x-1)\%N$ is lost. During recovery, we make sure the lost caches are repopulated to worker $x$.

Upon receiving the KV cache update from worker $(x-1)\%N$, for microbatch $j$ and generation step $t$, worker $x$ sends a message to the controller of the form $(x,j,t)$.  Therefore, the controller is aware of the KV cache replication status across all workers. In the event of failure, we follow a four-step process for recovery. First, worker $(x+1)\%N$ sends the replica KV cache it hosts to worker $x$ (repopulating $x$'s lost cache). Second, worker $(x-1)\%N$ sends its KV cache to $x$ (repopulating the lost replica at $x$). Third, the controller finds the microbatch $j$ and step $t$, that needs to be re-executed, since the cache of failed worker $x$ has not been replicated up to that point. Finally, as stage $x$ requires input from its preceding stages to re-execute a microbatch, the controller propagates $(j,t)$ to all workers, and Stage 1 resumes inference from microbatch $j$ and step $t$.

As a concrete example, consider the scenario in Figure~\ref{fig:ft_protocol}, where stage 2 fails. First, stage 3 will copy stage 2's KV cache replica back to stage 2. Second, stage 1 will copy its own KV cache to stage 2. Third, the controller identifies the microbatch $j$ and step $t$ that needs to be reexecuted. In that case, $j==1$ and $t==C$, since stage 2's KV cache for $1C$ had not been replicated before the failure. If stage 2 restarts from $1C$, it needs to get inputs (activations) from stage 1. Thus, finally, all stages execute $1C$.

%% file: content/evaluation.tex
\section{Evaluation}

\textbf{Setup} We use VMs with 2 A100-80GB GPUs, and inter-VM network bandwidth of 40 Gbps, and VMs with V100-16GB GPUs, and inter-VM network bandwidth of 32 Gbps. \\
\textbf{Models} We use HuggingFace versions of GPT2~\cite{Brown2020GPT}, OPT~\cite{zhang2022opt} and BLOOM~\cite{workshop2023bloom}, adapted for FasterTransformer. We use half-precision for all models. \\
\textbf{Experiments} Section \ref{sec:microbenchmarks} evaluates \dejavulib with microbenchmars. We evaluate the \dejavu disaggregation policy, microbatch swapping, and fault-tolerance functionality in sections \ref{sec:disaggregation_results}, \ref{sec:swapping_results} and \ref{sec:ft_results} respectively.   \\
\textbf{Baselines} We compare \dejavu with FasterTransformer, as it is the state-of-the-art framework that supports pipeline parallelism for LLM inference.
FasterTransformer does not allow for requests in a batch to finish earlier. Additionally, in a pipeline parallel setup, a batch is split into microbatches, based on the number of pipeline stages. A new microbatch cannot be scheduled until all microbatches in the current batch have been completed at the last pipeline stage. This leaves GPUs at earlier stages idle, until all microbatches are done at the final stage.
We modified FasterTransformer to allow scheduling at the microbatch level. Whenever a microbatch completed in any stage, it can be replaced by the next available microbatch.

\subsection{Microbenchmarks}\label{sec:microbenchmarks}

\begin{figure}[htp!]
\centering
    \includegraphics[width=\linewidth]{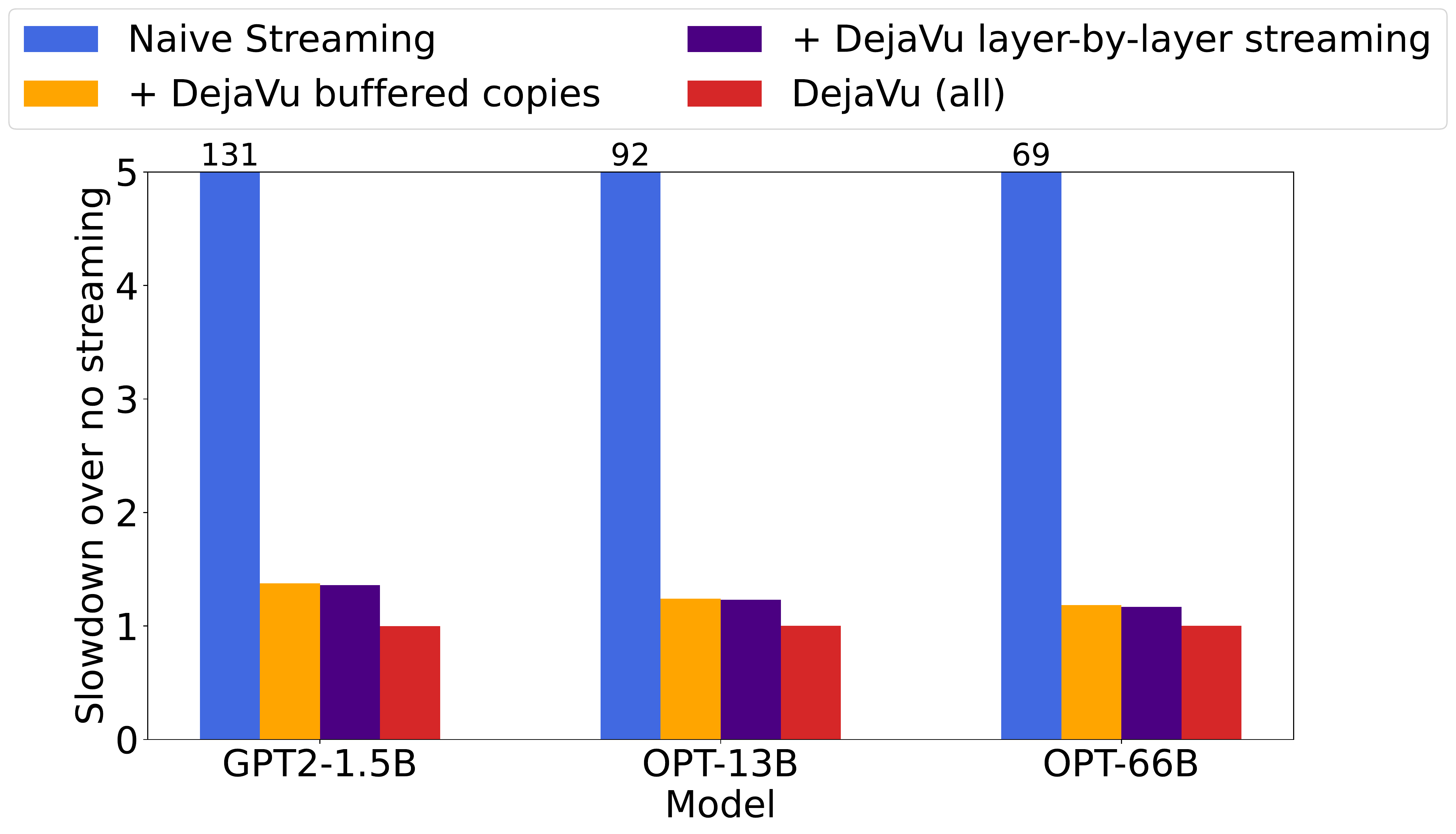}
\caption{Single-batch latency slowdown caused by KV streaming to remote CPU memory, when gradually applying the optimizations proposed by \dejavulib.}\label{fig:microbenchmark_breakdown}
\end{figure}

We evaluate the \dejavulib streaming mechanism, using requests with a prompt size of 500 tokens, generating 500 new tokens. We measure the time to complete a batch of requests, without streaming, and with streaming to local SSD, and remote CPU. We use only tensor parallelism when more than 1 GPU is employed (no pipeline parallelism). The \dejavulib streaming slowdown is within 2\% for local SSD and remote CPU memory (Appendix \ref{app_microbenchmarks}). The negligible overhead of \dejavulib is due to the optimizations described in \ref{sec:dejavulib}. Figure \ref{fig:microbenchmark_breakdown} provides a breakdown of the performance achieved by each of the optimizations. \textit{Baseline} stands for transferring all contiguous memory regions one by one. \textit{Buffered Copies} (Optimization (1) in \ref{sec:dejavulib}) has 95$\times$ improvement compared to baseline. The other two \dejavulib optimizations further improve streaming performance by 1.4$\times$.

\subsection{End-To-End Performance}\label{sec:e2e_results}

\subsubsection{Performance without failures}\label{sec:disaggregation_results}

\begin{figure}[htp!]
\centering
\subfloat[OPT-66B]{%
  \includegraphics[width=\linewidth]{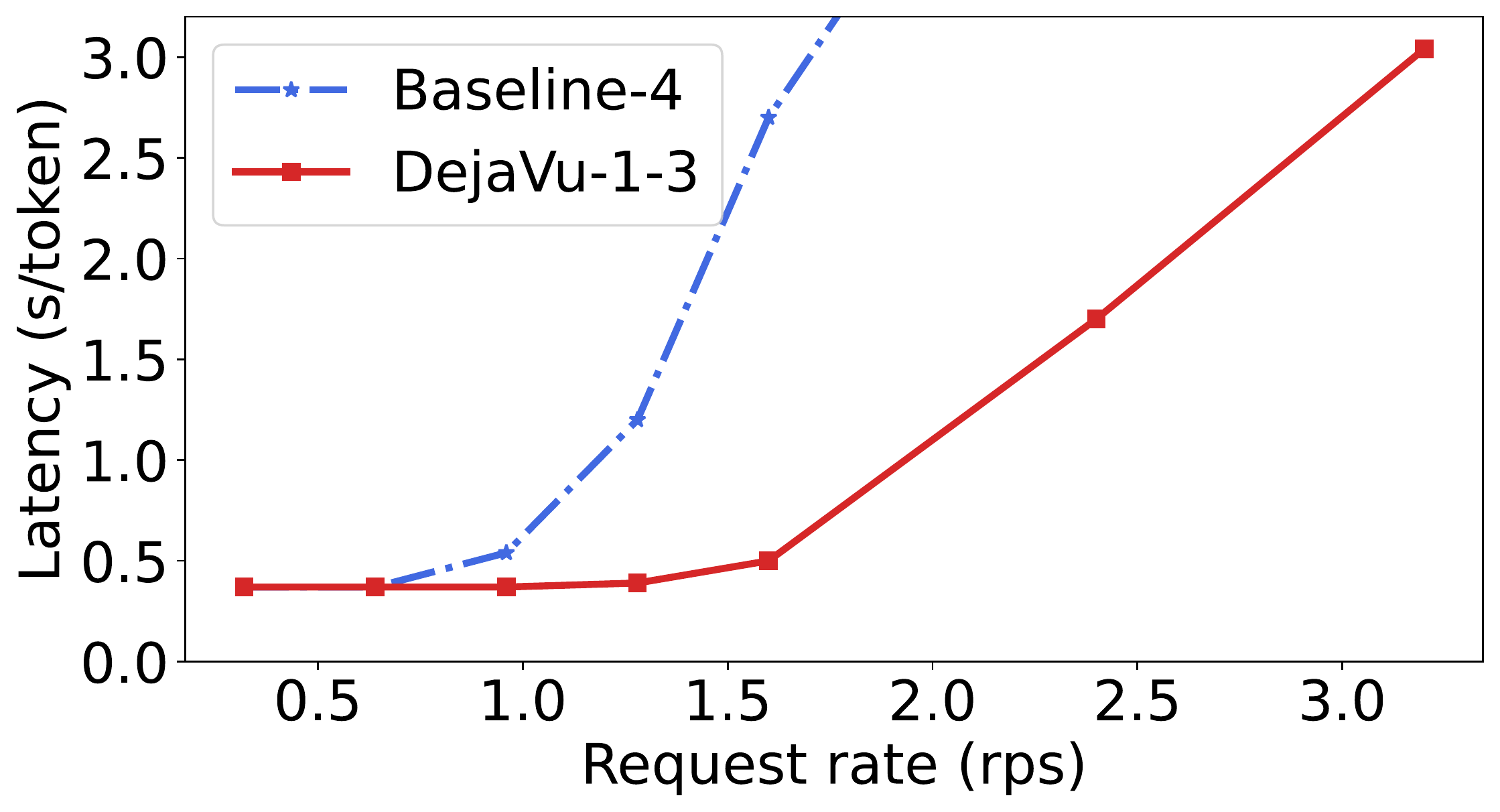}\label{fig:opt66-lmsys-dis-1000}%
}

\subfloat[BLOOM-176B]{%
  \includegraphics[width=\linewidth]{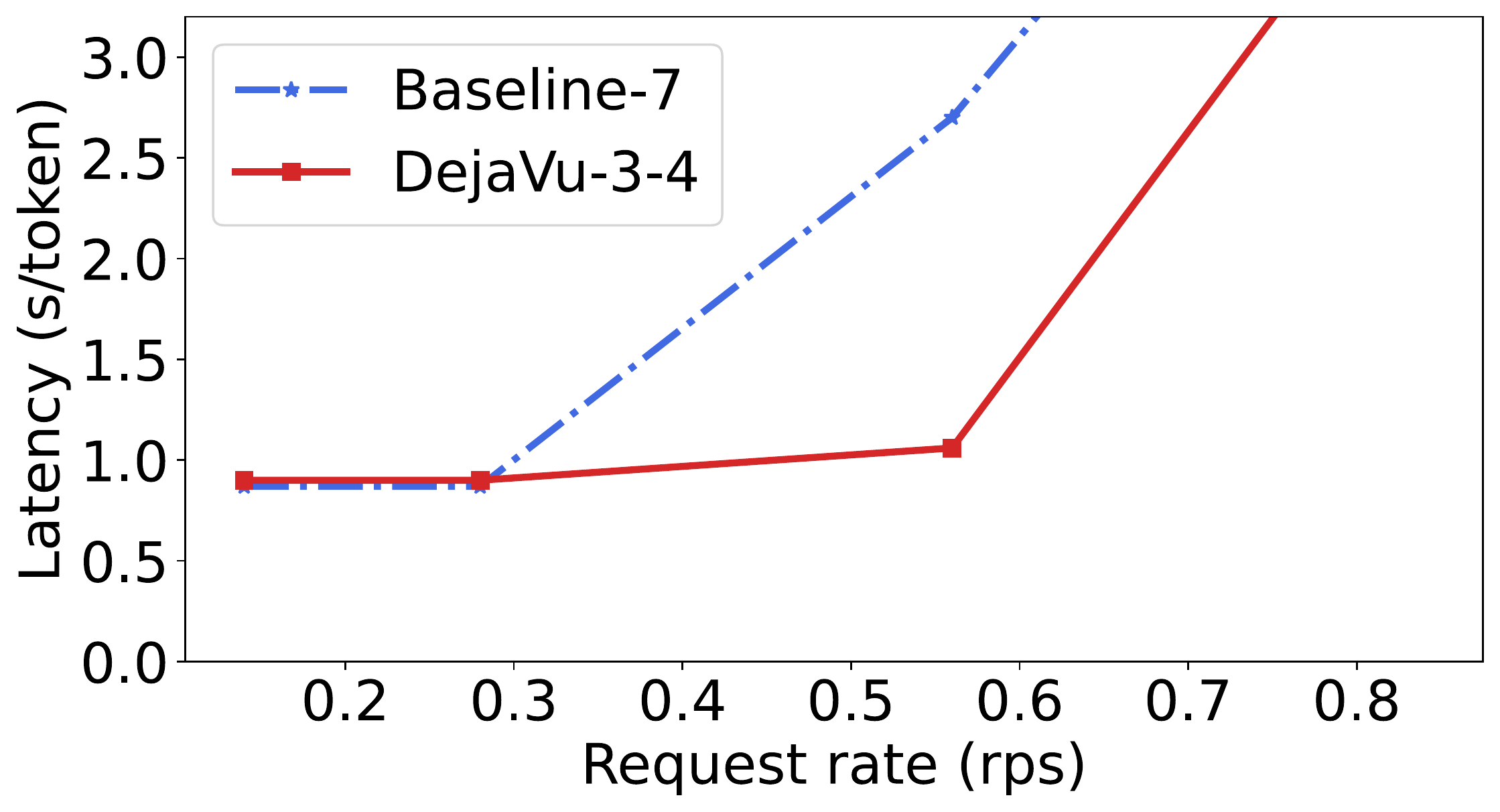}\label{fig:bloom_dis}%
}
\caption{E2E performance of the OPT-66B and BLOOM-176B model in the LMSys dataset. Baseline-X means that X machines were used in the pipeline. DejaVu-X-Y means that   X machines were used for prompt processing, and Y for token generation.}
\label{fig:dis}
\end{figure}

We now evaluate the performance of our disaggregated \dejavu system compared to the non-disaggregated baseline. We configure all our requests to a fixed prompt size (1000 tokens for Figure \ref{fig:dis}), and we sample the number of newly generated tokens from the LMSys dataset~\cite{zheng2023lmsyschat1m}, assuming all requests within a microbatch generate the same number of tokens. We use one client, which submits requests following a Poisson distribution in an open loop, with varying request rates. Similarly to Orca~\cite{GyeongIn2022Orca} and vLLM~\cite{Kwon2023VLLM} we report normalized latency (seconds/token) for each request rate. To compute the normalized latency for each request, we divide its end-to-end latency by the number of generated tokens.  Figure \ref{fig:dis} 
shows the median normalized latency for OPT-66B and BLOOM-175B. Since \dejavu targets pipeline parallel setups, we employ pipeline parallelism using multiple machines with a few GPUs each. Each pipeline stage is a VM with 2 GPUs running tensor model parallelism.  The legends in Figure \ref{fig:dis} show the pipeline parallelism depth.

As we increase the input request rate, the normalized latency increases, due to the systems' inability to maintain that high request rate, leading to queueing effects. \dejavu sustains low latency with up to 1.88$\times$, and 2$\times$ higher throughput than FasterTransformer baseline for the OPT-66B and BLOOM-176B models respectively. Since microbatches generate a variable number of tokens, new prompts are being injected, introducing bubbles in the pipeline of the baseline case, where all machines are dedicated to both prompt and token processing. \dejavu addresses this issue by allocating separate pipelines for prompt and token processing. Our planner selects the number of prompt processing and token generation pipelines as explained in section \ref{sec:disaggregation_impl}, to ensure that the token generation machines do not idle waiting for prompts to be processed. Disaggregation benefits are more noticeable with larger prompt sizes. Larger prompts result in extended prompt processing time, leading to larger bubbles in the baseline case, thereby downgrading performance. Despite the larger amount of data that needs to be streamed from the prompt processing to token generation machines with larger prompt sizes, \dejavu streaming optimizations (sec \ref{sec:dejavulib}) manage to fully hide the prompt streaming overhead. 

In appendix \ref{app_planner} we use our planner and simulator to evaluate various scenarios. Overall, we observe that \dejavu scales better than the baseline, leading to shorter makespan and cost for a given trace.

\subsubsection{Performance with microbatch swapping}\label{sec:swapping_results}

\begin{figure}[htp!]
\centering
    \includegraphics[width=\linewidth]{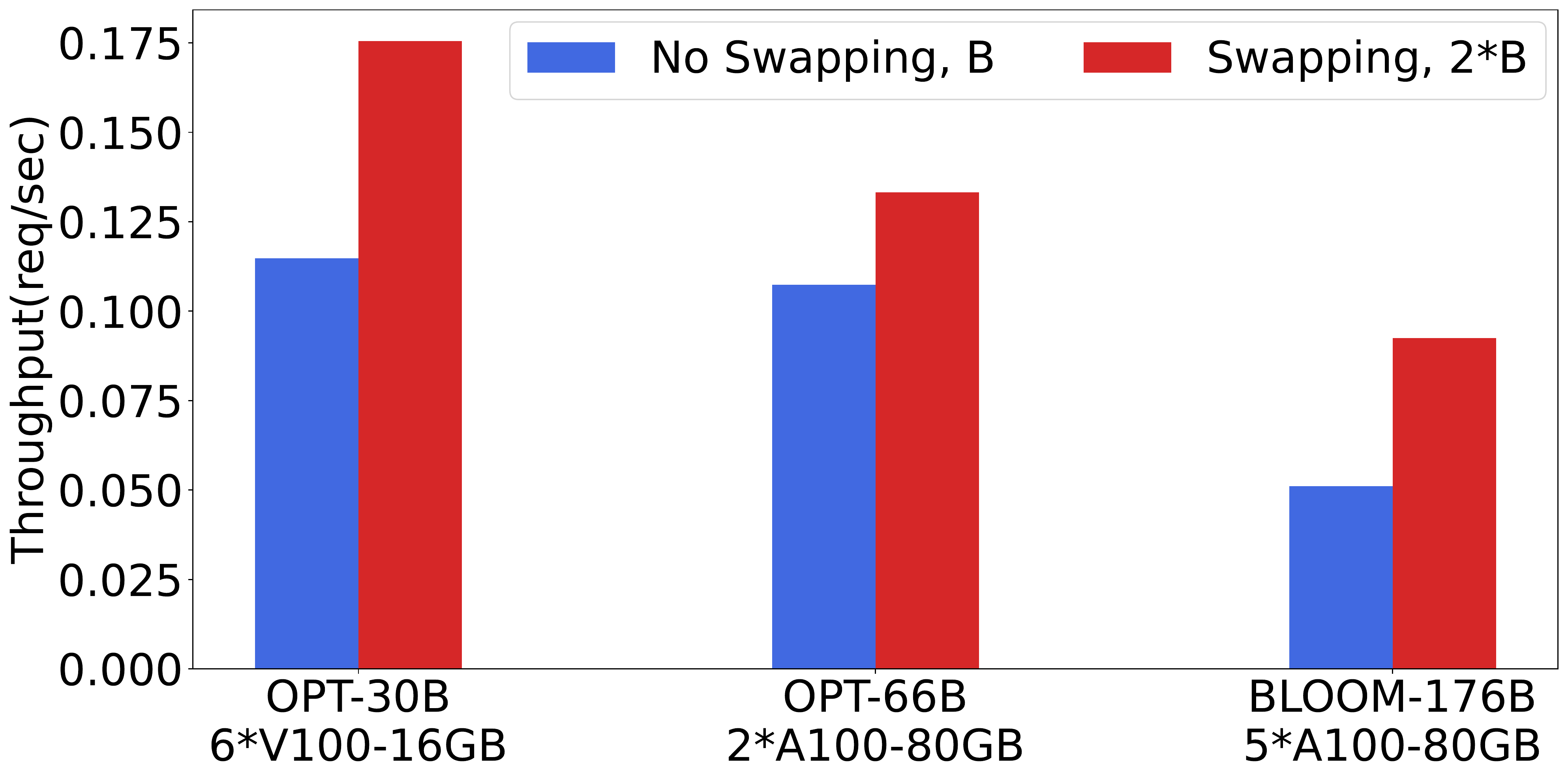}
    \caption{Benefit of microbatch swapping}\label{fig:swapping-2b}
\end{figure}

Swapping reduces the amount of GPU memory required for the KV cache, allowing larger batch sizes, and increasing system throughput. Figure \ref{fig:swapping-2b} illustrates this. For each model and set of GPUs (x-axis), we get the achieved system throughput with the largest feasible batch size without swapping $B$, and the achieved throughput with swapping enabled and batch size $2 \cdot B$. By accommodating larger batch size, we increase throughput by up to 1.8$\times$. However, the main bottleneck of swapping is the time needed to bring the KV cache back into the GPU. This depends on the size of the cache and the CPU-GPU PCIe bandwidth. Since each token generation step takes 10s-100s ms, the KV cache transferring should also be very fast. In Appendix \ref{app_swapping}, we formalize the benefits of microbatch swapping and evaluate the mechanism varying the sequence length and batch size.

\subsubsection{Performance with failures}\label{sec:ft_results}
\begin{figure}[t]
\centering
    \includegraphics[width=\linewidth]{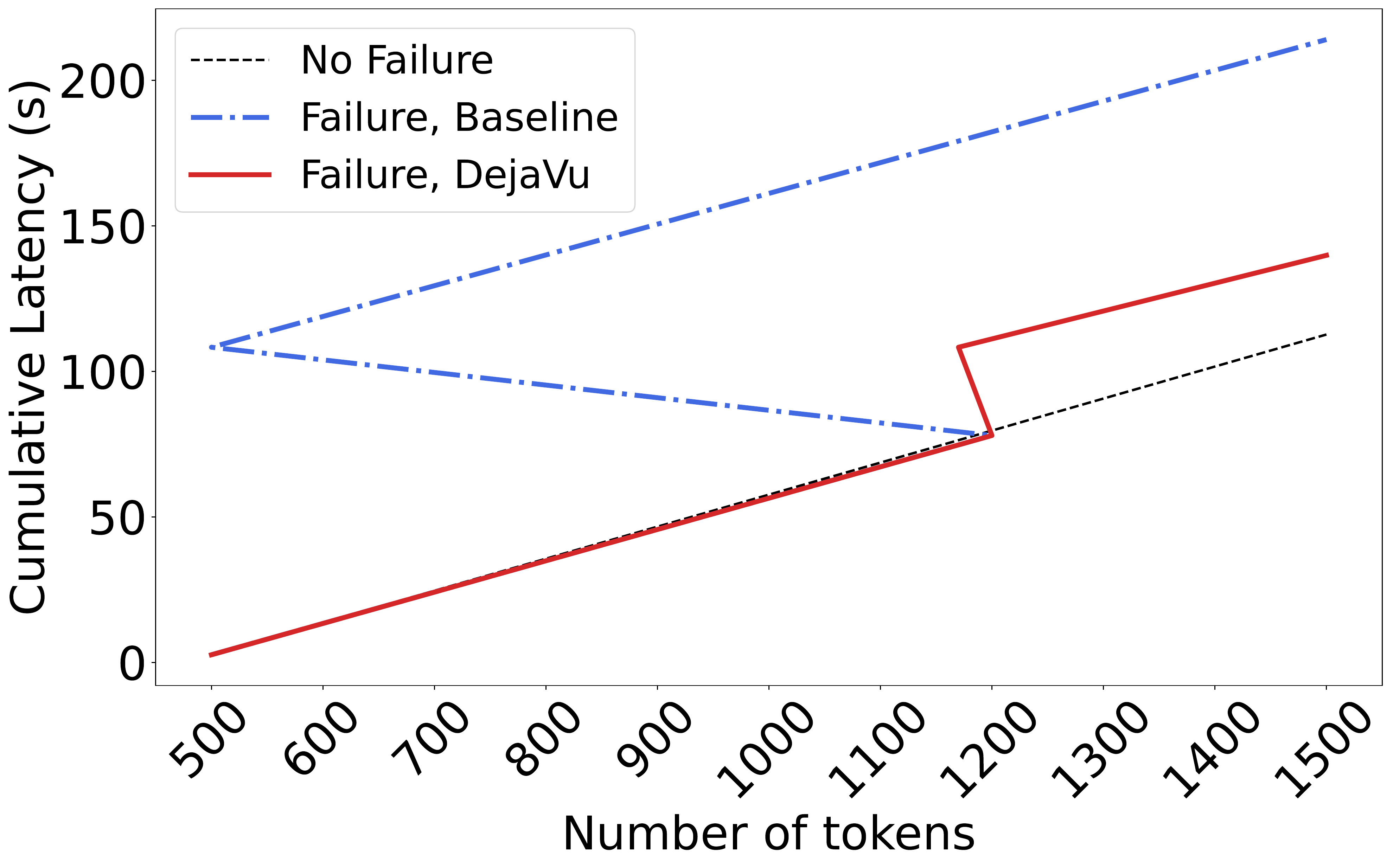}
    \caption{Effect on cumulative latency for a single microbatch when a failure occurs at token generation step 1200. }\label{fig:single-failure}
\end{figure}

In this section, we evaluate the performance of \dejavu in the event of failures. We serve the OPT-66B model in a cluster of 4 machines, using pipeline parallelism with 4 stages. Each stage in the pipeline does both prompt processing and token generation.
%i.e. we do not apply prompt-token disaggregation.
We use one client that submits homogeneous requests to the workers. Each request has a prompt size of 500 tokens and generates 1000 extra tokens. 
%In Figure \ref{fig:single-failure}, we show a larger-scale representation of the toy example presented in Figure \ref{fig:failure}. 
We incur a pipeline stage failure at token generation step 1200. Figure \ref{fig:single-failure} depicts the cumulative latency of one of the active microbatches at the moment of failure. A single failure led to 1.91$\times$ increase in the latency of a set of microbatches. In contrast, with \dejavu the increase due to failure is 1.24$\times$. 

\begin{figure}[t]
\centering
    \includegraphics[width=\linewidth]{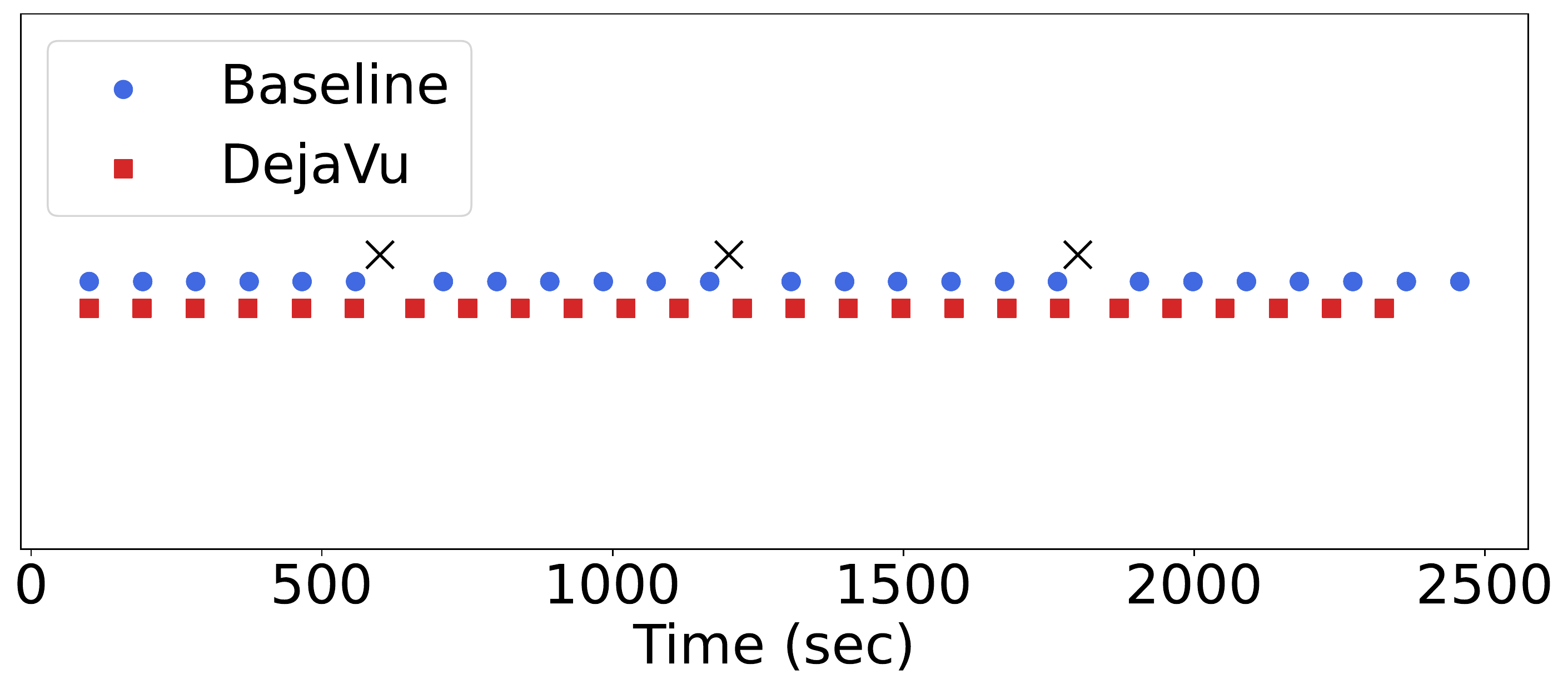}
    \caption{Request completions over time. We introduce failures after 600, 1200, and 1800 sec (marked with black X). }\label{fig:trace-failure}
\end{figure}

In Figure \ref{fig:trace-failure} we introduce failures at various timestamps while serving a set of requests. In the case of baseline, all workers need to restart, and the processing of the microbatches that were active at the point of failure starts from scratch. With \dejavu, due to the lightweight cache streaming protocol, token generation just restarts from the latest replicated step, leading to 1.16$\times$ shorter runtime.

%% file: content/relatedwork.tex
\section{Related Work}
\textbf{Serving Systems for LLMs} The widespread adoption of LLMs led to multiple LLM serving systems, such as FasterTransformer~\cite{NVIDIA21FasterTransformer}, TensorRT-LLM~\cite{NVIDIA21TensorRT}, and DeepSpeed Inference~\cite{aminabadi2022deepspeed}. Orca~\cite{GyeongIn2022Orca} introduces iteration-level scheduling allowing requests at a batch to be at different phases (prompt processing or token generation), but overlooks the potential negative impact on throughput caused by the difference between prompt processing and per-token generation time. We are working on integrating iteration-level scheduling in \dejavu. 
vLLM~\cite{Kwon2023VLLM} reduces KV cache overprovisioning by using dynamic memory allocation, and swapping KV cache blocks to the CPU under GPU memory pressure for individual requests. FlexGen~\cite{sheng2023flexgen} proposes a mechanism to serve LLMs with limited GPU memory, leveraging CPU memory and disk, and swapping model weights and cache values as needed. \footnote{Swapping has also been used in ML training to deal with limited GPU memory~\cite{Li2022Harmony, Rajbhandari21Zero}}
In contrast to these works, \dejavu targets pipeline parallel inference and employs swapping at the level of microbatches. Recently, systems for co-scheduling and batching multiple LoRA models have been proposed~\cite{chen2023punica, sheng2023slora}. H2O~\cite{zhang2023h2o} and LESS~\cite{dong2024less} observe sparsity in the KV cache and aim to evict KV cache entries (thus reducing KV cache size) without harming inference quality. These works are complementary to \dejavu.

\textbf{Differences in prompt and token processing}  Some recent works, developed concurrently with \dejavu, also address the discrepancy in prompt and token processing times. Sarathi~\cite{agrawal2023sarathi} proposes partitioning prefill requests into smaller chunks and merging them with decodes. Splitwise~\cite{patel2023splitwise} and  DistServe~\cite{zhong2024distserve} propose separating prompt from token processing. Splitwise~\cite{patel2023splitwise} employs disaggregation to reduce power consumption and cost, by using heterogeneous GPUs for each phase independently.  Splitwise is primarily simulation-based and supports execution modes with limited parallelism; models fit on a single GPU or run tensor-model parallel across GPUs, and they do no consider pipeline parallel serving (required for recent massive LLMs). DistServe employs distinct batching and parallelism for prompt processing and token generation, based on model characteristics and simulation findings. \dejavu uses disaggregation to minimize bubbles in pipeline parallel setups and optimizes machine allocation for prompt and token pipelines to maximize system throughput.

\textbf{LLM serving on preemptible resources} SpotServe~\cite{miao2023spotserve} is a framework for serving LLMs over spot cloud resources. SpotServe utilizes the grace period (e.g. 30 sec in AWS) before a VM is preempted, to optionally migrate KV cache contents to avoid restarting inference from scratch. However, this approach cannot protect from sudden failures which can harm request latency (see Figure \ref{fig:single-failure}). In contrast, \dejavu uses a token-level KV cache replication strategy with minimal overhead, offering continuous fault tolerance and seamless recovery from any failure.

Overall, \dejavu stands out by being comprehensive in addressing the key LLM-serving challenges we highlight in the paper; \dejavulib, its KV cache streaming library, is designed for high-performance and versatility to address these challenges. 
\dejavu offers high-throughput, fault-tolerance, and seamless recovery upon failures, as well as opportunities for memory savings in LLM inference. 

%% file: content/conclusion.tex
\section{Conclusion}

\dejavu is a system for efficient and fault-tolerant LLM serving at scale. It decouples prompt processing from token generation to mitigate pipeline bubbles caused by the differences in prompt processing and per-token generation times. Additionally, it optimizes memory utilization in pipeline parallel setups by implementing microbatch-level swapping to and from CPU memory. Finally, it employs cache replication and failure handling mechanisms to provide seamless recovery and minimize redundant work in the event of failures. \dejavu leverages \dejavulib, a modular library that allows for KV cache streaming under various setups with minimal overhead. \dejavu improves LLM serving throughput by up to 2$\times$ compared to state-of-the-art systems.

%% file: content/appendix.tex
\newpage

\appendix
\onecolumn

\section{Prompt processing and token generation time}\label{app_llm_profiling}

\begin{figure}[htp!]
\centering
    \includegraphics[scale=0.175]{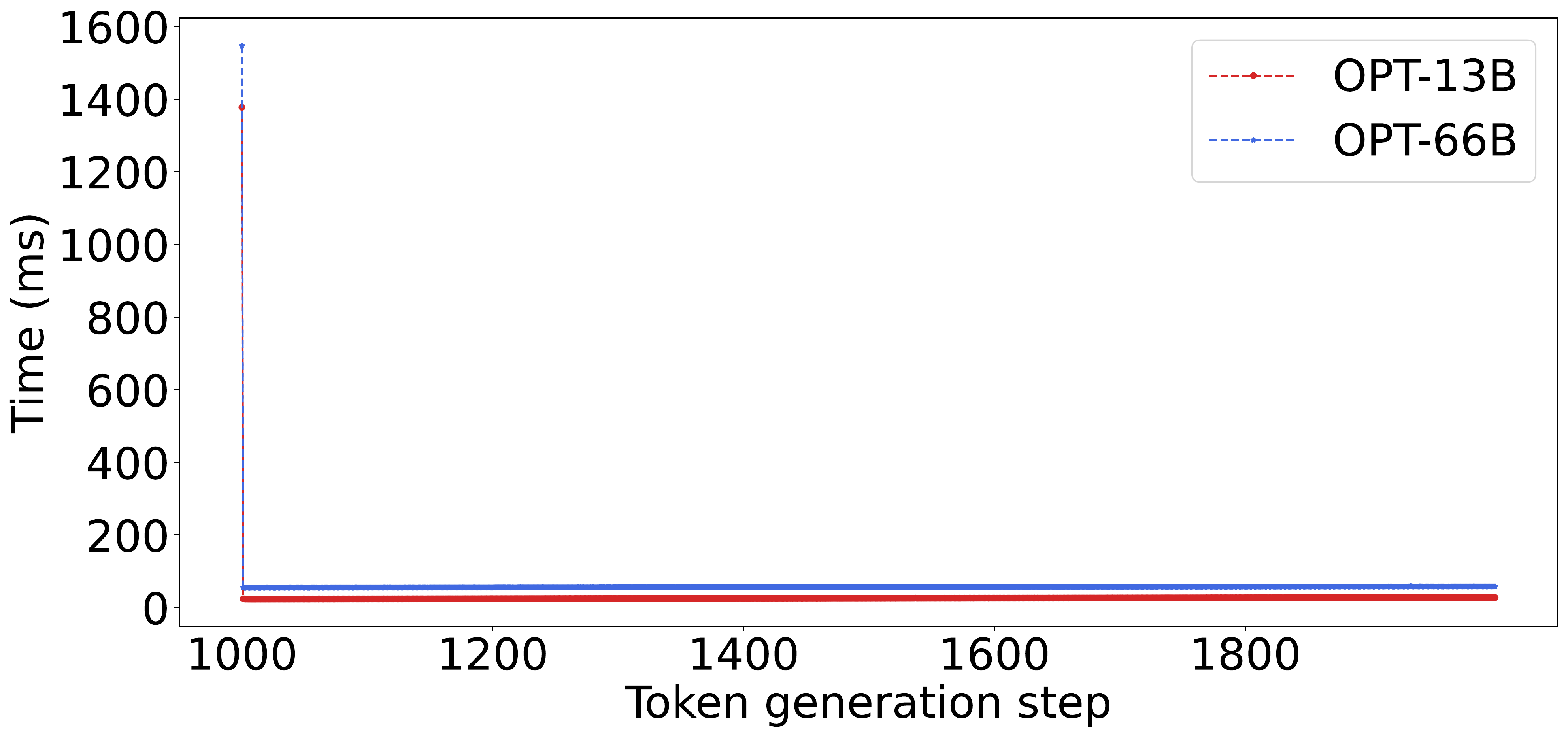}
    \caption{Prompt processing and token generation time for OPT-13B and OPT-66B serving with batch size 8}\label{fig:prompt_token_time}
\end{figure}

Figure \ref{fig:prompt_token_time} shows the per-token generation time when serving the OPT-13B and OPT-66B model, with prompt size 1000, and generating 1000 extra tokens. We observe that the time to generate the first token (i.e. prompt processing time) is much higher than the time to generate each subsequent token (which is nearly constant).

\begin{figure*}[htp!]
\subfloat[Batch size 1]{%
  \includegraphics[width=0.5\textwidth]{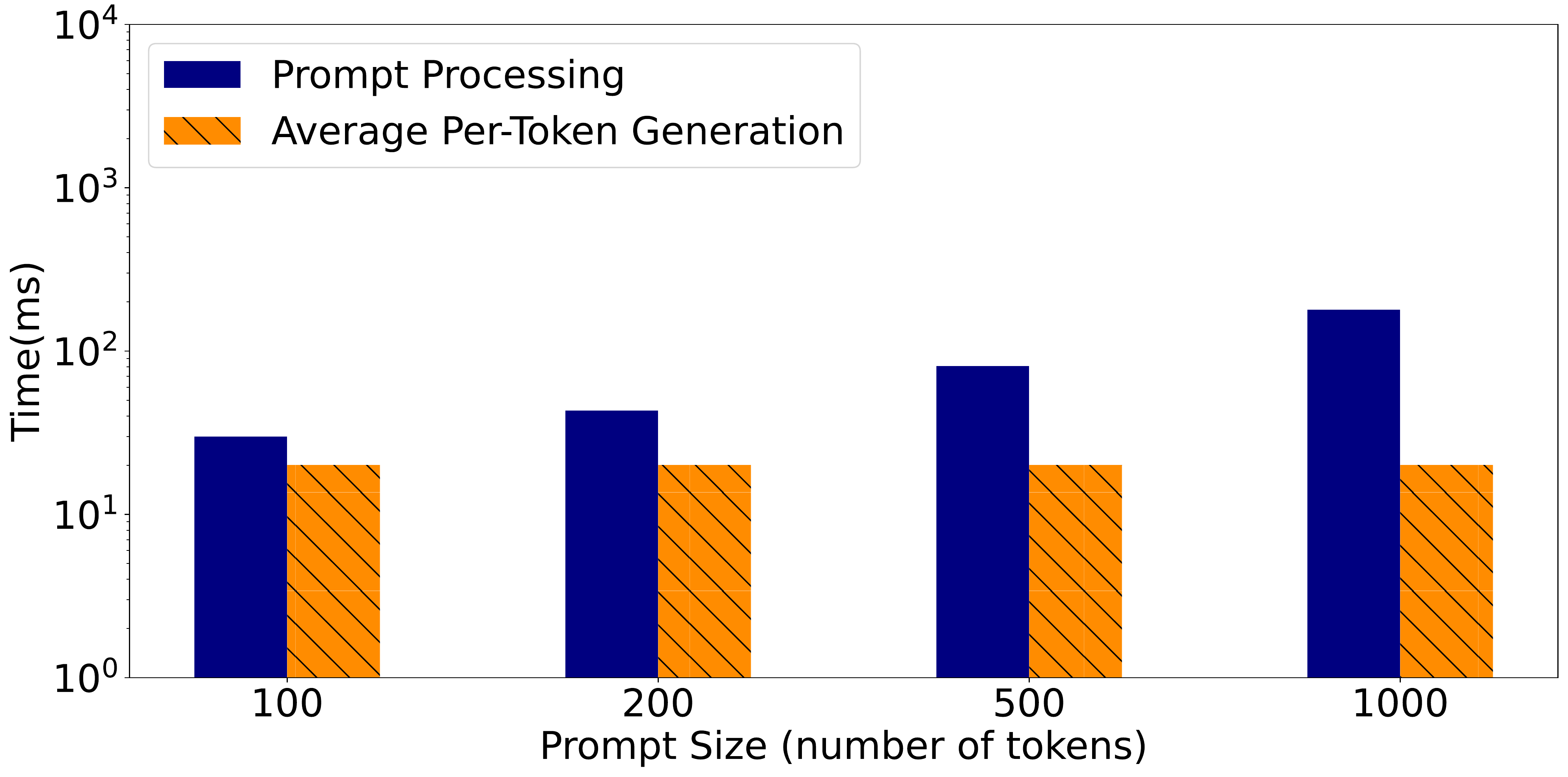}%
}
\subfloat[Batch size 2]{%
  \includegraphics[width=0.5\textwidth]{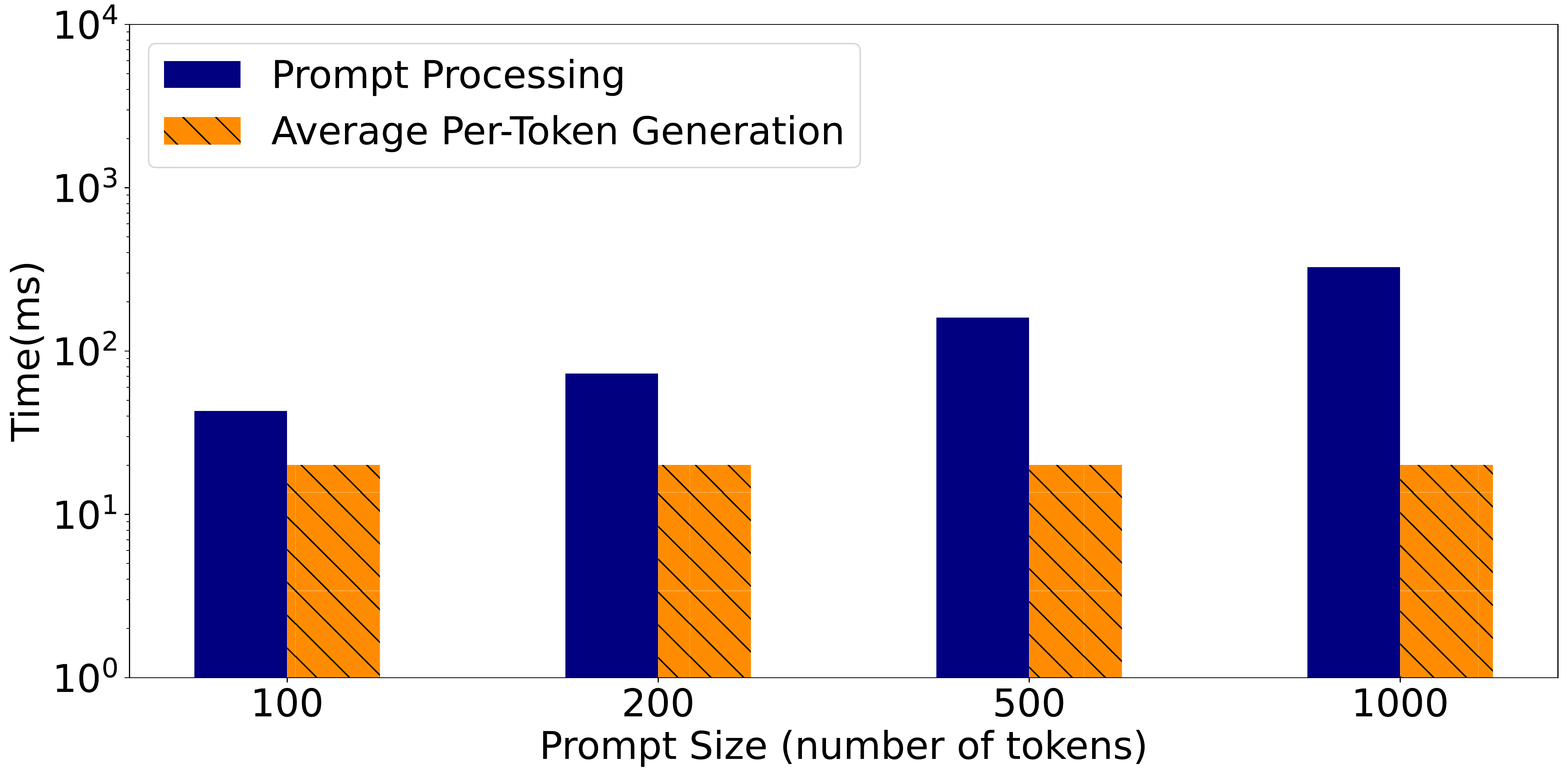}%
}

\subfloat[Batch size 4]{%
  \includegraphics[width=0.5\textwidth]{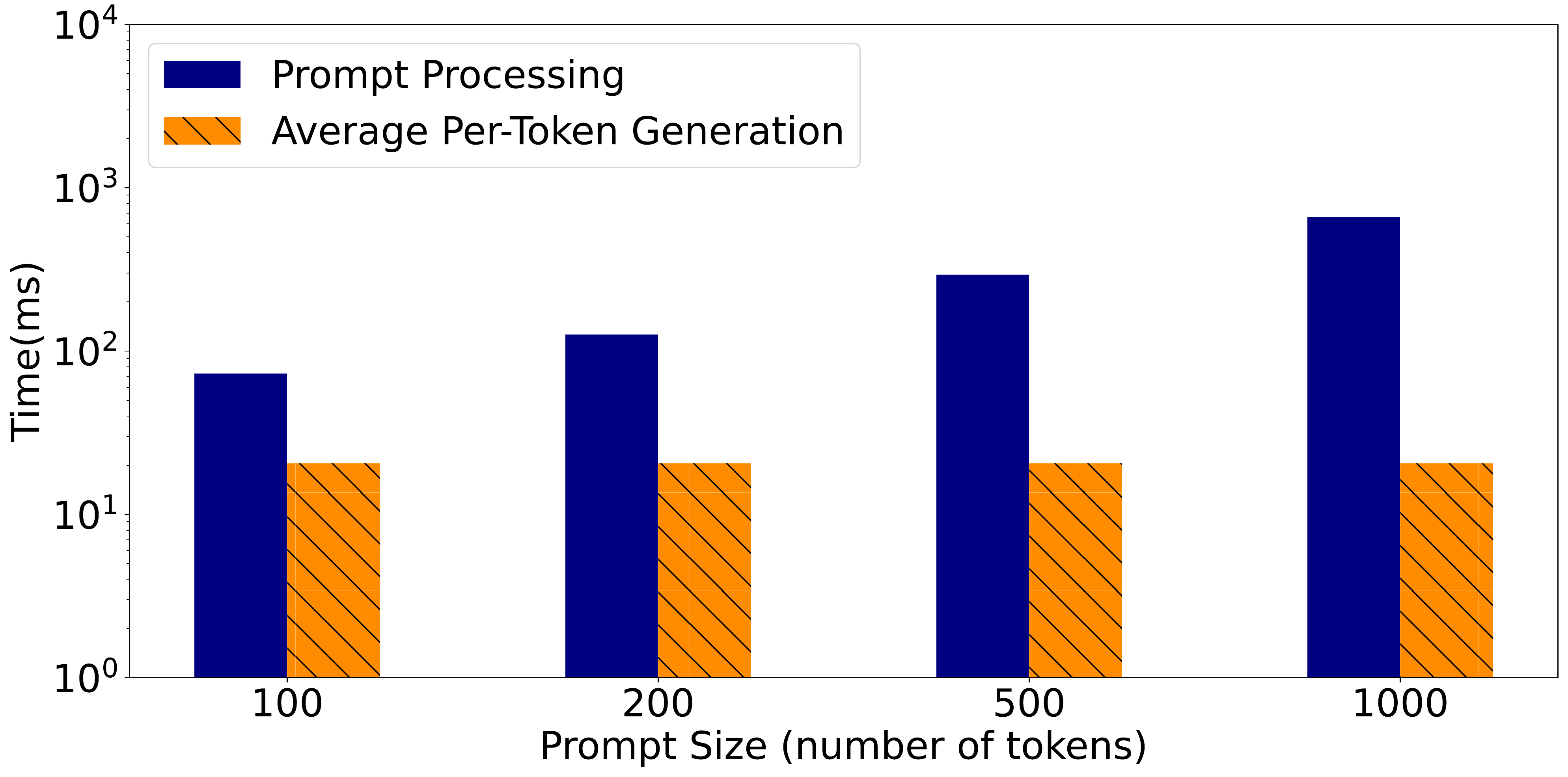}%
}
\subfloat[Batch size 8]{%
  \includegraphics[width=0.5\textwidth]{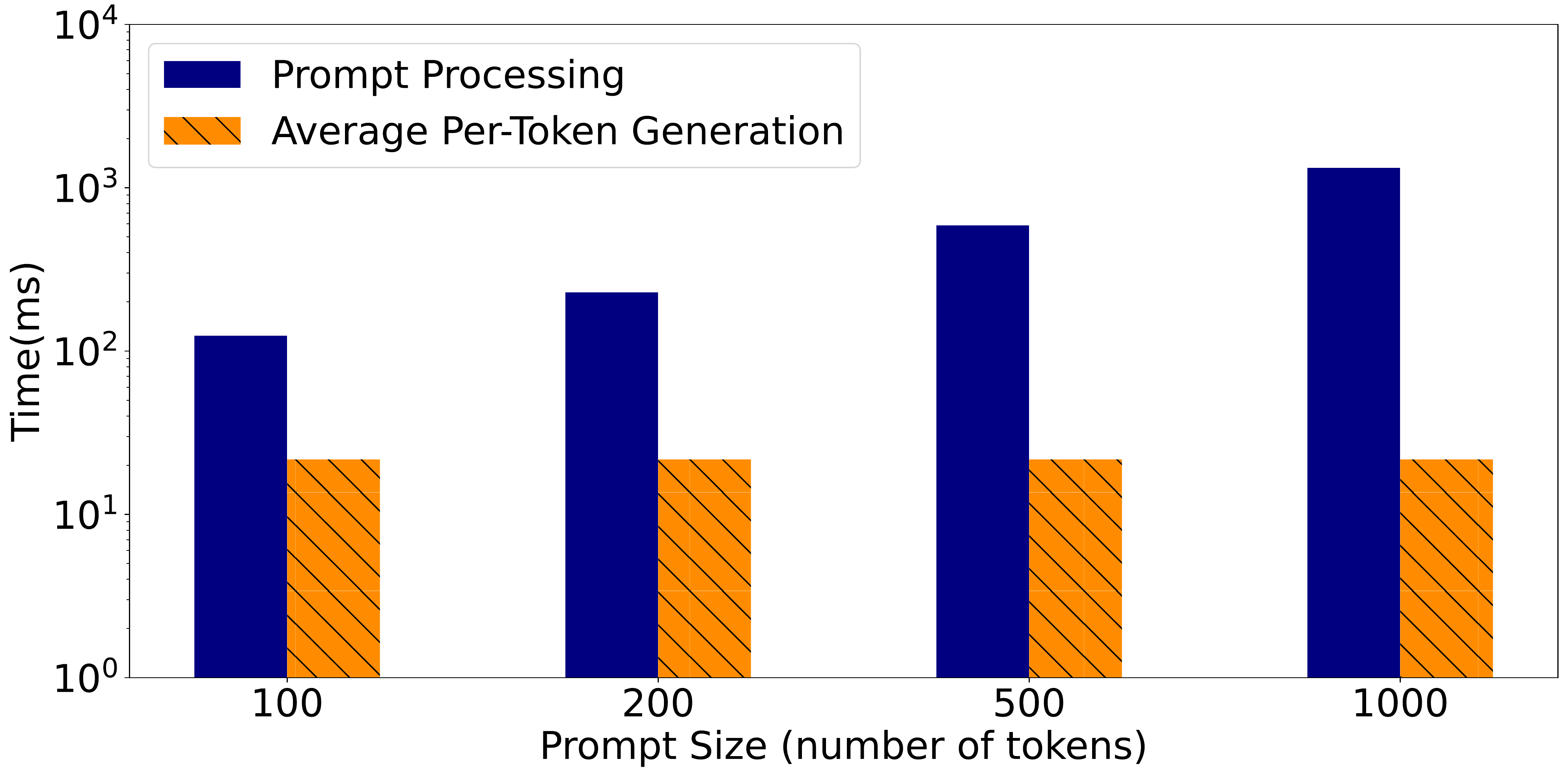}%
}
\caption{Prompt processing and token generation times for the OPT-13B model}
\label{fig:prompt_token_opt13}
\end{figure*}

\begin{figure*}[htp!]
\subfloat[Batch size 1]{%
  \includegraphics[width=0.5\textwidth]{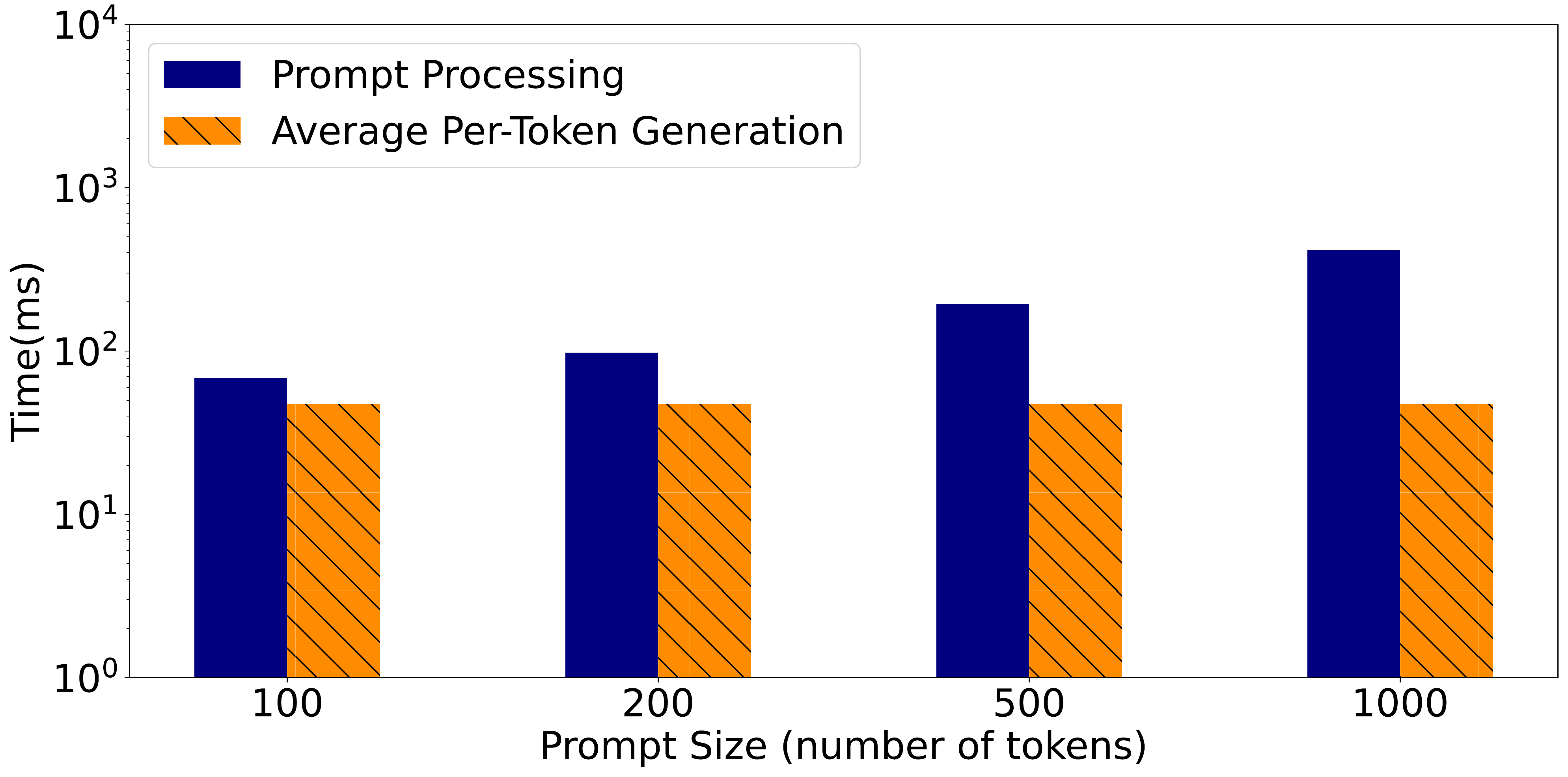}%
}
\subfloat[Batch size 2]{%
  \includegraphics[width=0.5\textwidth]{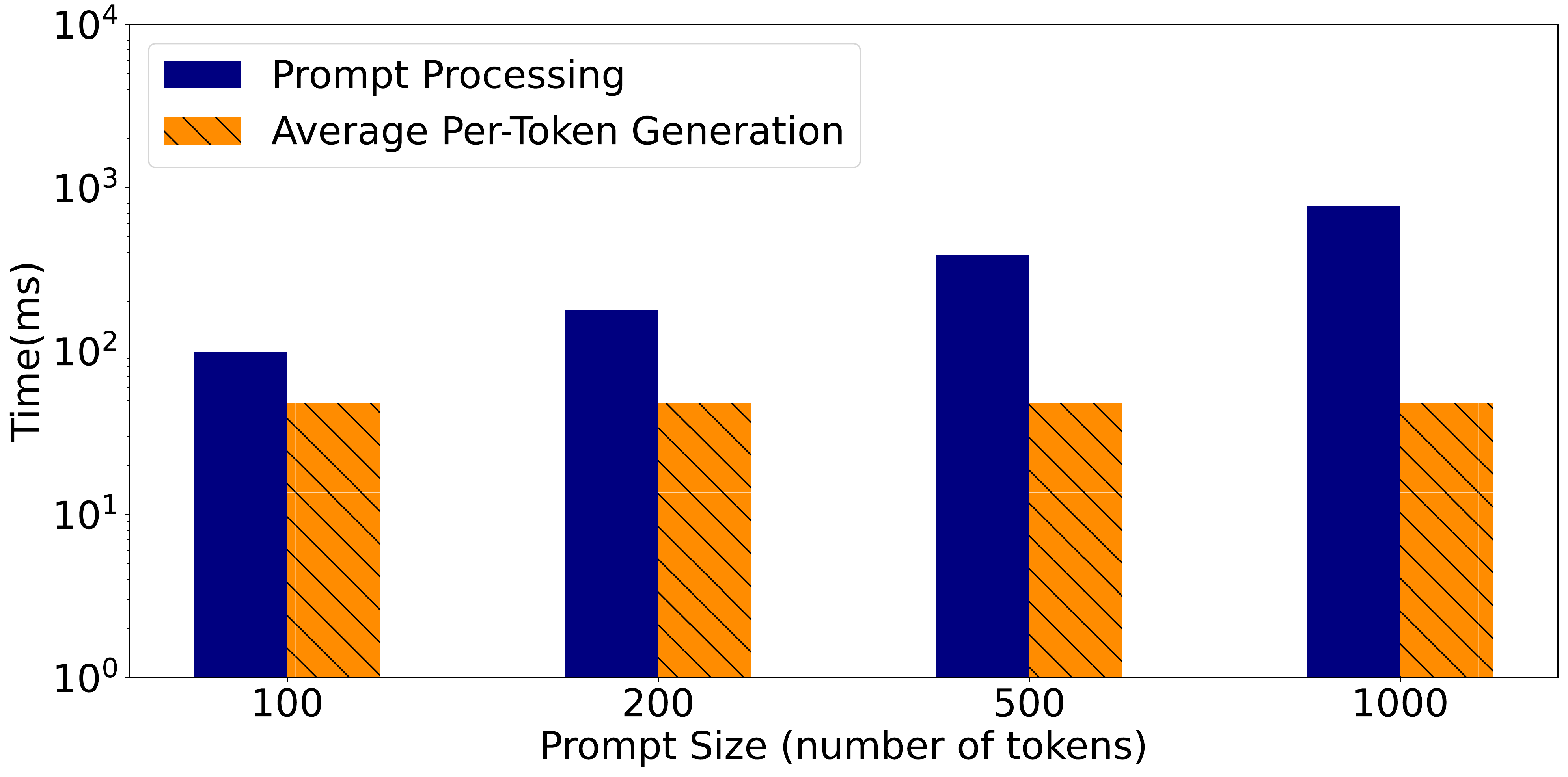}%
}

\subfloat[Batch size 4]{%
  \includegraphics[width=0.5\textwidth]{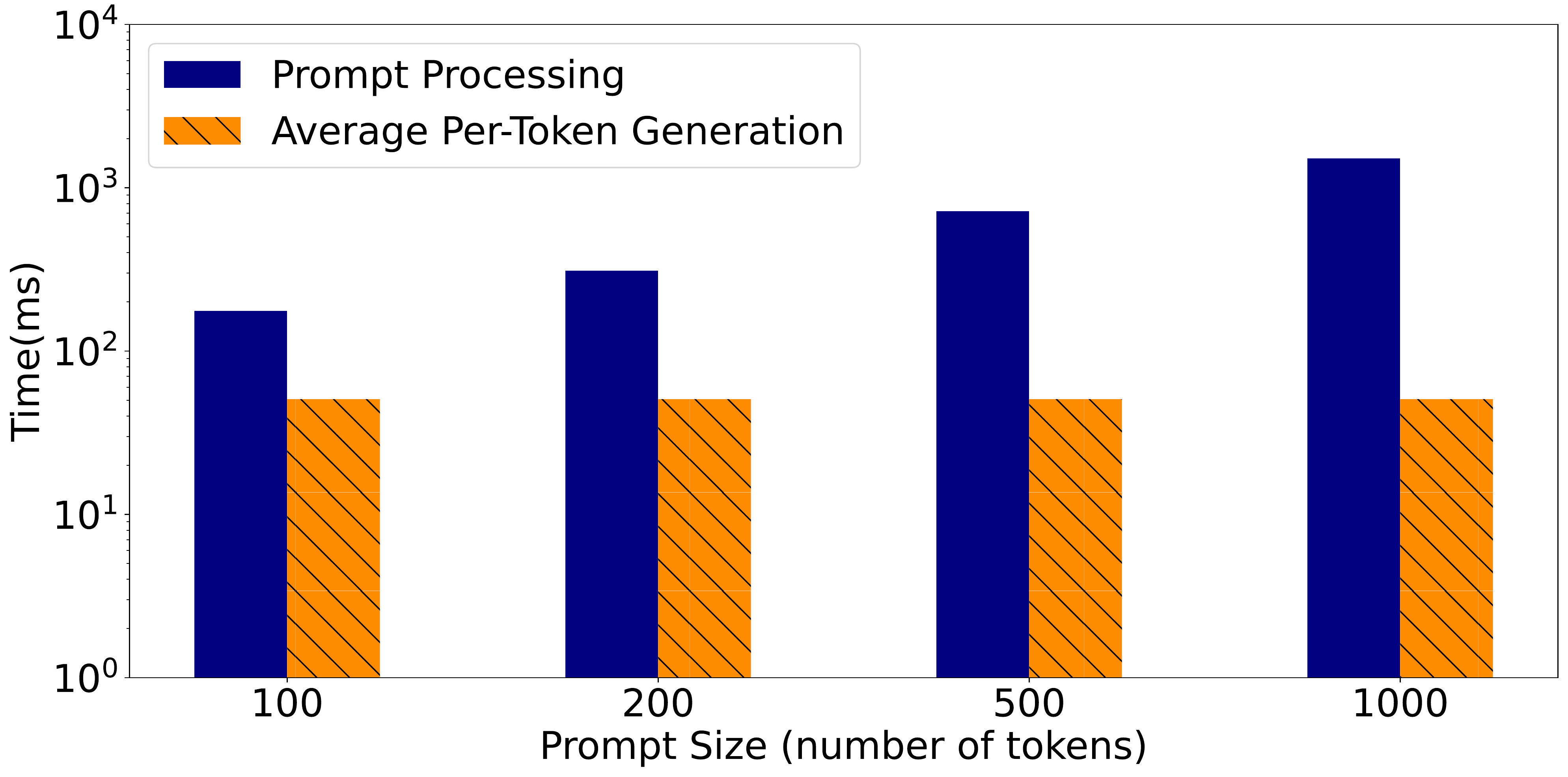}%
}
\subfloat[Batch size 8]{%
  \includegraphics[width=0.5\textwidth]{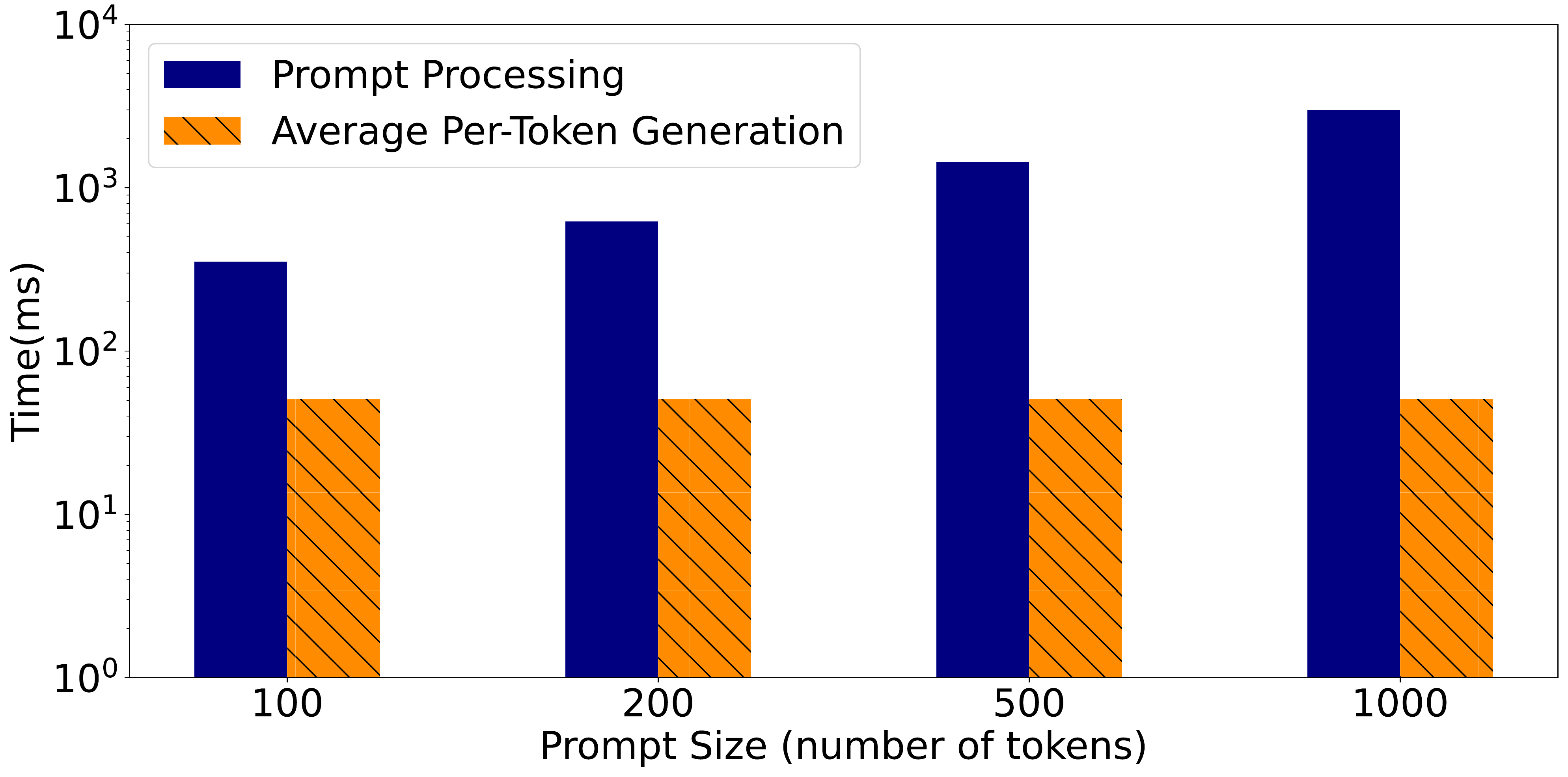}%
}
\caption{Prompt processing and token generation times for the OPT-66B model}
\label{fig:prompt_token_opt66}
\end{figure*}

\begin{figure*}[htp!]
\subfloat[Batch size 1]{%
  \includegraphics[width=0.5\textwidth]{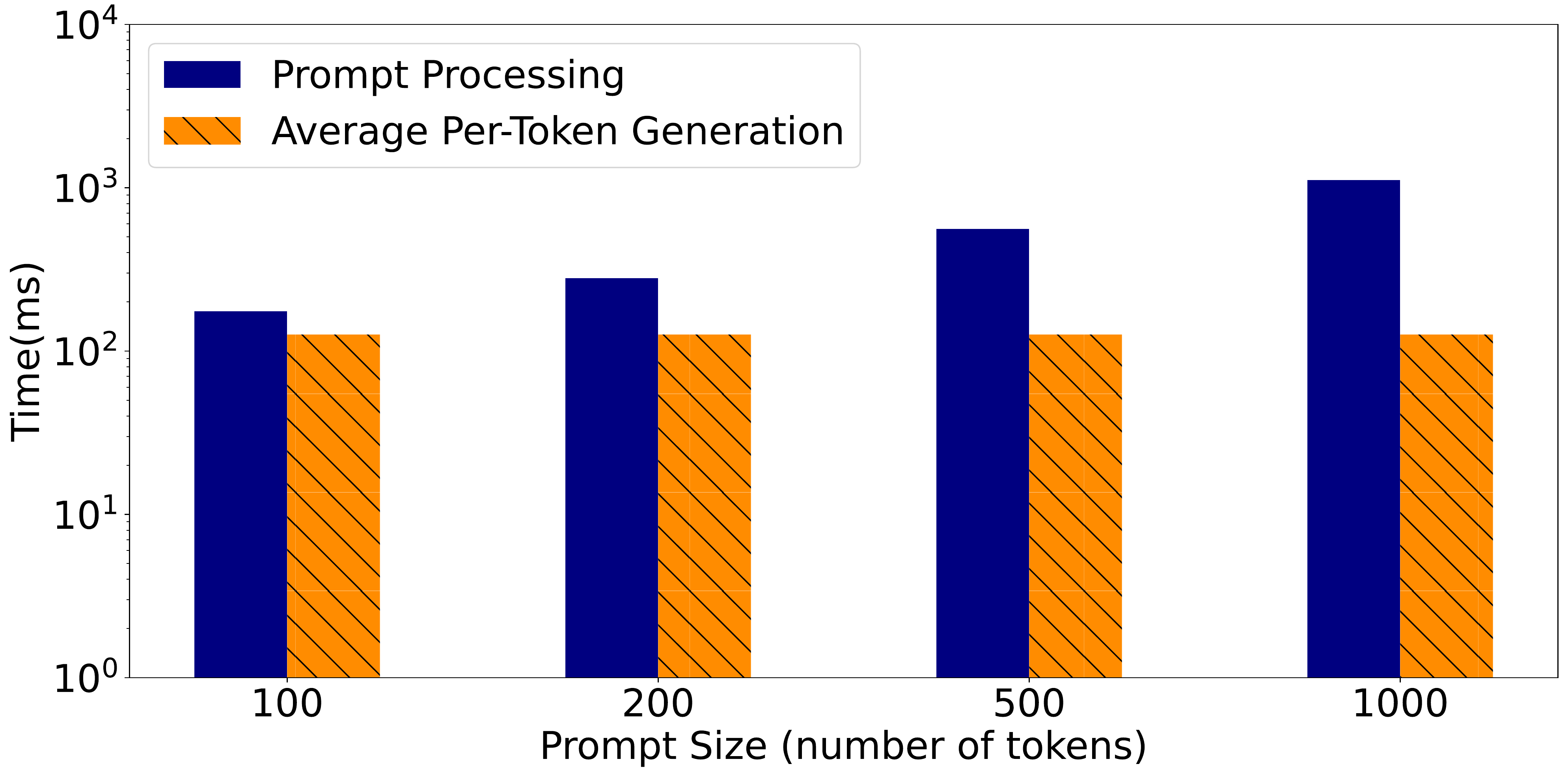}%
}
\subfloat[Batch size 2]{%
  \includegraphics[width=0.5\textwidth]{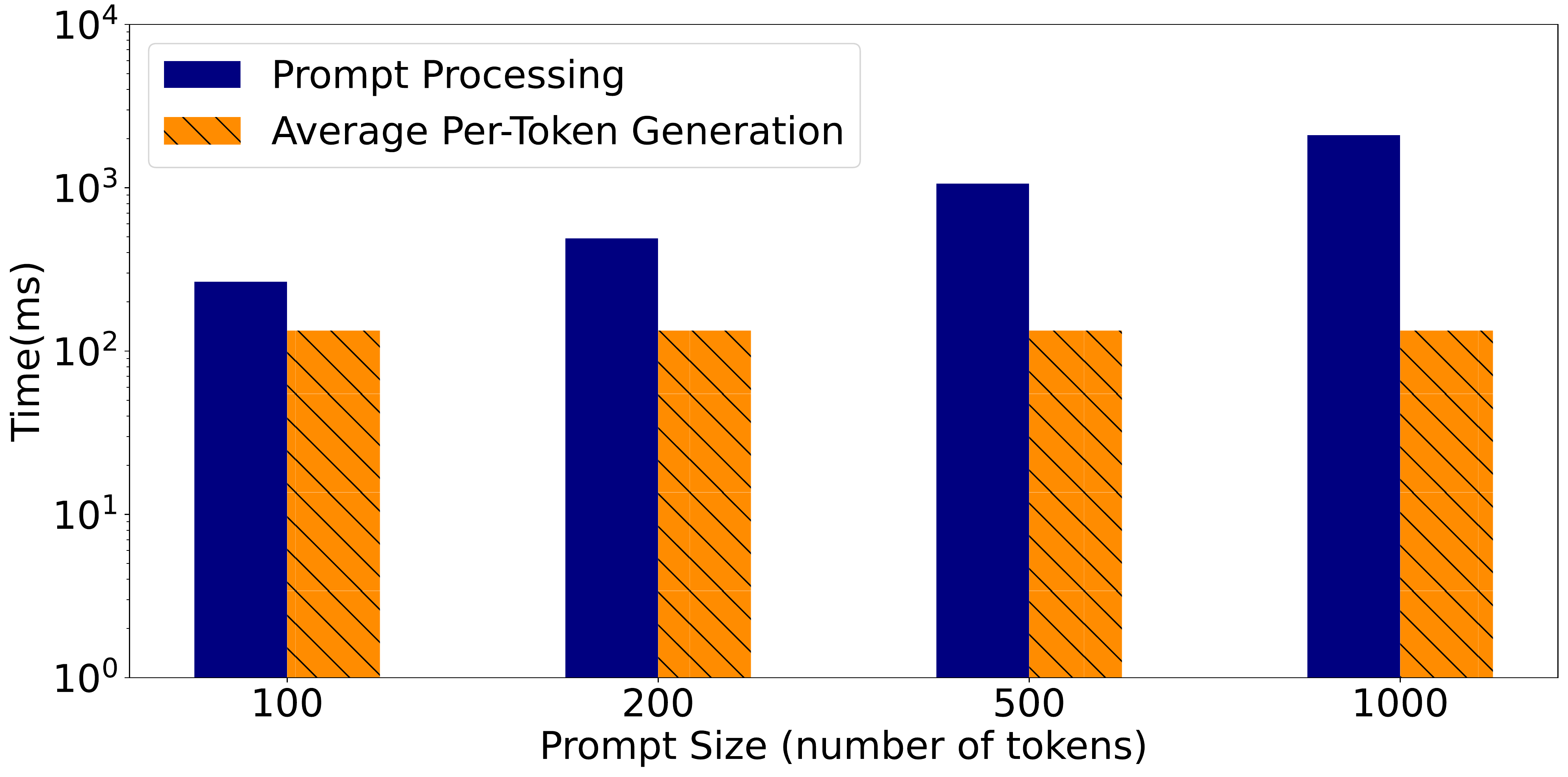}%
}

\subfloat[Batch size 4]{%
  \includegraphics[width=0.5\textwidth]{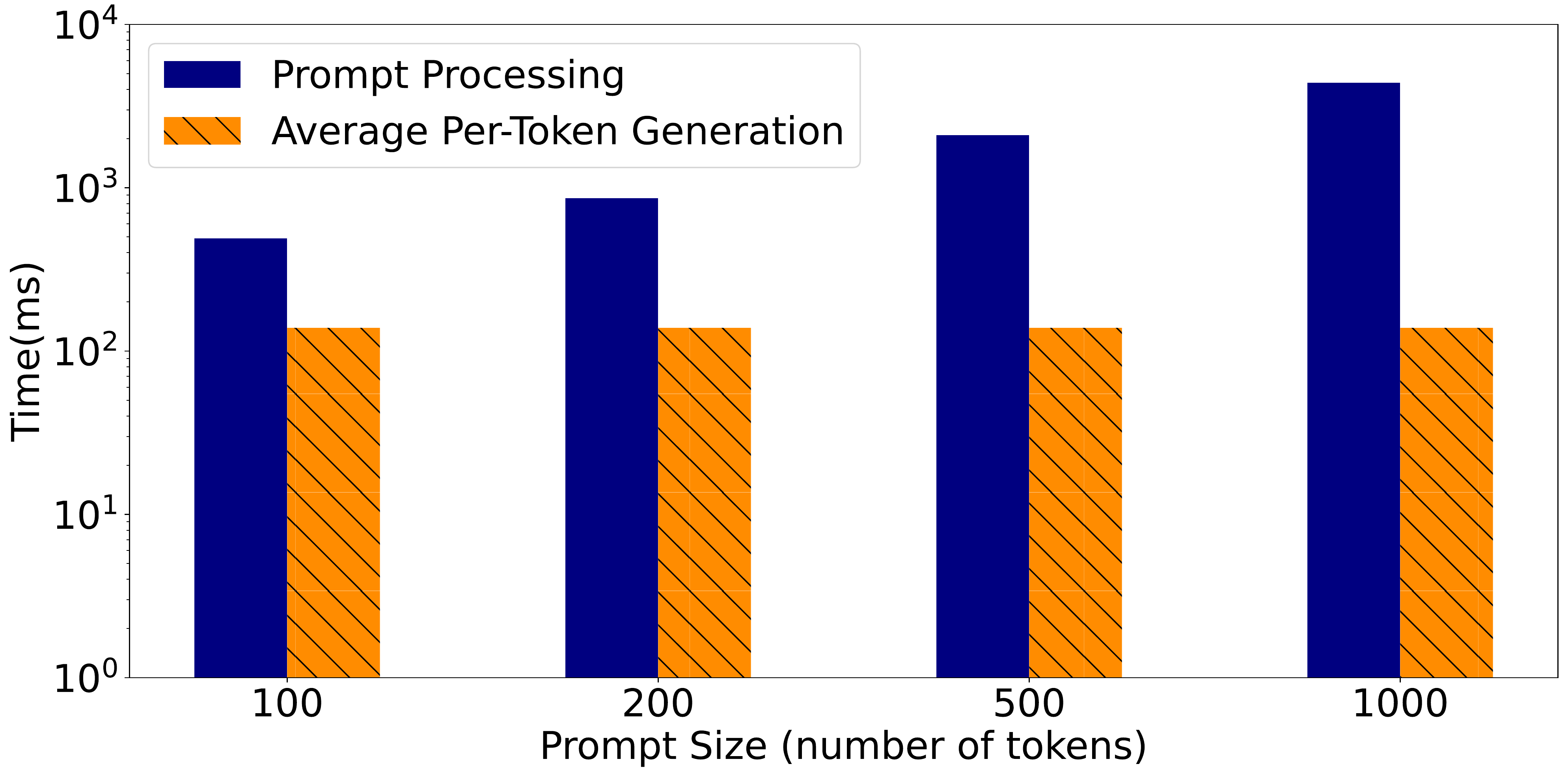}%
}
\subfloat[Batch size 8]{%
  \includegraphics[width=0.5\textwidth]{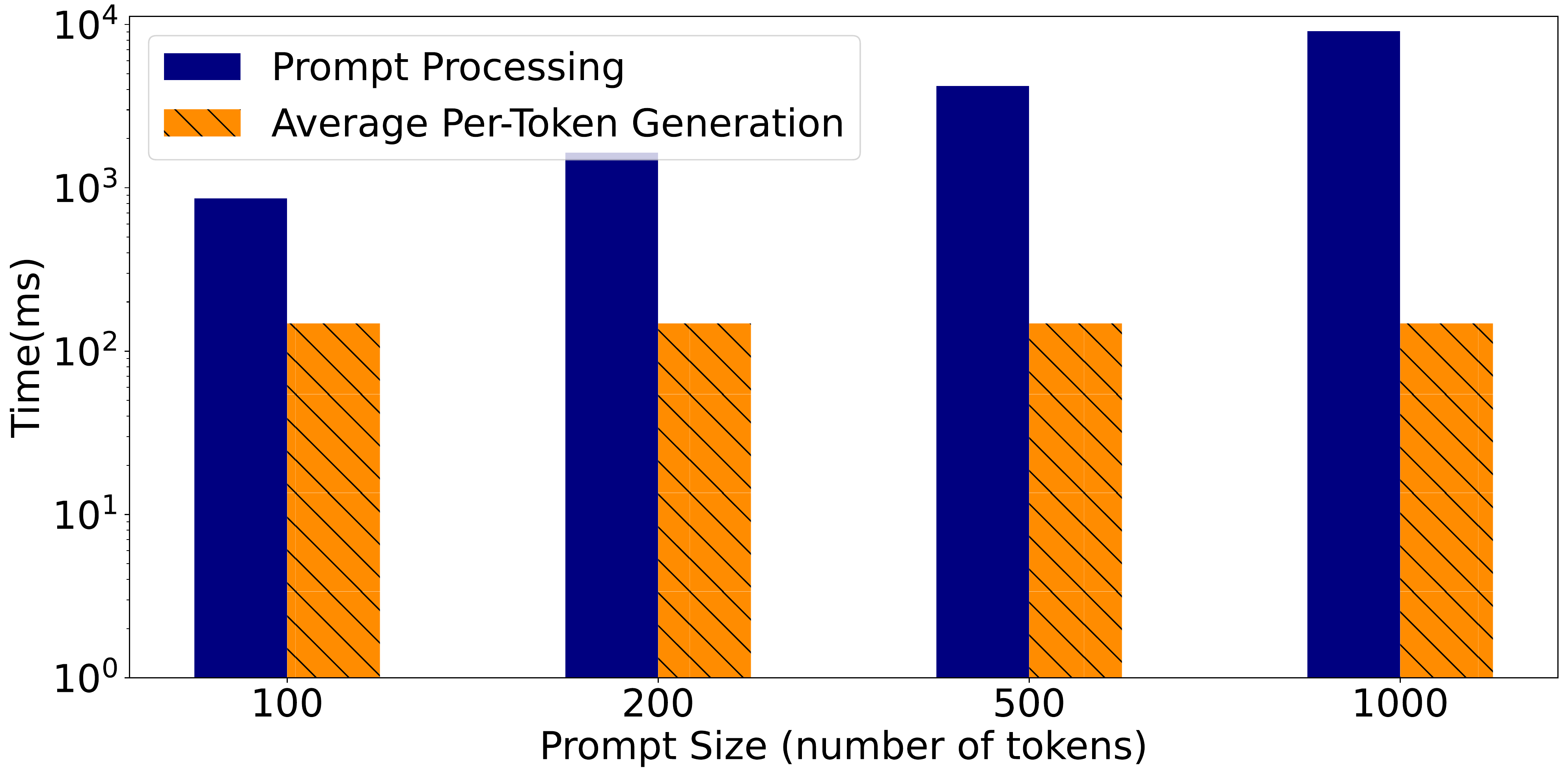}%
}
\caption{Prompt processing and token generation times for the BLOOM-176B model}
\label{fig:prompt_token_bloom176}
\end{figure*}

Figures \ref{fig:prompt_token_opt13}, \ref{fig:prompt_token_opt66}, and \ref{fig:prompt_token_bloom176} show average per-token generation time, and prompt processing time with different prompt sizes and batch sizes. Prompt processing time scales almost linearly with batch size and prompt length while being up to 106$\times$ higher than per-token generation latency. 

\clearpage

\section{\dejavu Planner}\label{app_planner}

In this section, we evaluate how disaggregation performs under various scenarios. Due to limited budget, we use our simulator to model a large number of machines. We use different types of models (OPT~\cite{zhang2022opt} and BLOOM~\cite{workshop2023bloom}), different types of GPUs (V100-16GB, and A100-80GB) and different number of GPUs per machine (i.e. varying degrees of tensor parallelism). As in section \ref{sec:disaggregation_results}, we use a modified version of the LMSys dataset for the number of generated tokens per microbatch~\cite{zheng2023lmsyschat1m}, and keep the prompt size equal to 1000 tokens. 

\begin{figure*}[htp!]
\subfloat[Makespan]{%
  \includegraphics[width=0.5\textwidth]{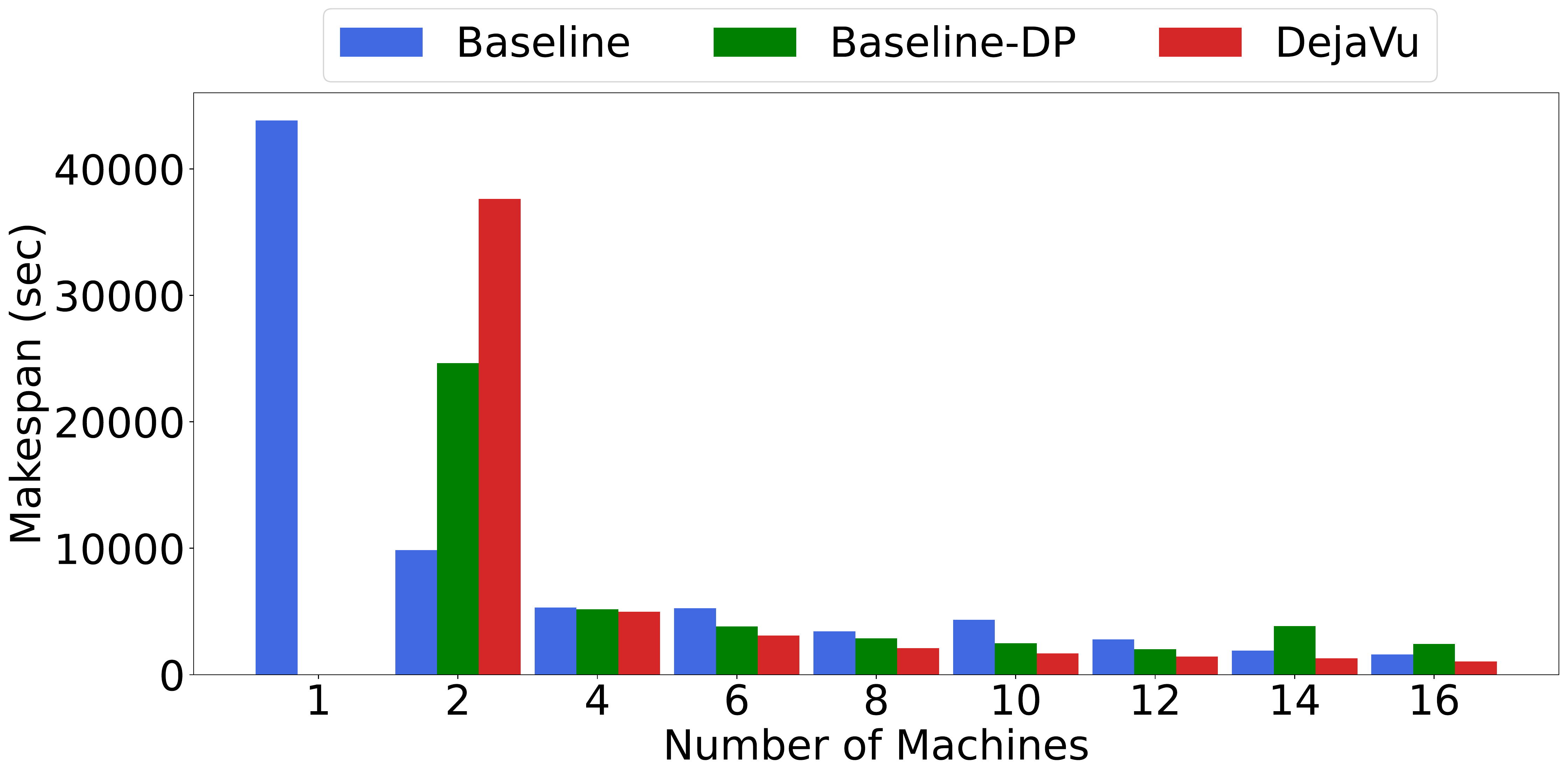}%
}
\subfloat[Normalized Cost]{%
  \includegraphics[width=0.5\textwidth]{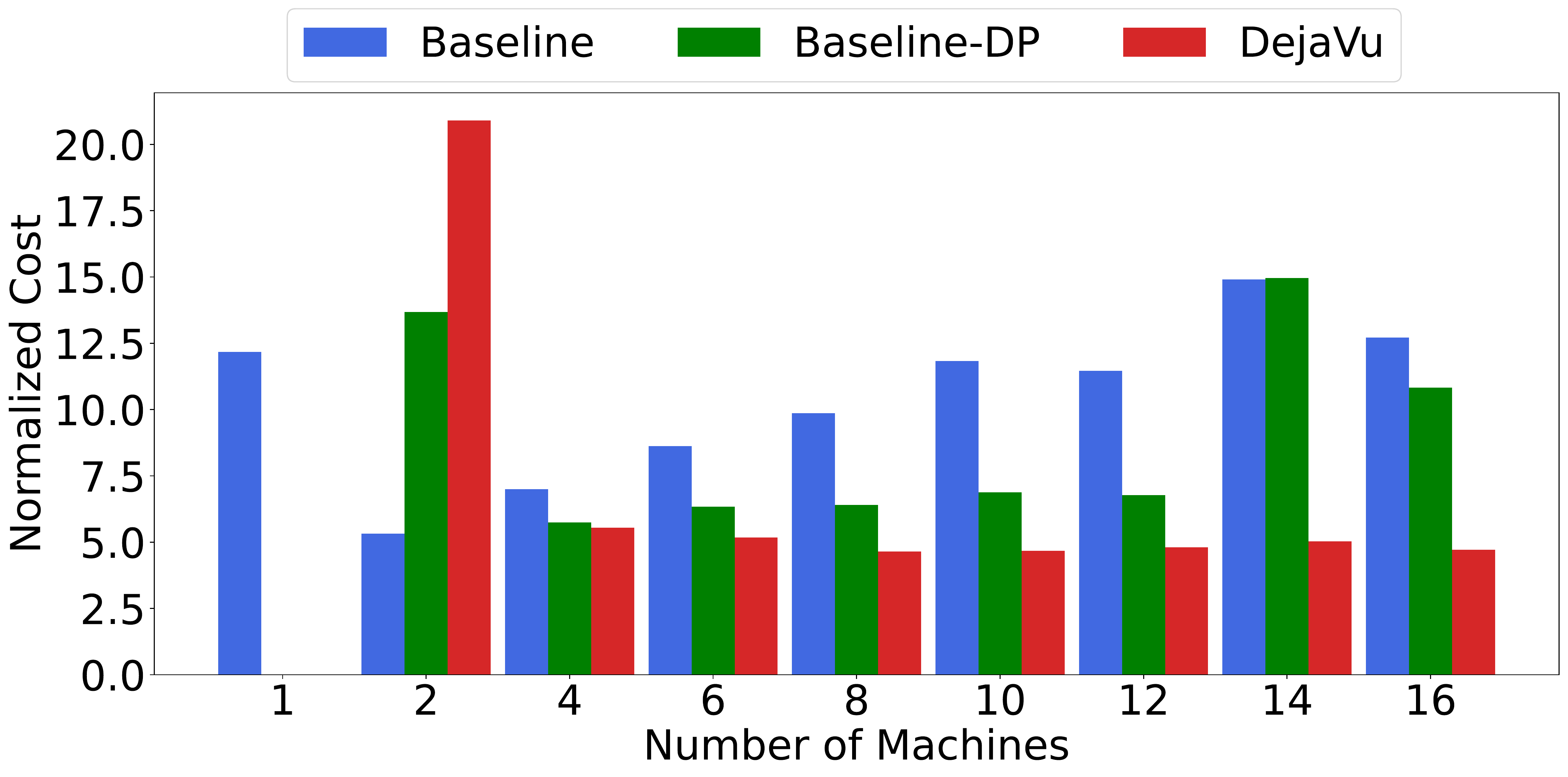}%
}
\caption{Makespan and cost for the best configurations found with different policies for the OPT-66B model on 2-A100-80GB machines. The best configurations found for each use case are shown in table \ref{table:opt66_2a100_planner}}
\label{fig:opt66_2_planner}
\end{figure*}

\begin{figure*}[htp!]
\subfloat[Makespan]{%
  \includegraphics[width=0.5\textwidth]{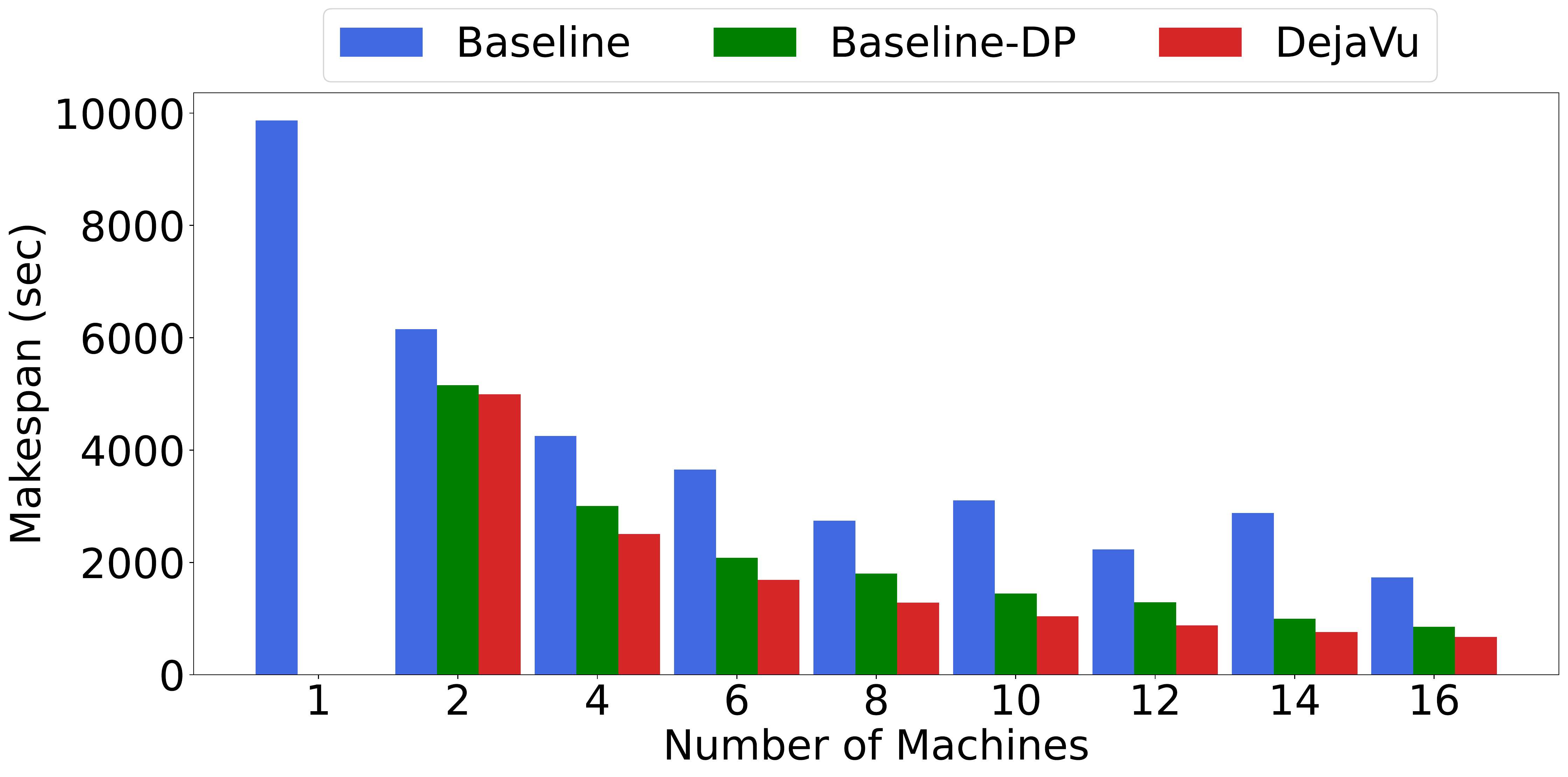}%
}
\subfloat[Normalized Cost]{%
  \includegraphics[width=0.5\textwidth]{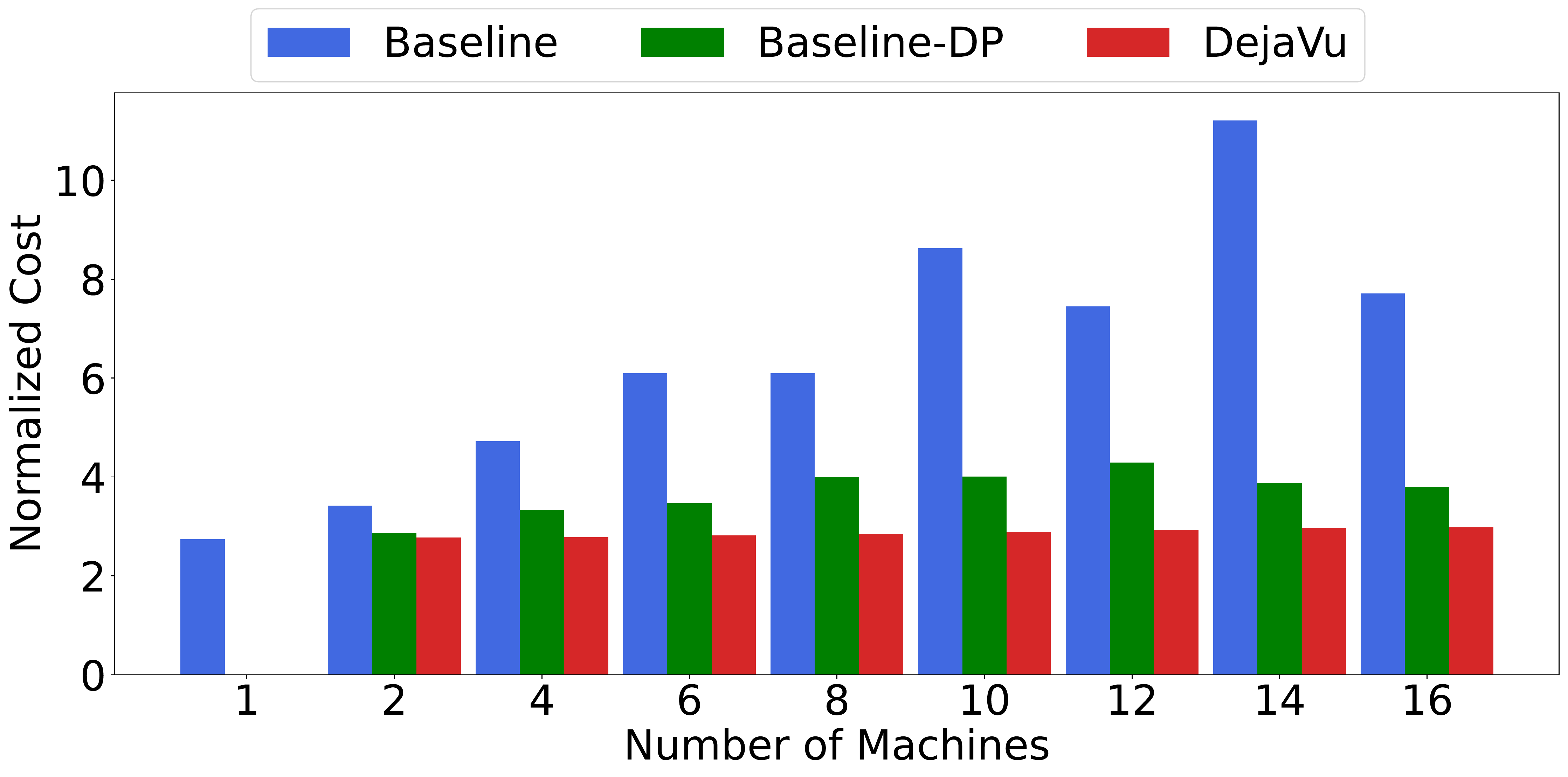}%
}
\caption{Makespan and cost for the best configurations found with different policies for the OPT-66B model on 4-A100-80GB machines. The best configurations found for each use case are shown in table \ref{table:opt66_4a100_planner}}
\label{fig:opt66_4_planner}
\end{figure*}

\begin{figure*}[htp!]
\subfloat[Makespan]{%
  \includegraphics[width=0.5\textwidth]{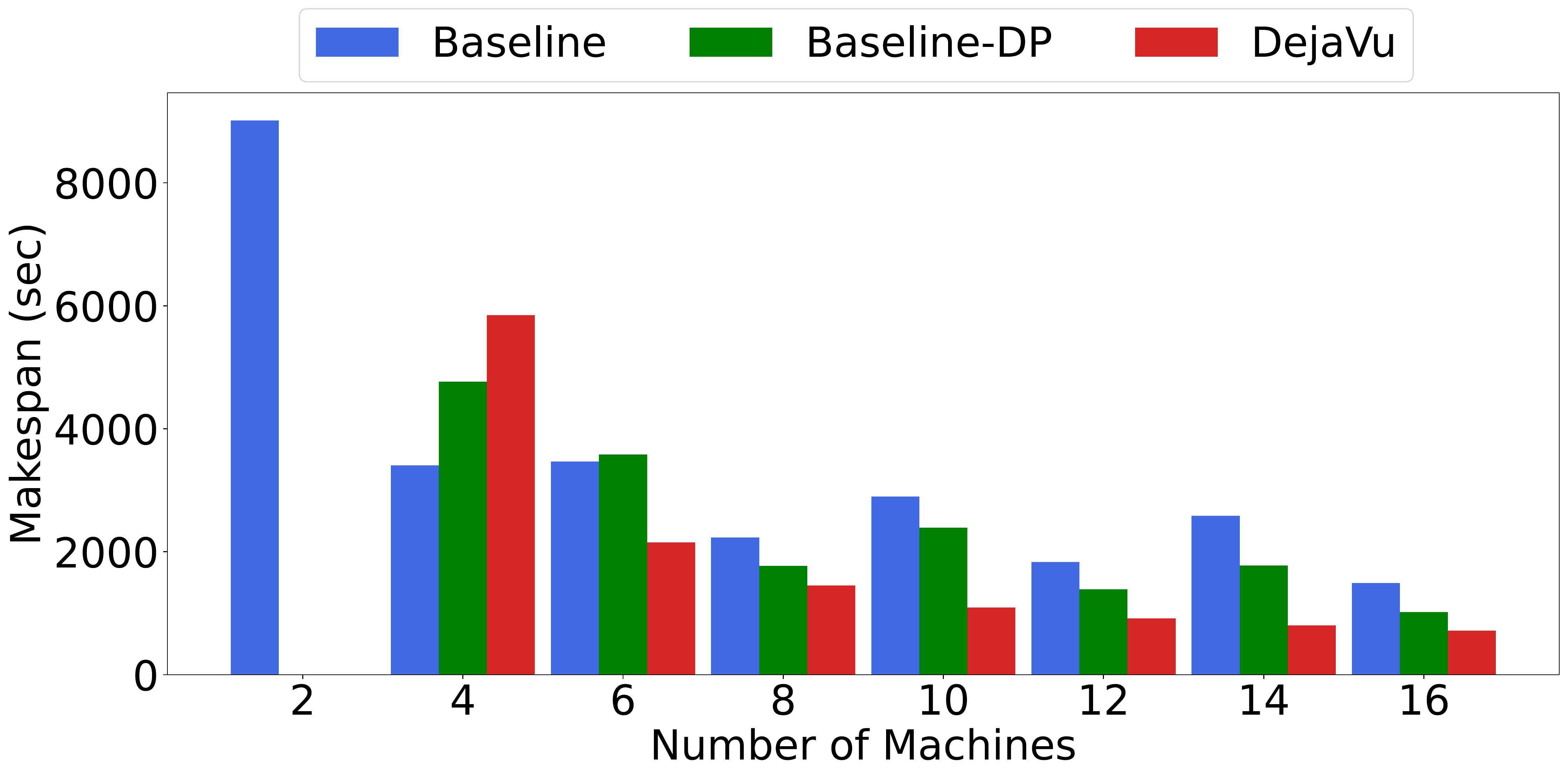}%
}
\subfloat[Normalized Cost]{%
  \includegraphics[width=0.5\textwidth]{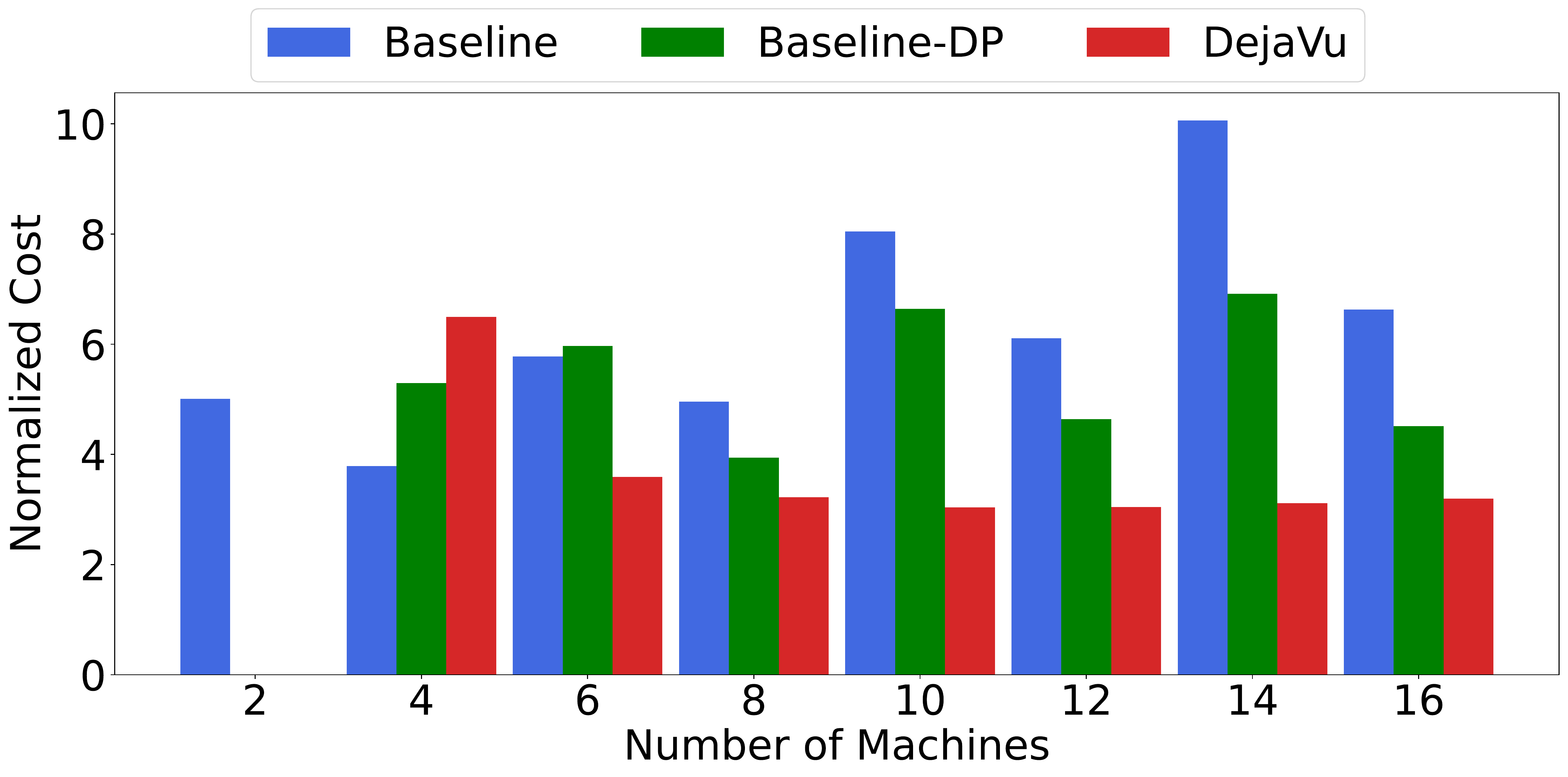}%
}
\caption{Makespan and cost for the best configurations found with different policies for the OPT-30B model on 4-V100-16GB machines. The best configurations found for each use case are shown in table \ref{table:opt30_4v100_planner}}
\label{fig:opt30_planner}
\end{figure*}

\begin{figure*}[htp!]
\subfloat[Makespan]{%
  \includegraphics[width=0.5\textwidth]{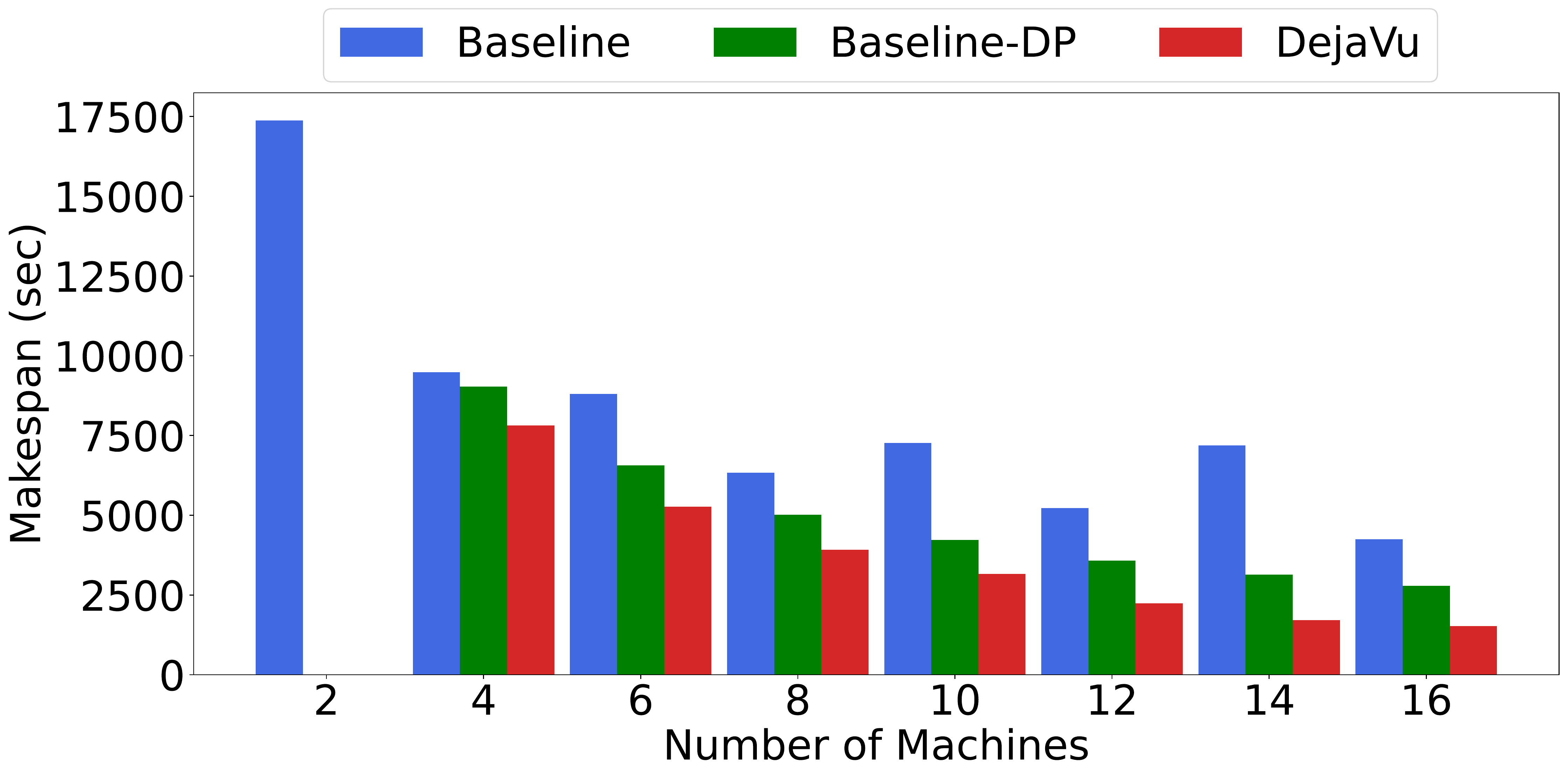}%
}
\subfloat[Total Cost]{%
  \includegraphics[width=0.5\textwidth]{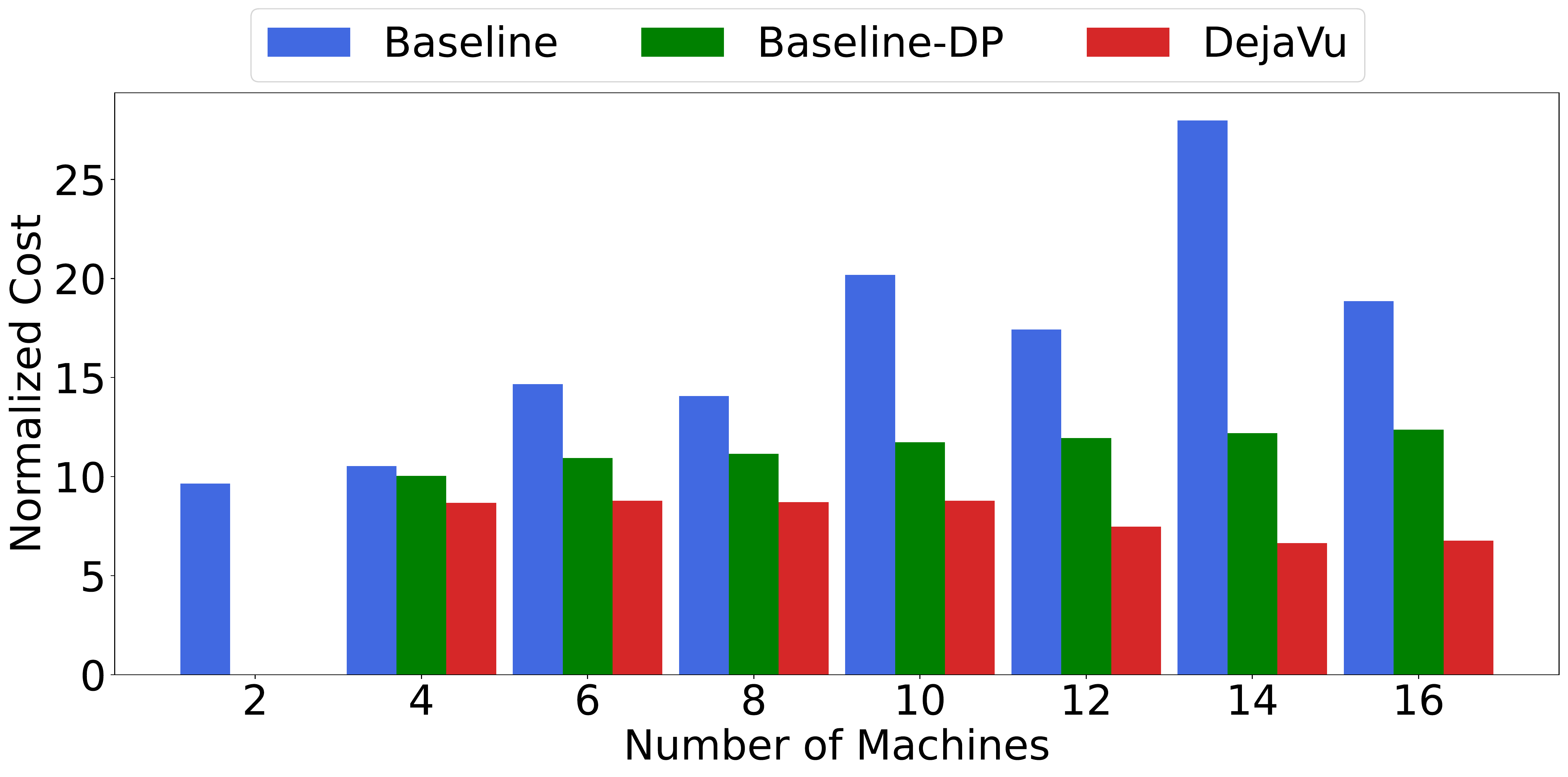}%
}
\caption{Makespan and cost for the best configurations found with different policies for the BLOOM-176B model on 4-A100-80GB machines. The best configurations found for each use case are shown in table \ref{table:bloom_planner}}
\label{fig:bloom_planner}
\end{figure*}

Figures \ref{fig:opt66_2_planner}-\ref{fig:bloom_planner} show the Makespan and Normalized total cost (with respect to the hourly cost of a single VM) for various numbers of available machines for our trace. For $D$ available machines, we simulate 3 different scenarios:
\begin{enumerate}
    \item \textbf{Baseline (Tensor Model + Pipeline Parallel)}: all $D$ machines run pipeline parallelism (all GPUs within each machine run Tensor Model Parallelism). We vary the microbatch size $b$, and pick the microbatch size that leads to the shortest makespan for our trace of requests.
    \item \textbf{Baseline-DP (Tensor Model + Pipeline + Data Parallel)}: we use $d$ pipelines, serving requests in parallel. Each pipeline has depth $\frac{D}{d}$. We also vary the microbatch size $b$, that each pipeline is using. We pick the combination $d$-$b$ that leads to the shortest makespan.
    \item \textbf{\dejavu  (Tensor Model + Pipeline Parallel)}: we use $D_p$ machines for prompt processing (with pipeline depth $D_p$), and $D_t$ machines for token generation (with pipeline depth $D_t$), where $D=D_p + D_t$.  We also vary the microbatch size $b$, that each pipeline is using. We pick the combination $D_p$-$b$ that leads to teh shortest makespan.
\end{enumerate}
Tables \ref{table:opt66_2a100_planner}-\ref{table:bloom_planner} show the best configurations found for each use case. For the \textit{Baseline}, $(Yp, Zb)$ indicates that a pipeline of depth $Y$ was used, with microbatch size $Z$. For the \textit{Baseline-DP},  $(Xd, Yp, Zb)$ indicates that $X$ pipelines were used, each with depth $Y$ and microbatch size $Z$. For \textit{\dejavu},  $((Y_1d, Z_1b), (Y_2d, Z_2b))$ means that the pipeline used for prompt processing has depth $Y_1$ and microbatch size $Z_1$, and the pipeline used for token generation has depth $Y_2$ and microbatch size $Z_2$.

\begin{figure*}[htp!]
\subfloat[Without early stopping]{%
  \includegraphics[width=0.5\textwidth]{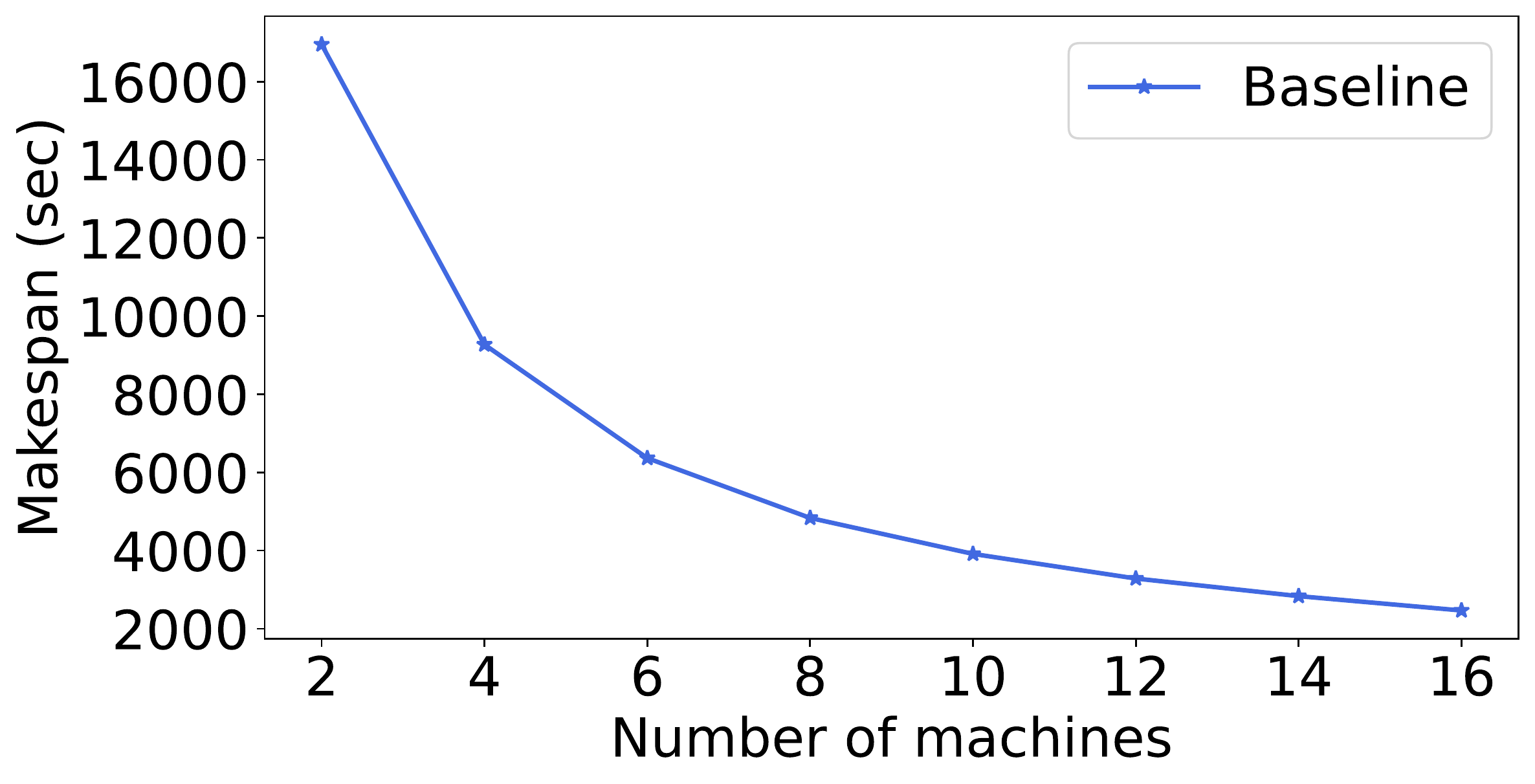}%
\label{fig:bloom_noes}}
\subfloat[With early stopping]{%
  \includegraphics[width=0.5\textwidth]{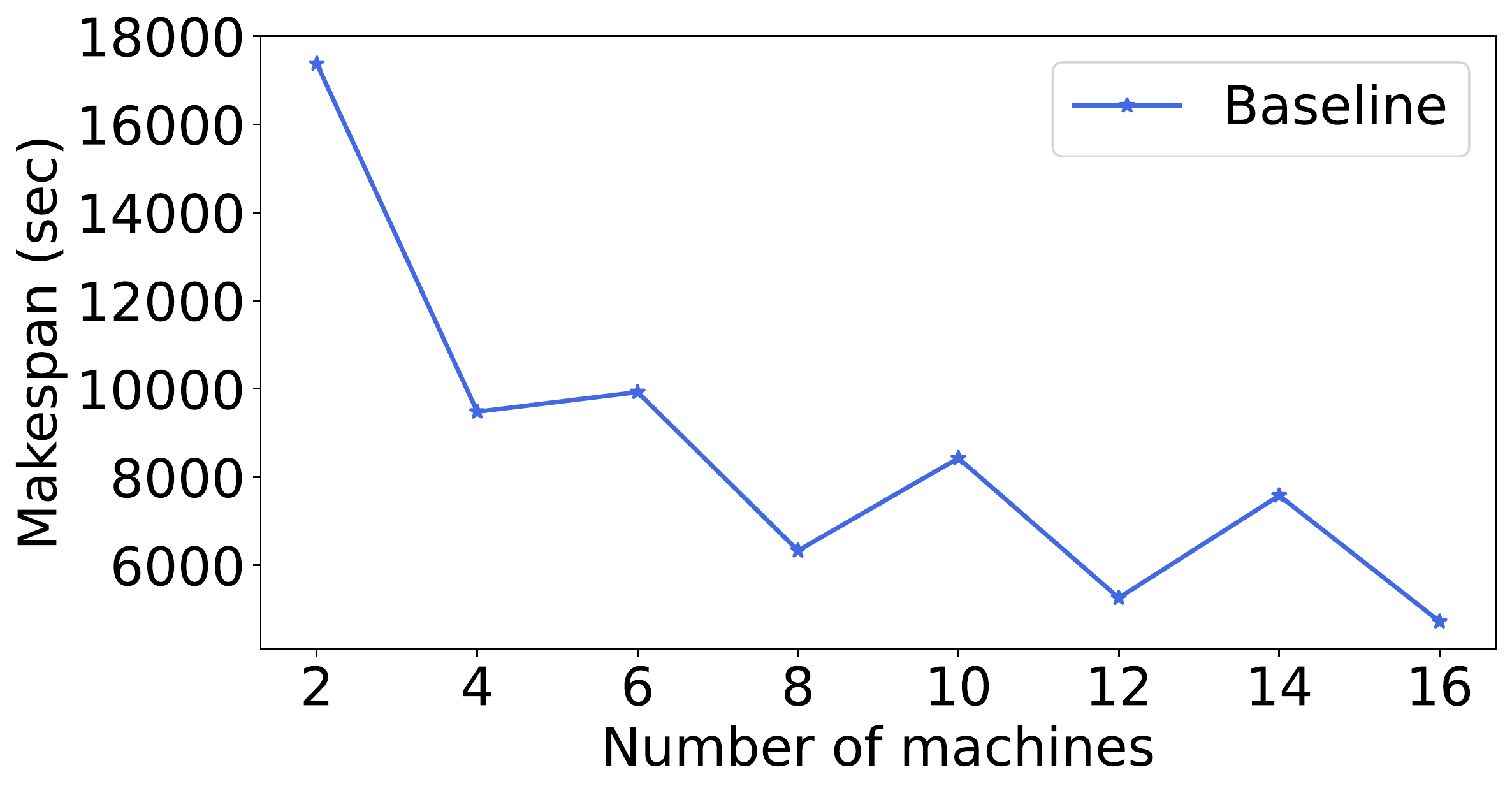}%
\label{fig:bloom_es}}
\caption{Makespan for BLOOM-176B with and without early stopping, varying the number of available machines. We keep the microbatch size constant at 16 requests.}
\label{fig:bloom_analysis}
\end{figure*}

\begin{figure*}[htp!]
\subfloat[Without early stopping]{%
  \includegraphics[width=0.5\textwidth]{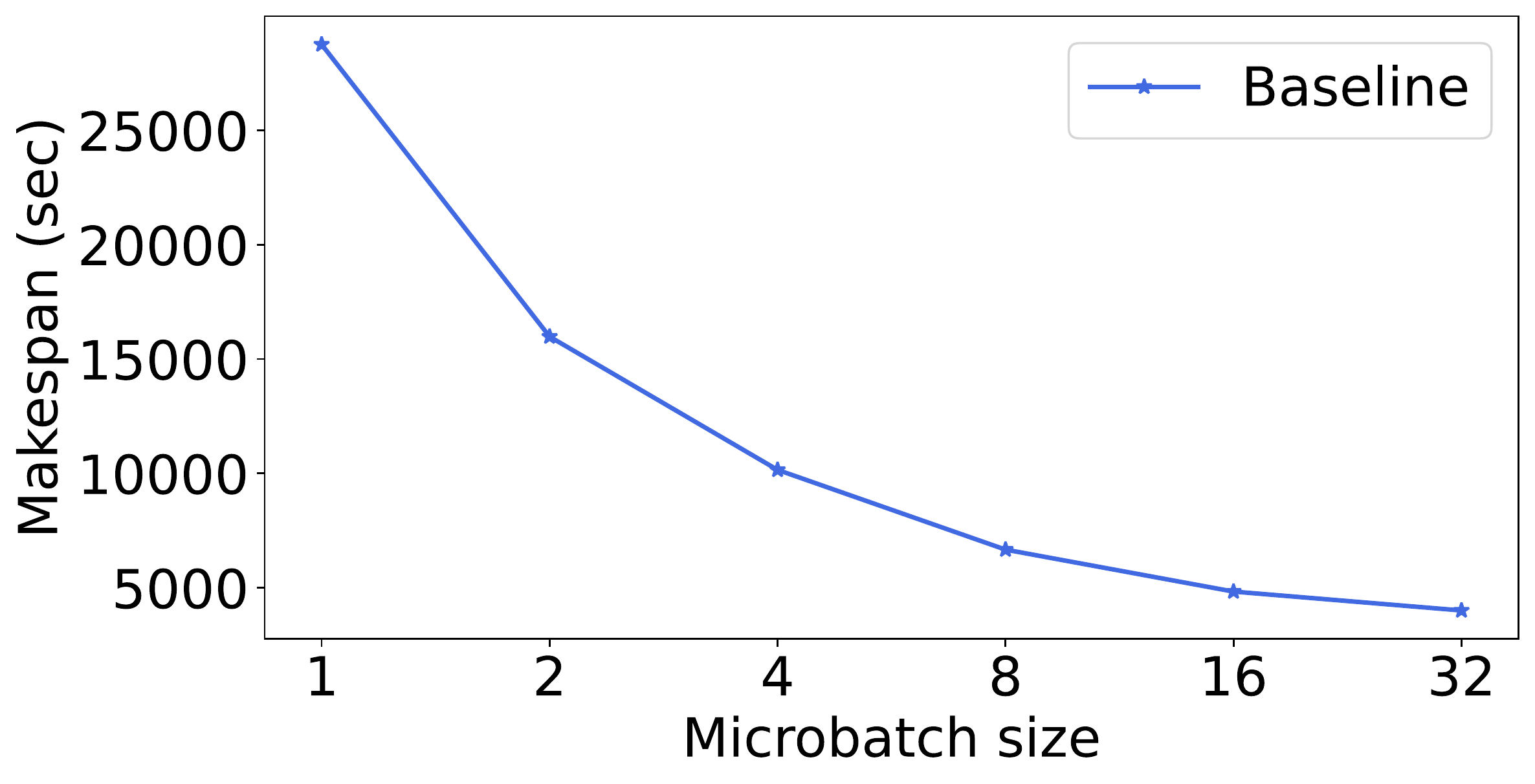}%
\label{fig:bloom_bs_noes}}
\subfloat[With early stopping]{%
  \includegraphics[width=0.5\textwidth]{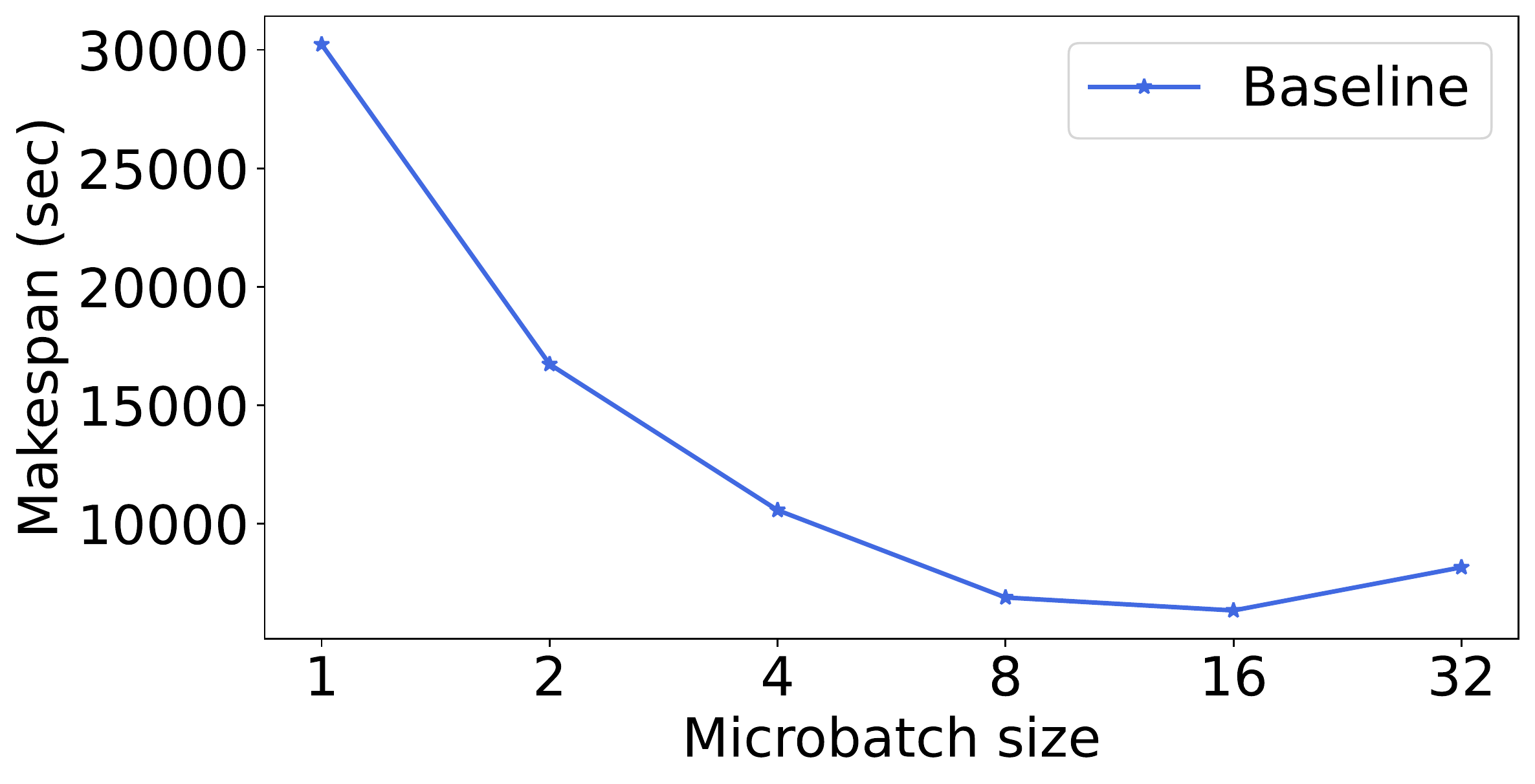}%
}
\caption{Makespan for BLOOM-176B on 8 machines, with and without early stopping, varying the microbatch size.}
\label{fig:bloom_bs_analysis}
\end{figure*}

Overall, \textit{Baseline} (i.e. using tensor model and pipeline parallelism) has a shorter makespan with a larger number of available machines (and subsequently the pipeline depth) and microbatch size (which can also be explained from formula \ref{formula:basethr}). However, we observe 3 trends: First, the scalability of \textit{Baseline} diminishes as pipeline depth increases, i.e. an Y$\times$ increase in the pipeline depth does not correspond to a Y$\times$ decrease in makespan. This becomes evident from the normalized cost Figures \ref{fig:opt66_2_planner}-\ref{fig:bloom_planner}, where the cost increases with the number of machines. Second, with a real-world trace such as the LMSys~\cite{zheng2023lmsyschat1m} dataset, the number of "non-overlapping" early stops, and their impact on the pipeline depends on the trace and the pipeline depth. For example, in Figure \ref{fig:bloom_analysis}, we plot the makespan for \textit{Baseline} for the BLOOM model, with and without early stops for a fixed microbatch size. Without early stops, as we increase the number of machines, the makespan decreases. With early stops, we see sudden increases in makespan with 6,10, and 14 machines, due to a higher number of non-overlapping early stops. Third, although a larger microbatch size can reduce the makespan, it also increases proportionally the time for prompt computation ($Y$). However, the time per token only slightly increases (see Figures \ref{fig:prompt_token_opt13}-\ref{fig:prompt_token_bloom176}). Consequently, larger microbatch size leads to larger differences between prompt processing and token generation time, which increases the pipeline bubbles in the case of requests exiting early. Thus, a larger microbatch size is not always beneficial, as shown in Figure \ref{fig:bloom_bs_analysis}. 

Introducing Data Parallelism is beneficial for performance. Overall, \textit{Baseline-DP} outperforms \textit{Baseline} by 2.29$\times$. However, the varying number of generated token at the LMSys trace can cause imbalance to the different data parallel pipelines, both in the total number of tokens they generate, and also in the number of early stops they encounter. 

\dejavu can alleviate all of these issues by disaggregating prompt processing from token generation, and eliminating the impact of early stops. Overall, when dedicating the same number of machines to all baselines, \dejavu leads to 4.2$\times$ and 2.22$\times$ shorter makespan compared to \textit{Baseline}, and \textit{Baseline-DP} respectively.

\begin{table*}[h]
\centering
\caption{Best configurations found for Figure~\ref{fig:opt66_2_planner}.}
\vskip 0.15in
\begin{center}
\begin{small}
\begin{tabular}{c|c|c|c}
\toprule
\textbf{Number of machines} & \textbf{Baseline} & \textbf{Baseline-DP} & \textbf{\dejavu}\\
\midrule
1 & (1$p$, 4$b$) & & \\
\hline
2 & (2$p$, 16$b$) & (2$d$, 1$p$, 4$b$) & ((1$p$, 4$b$), (1$p$, 4$b$))\\
\hline
4 & (4$p$, 16$b$) & (2$d$, 2$p$, 16$b$) & ((2$p$, 2$b$), (2$p$, 2$b$))\\
\hline
6 & (6$p$, 16$b$) & (3$d$, 2$p$, 16$b$) & ((2$p$, 16$b$), (4$p$, 16$b$))\\
\hline
8 & (8$p$, 16$b$) & (2$d$, 4$p$, 16$b$) & ((3$p$, 16$b$), (5$p$, 16$b$))\\
\hline
10 & (10$p$, 16$b$) & (5$d$, 2$p$, 16$b$) & ((4$p$, 16$b$), (6$p$, 16$b$))\\
\hline
12 & (12$p$, 16$b$) & (3$d$, 4$p$, 16$b$) & ((5$p$, 16$b$), (7$p$, 16$b$))\\
\hline
14 & (14$p$, 16$b$) & (7$d$, 2$p$, 16$b$) & ((5$p$, 16$b$), (9$p$, 16$b$))\\
\hline
16 & (16$p$, 8$b$) & (4$d$, 4$p$, 16$b$) & ((6$p$, 16$b$), (10$p$, 16$b$))\\
\bottomrule
\end{tabular}
\end{small}
\end{center}
\vskip -0.1in
\label{table:opt66_2a100_planner}
\end{table*}

\begin{table*}[h]
\centering
\caption{Best configurations found for Figure~\ref{fig:opt66_4_planner}.}
\vskip 0.15in
\begin{center}
\begin{small}
\begin{tabular}{c|c|c|c}
\toprule
\textbf{Number of machines} & \textbf{Baseline} & \textbf{Baseline-DP} & \textbf{\dejavu}\\
\midrule
1 & (1$p$, 32$b$) & & \\
\hline
2 & (2$p$, 32$b$) & (2$d$, 1$p$, 32$b$) & ((1$p$, 32$b$), (1$p$, 32$b$))\\
\hline
4 & (4$p$, 16$b$) & (4$d$, 1$p$, 32$b$) & ((2$p$, 32$b$), (2$p$, 32$b$))\\
\hline
6 & (6$p$, 32$b$) & (6$d$, 1$p$, 32$b$) & ((3$p$, 32$b$), (3$p$, 32$b$))\\
\hline
8 & (8$p$, 16$b$) & (8$d$, 1$p$, 32$b$) & ((4$p$, 32$b$), (4$p$, 32$b$))\\
\hline
10 & (10$p$, 32$b$) & (10$d$, 1$p$, 32$b$) & ((5$p$, 32$b$), (5$p$, 32$b$))\\
\hline
12 & (12$p$, 16$b$) & (12$d$, 1$p$, 32$b$) & ((6$p$, 32$b$), (6$p$, 32$b$))\\
\hline
14 & (14$p$, 32$b$) & (14$d$, 1$p$, 32$b$) & ((7$p$, 32$b$), (7$p$, 32$b$))\\
\hline
16 & (16$p$, 8$b$) & (16$d$, 1$p$, 32$b$) & ((8$p$, 32$b$), (8$p$, 32$b$))\\
\bottomrule
\end{tabular}
\end{small}
\end{center}
\vskip -0.1in
\label{table:opt66_4a100_planner}
\end{table*}

\begin{table*}[h]
\centering
\caption{Best configurations found for Figure~\ref{fig:opt30_planner}.}
\vskip 0.15in
\begin{center}
\begin{small}
\begin{tabular}{c|c|c|c}
\toprule
\textbf{Number of machines} & \textbf{Baseline} & \textbf{Baseline-DP} & \textbf{\dejavu}\\
\midrule
2 & (2$p$, 8$b$) &  & \\
\hline
4 & (4$p$, 16$b$) & (2$d$, 2$p$, 8$b$) & ((2$p$, 8$b$), (2$p$, 8$b$))\\
\hline
6 & (6$p$, 16$b$) & (3$d$, 2$p$, 8$b$) & ((2$p$, 16$b$), (4$p$, 16$b$))\\
\hline
8 & (8$p$, 16$b$) & (2$d$, 4$p$, 16$b$) & ((3$p$, 16$b$), (5$p$, 16$b$))\\
\hline
10 & (10$p$, 16$b$) & (5$d$, 2$p$, 8$b$) & ((4$p$, 16$b$), (6$p$, 16$b$))\\
\hline
12 & (12$p$, 16$b$) & (3$d$, 4$p$, 16$b$) & ((5$p$, 16$b$), (7$p$, 16$b$))\\
\hline
14 & (14$p$, 16$b$) & (7$d$, 2$p$, 8$b$) & ((6$p$, 16$b$), (8$p$, 16$b$))\\
\hline
16 & (16$p$, 8$b$) & (4$d$, 4$p$, 16$b$) & ((7$p$, 16$b$), (9$p$, 16$b$))\\
\bottomrule
\end{tabular}
\end{small}
\end{center}
\vskip -0.1in
\label{table:opt30_4v100_planner}
\end{table*}

\begin{table*}[h]
\centering
\caption{Best configurations found for Figure~\ref{fig:bloom_planner}.}
\vskip 0.15in
\begin{center}
\begin{small}
\begin{tabular}{c|c|c|c}
\toprule
\textbf{Number of machines} & \textbf{Baseline} & \textbf{Baseline-DP} & \textbf{\dejavu}\\
\midrule
2 & (2$p$, 16$b$) &  & \\
\hline
4 & (4$p$, 16$b$) & (2$d$, 2$p$, 16$b$) & ((2$p$, 16$b$), (2$p$, 16$b$))\\
\hline
6 & (6$p$, 32$b$) & (3$d$, 2$p$, 16$b$) & ((3$p$, 16$b$), (3$p$, 16$b$))\\
\hline
8 & (8$p$, 16$b$) & (2$d$, 4$p$, 16$b$) & ((4$p$, 16$b$), (4$p$, 16$b$))\\
\hline
10 & (10$p$, 32$b$) & (5$d$, 2$p$, 16$b$) & ((5$p$, 16$b$), (5$p$, 16$b$))\\
\hline
12 & (12$p$, 16$b$) & (6$d$, 2$p$, 16$b$) & ((6$p$, 32$b$), (6$p$, 32$b$))\\
\hline
14 & (14$p$, 32$b$) & (7$d$, 2$p$, 16$b$) & ((8$p$, 32$b$), (6$p$, 32$b$))\\
\hline
16 & (16$p$, 8$b$) & (4$d$, 4$p$, 16$b$) & ((10$p$, 32$b$), (6$p$, 32$b$))\\
\bottomrule
\end{tabular}
\end{small}
\end{center}
\vskip -0.1in
\label{table:bloom_planner}
\end{table*}

\clearpage
\section{Traces of disaggregation}\label{app_disag_trace}

\begin{figure*}[htp!]

\subfloat[Prompt processing and token generation at the same pipeline.]{%
  \includegraphics[height=3cm,width=\linewidth]{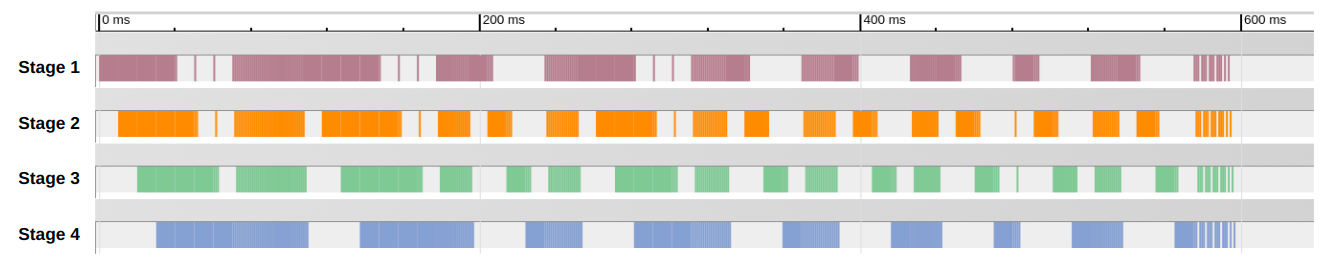}\label{fig:centralized}%
}

\subfloat[Using a different pipeline for prompt processing and token generation.]{%
  \includegraphics[clip,width=\linewidth]{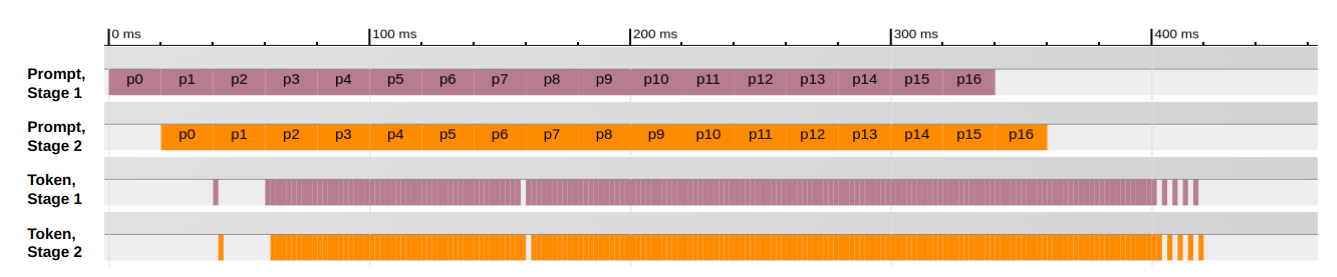}\label{fig:disaggregated}%
}
\caption{Comparison between dedicating 4 machines to both prompt processing and token generation, vs using 2 machines for prompt processing and 2 machines for token generation. Prompt processing latency is 10 $\times$ higher than per-token generation latency.}
\label{fig:disaggregated_comparison}
\end{figure*}

\section{Microbenchmarks}\label{app_microbenchmarks}

\begin{figure*}[htp!]
\subfloat[\dejavu slowdown when streaming to remote CPU]{%
  \includegraphics[scale=0.25]{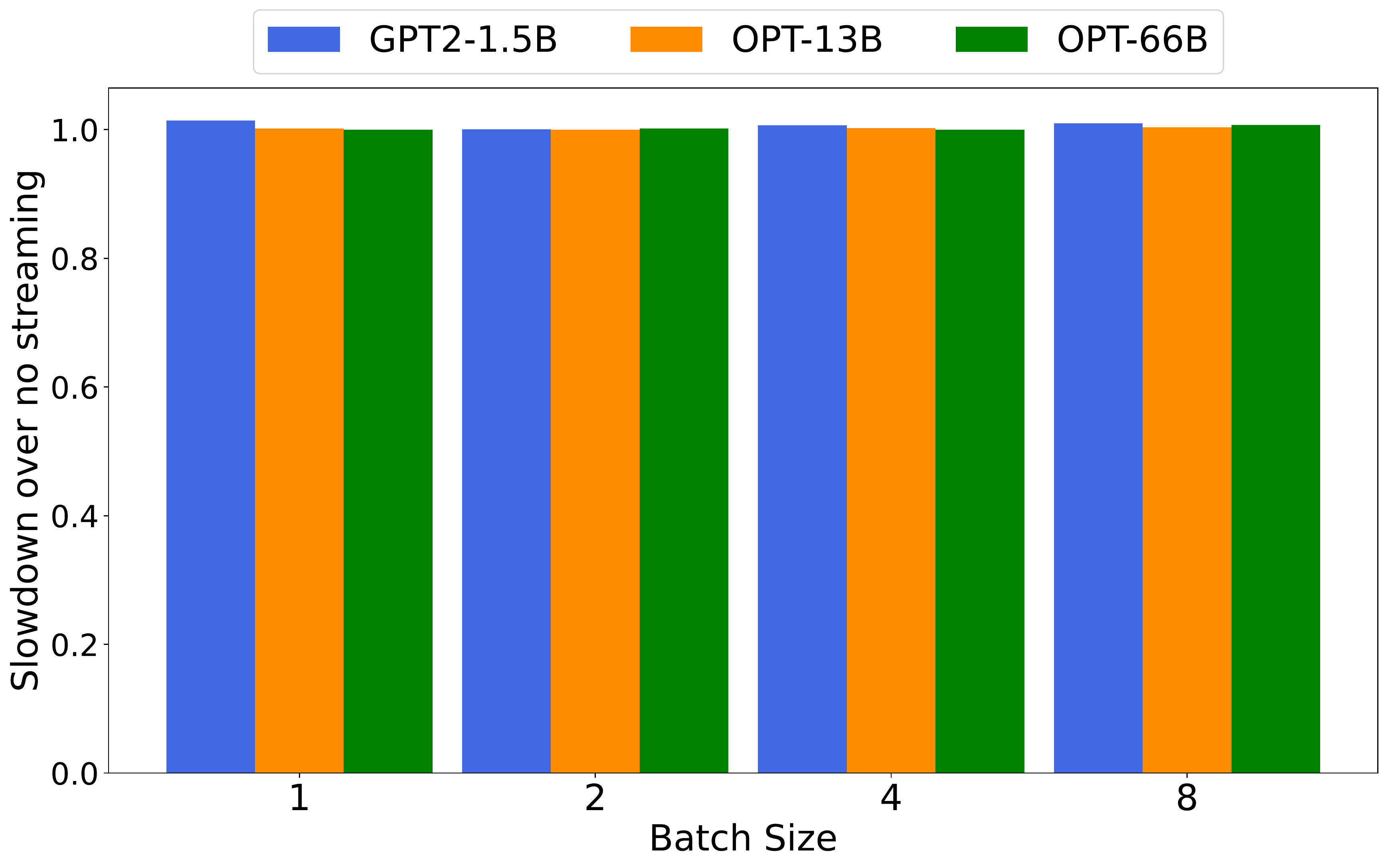}\label{fig:microbenchmark_remote}%
  }
\subfloat[\dejavu slowdown when streaming to local SSD]{%
 \includegraphics[scale=0.25]{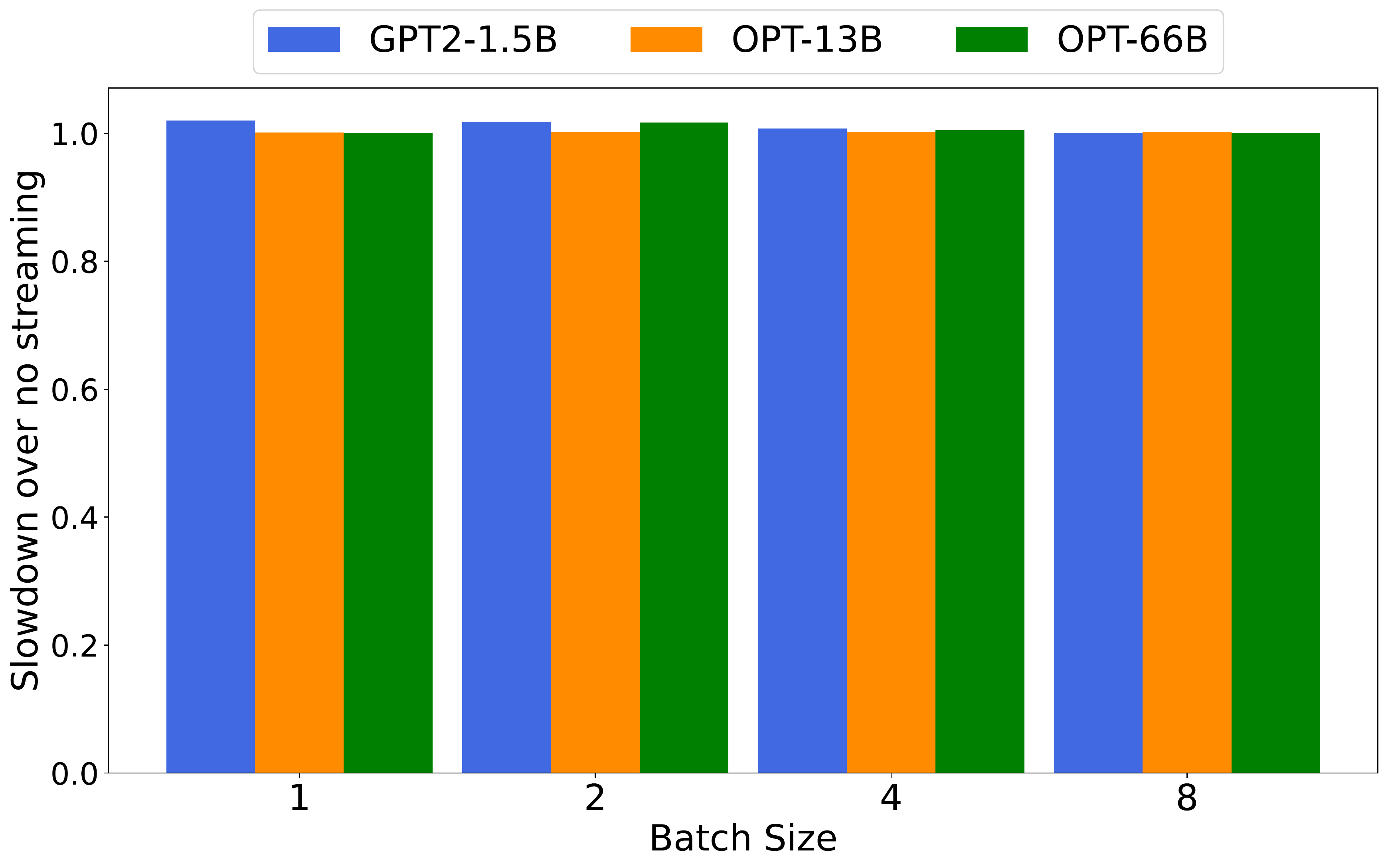}
   \label{fig:microbenchmark_local}%
}
\caption{Single-batch slowdown of \dejavu KV cache streaming}
\label{fig:microbenchmark}
\end{figure*}

Figure \ref{fig:microbenchmark} shows the slowdown of \dejavulib when streaming to remote CPU memory for a single batch (i.e. no pipeline parallelism) that contains requests with prompt size 500, and generating 500 extra tokens. The slowdown compared to no streaming is always within 2\%. \dejavulib might cause some slowdown in prompt processing time, due to streaming that cannot be hidden without pipeline parallelism. However, since many tokens are generated, this slowdown is negligible.

\section{Understanding the benefits of microbatch swapping}\label{app_swapping}

The performance of microbatch swapping heavily depends on the time needed to swap the KV cache back in the GPU. In this section, we formalize the performance of pipeline parallel inference with and without microbatch swapping and investigate where it makes sense to use swapping or not.

Since the amount of GPU memory needed for the KV cache, when the microbatch swapping optimization is enabled, is smaller than without swapping, we can use larger batch sizes, which are beneficial for the inference throughput. Assume a case where with a given set of GPUs, we can fit microbatch size $B$ without swapping, and microbatch size $2 \cdot B$ with swapping. We also assume that the time for token generation $t$ is constant with both microbatch sizes (which has been validated experimentally). The time for prompt processing with microbatch size $B$ is $P$, while for microbatch size $2 \cdot B$ is $2 \cdot P$.

Microbatch swapping will lead to better throughput when:

$$ 2 \cdot (P + N \cdot t) \geq 2 \cdot P + \sum_{i=p}^{i=N}{\max(t,transf_{i})} $$
$$ 2 \cdot N \cdot t \geq \sum_{i=p}^{i=N+p}{\max(t,transf_{i})} $$

where $transf_{i}$ stands for the time need to transfer the KV cache back in GPU from host memory.
We have that $transf_{i} = \frac{i \cdot B \cdot C_i}{pciebw}$, where $C_i$ is the single-token, single-request KV cache size, and $pciebw$ is the PCIe bandwidth.
Thus, we have that:

$$ 2 \cdot N \cdot t \geq \sum_{i=p}^{i=N+p}{\max(t, \frac{i \cdot B \cdot C_i}{pciebw})} $$

\subsection{Experimental Analysis}

For a given model, the main factors that affect the performance of microbatch swapping are the batch size and the number of generated tokens. For our analysis, we use OPT-30B, OPT-66B, BLOOM-176B, and 2 types of GPUs: V100-16GB, and A100-80GB. As expected, with larger batch sizes or sequence lengths, the overhead of streaming the KV cache in becomes high, that swapping is not beneficial anymore.

\subsubsection{Constant batch size, vary the number of generated tokens}

\begin{figure*}[htp!]
\centering
\subfloat[OPT-30B, 6 V100-16GB, B=1]{%
  \includegraphics[width=0.5\linewidth]{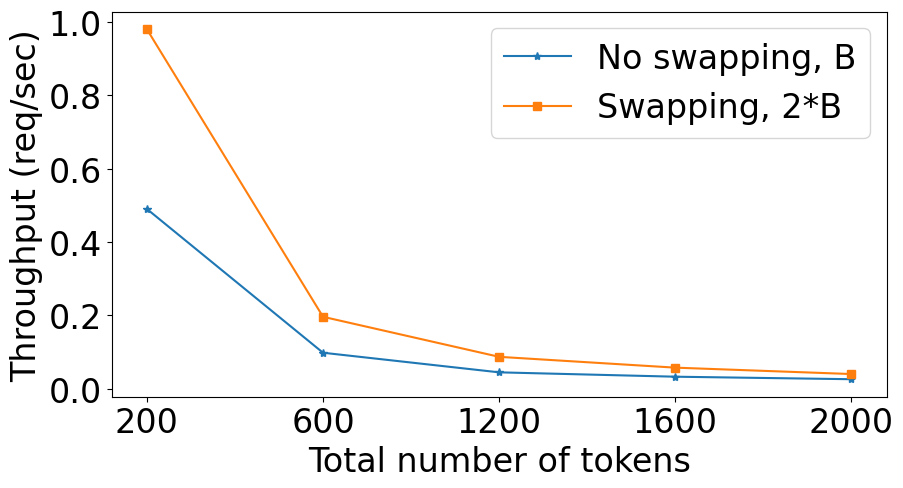}%
}
\subfloat[OPT-66B, 2 A100-80GB, B=1]{%
  \includegraphics[width=0.5\linewidth]{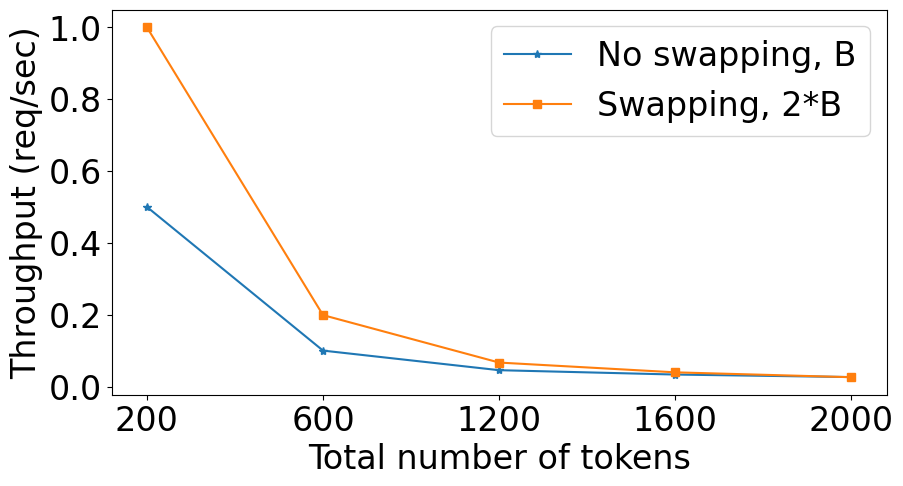}%
}
\caption{Throughput with and without swapping, varying the sequence length}
\label{fig:swapping-SeqLen}
\end{figure*}

In Figure \ref{fig:swapping-SeqLen}, We keep the batch size constant, vary the sequence length, and measure the throughput in each setup.

\subsubsection{Constant number of generated tokens, vary batch size}

\begin{figure*}[htp!]
\subfloat[N=200]{%
  \includegraphics[width=0.33\textwidth]{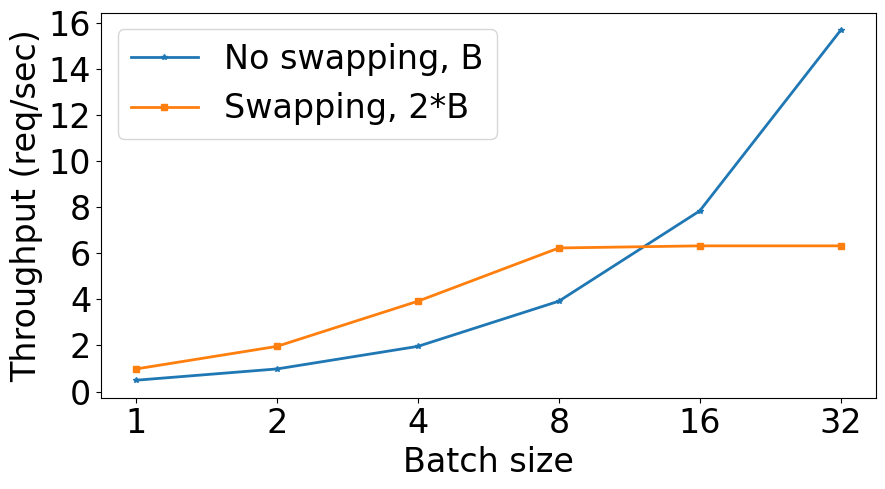}%
}
\subfloat[N=1200]{%
  \includegraphics[width=0.33\textwidth]{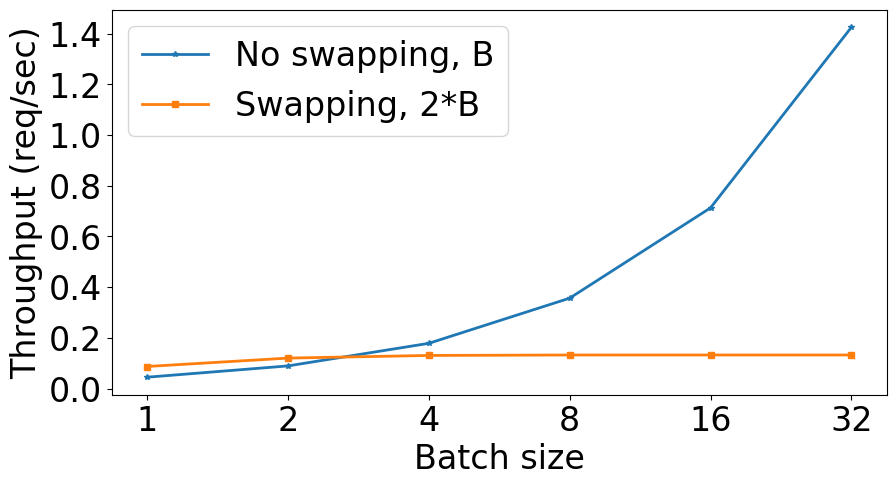}%
}
\subfloat[N=2000]{%
  \includegraphics[width=0.33\textwidth]{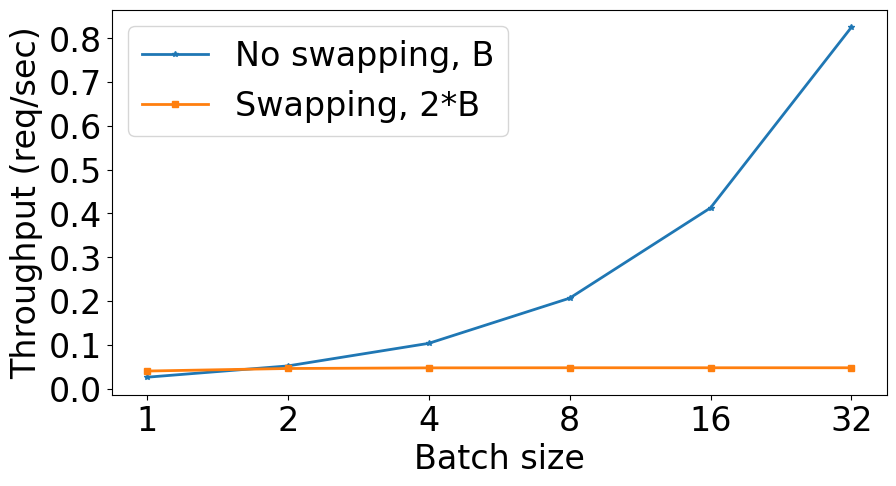}%
}
\caption{Throughput for OPT-30B, 6 V100-16GB}
\label{fig:swapping-OPT-30-Bsize}
\end{figure*}

\begin{figure*}[ht]
\subfloat[N=200]{%
  \includegraphics[width=0.33\textwidth]{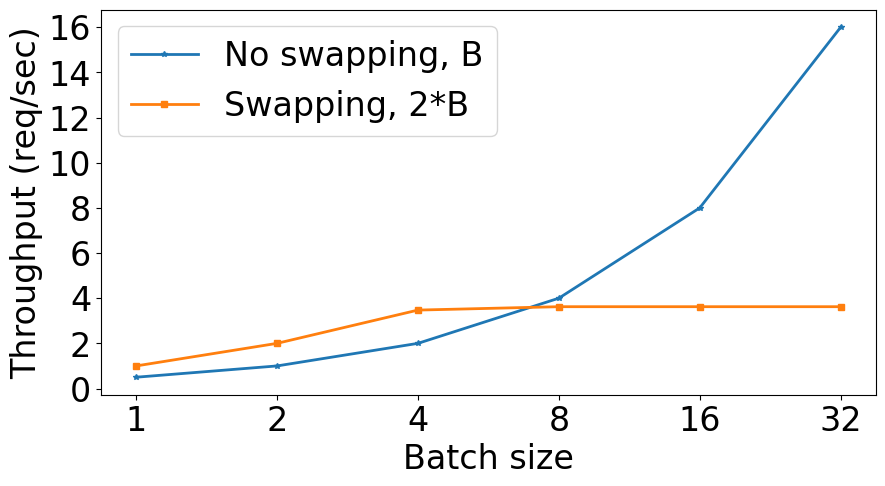}%
}
\subfloat[N=1200]{%
  \includegraphics[width=0.33\textwidth]{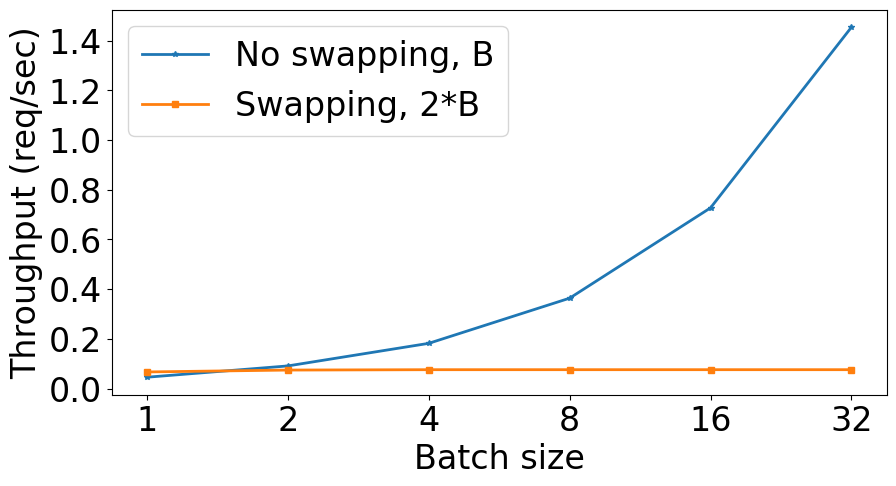}%
}
\subfloat[N=2000]{%
  \includegraphics[width=0.33\textwidth]{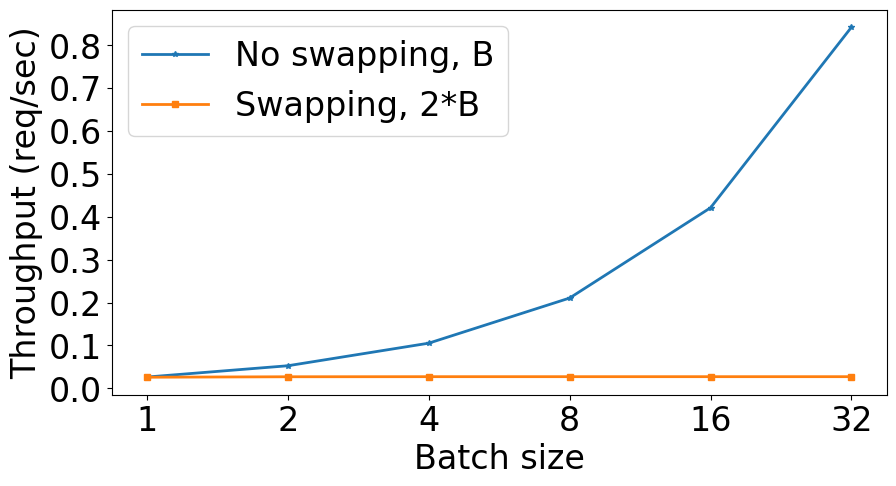}%
}
\caption{Throughput for OPT-66B, 2 A100-80GB}
\label{fig:swapping-OPT-66-Bsize}
\end{figure*}

\begin{figure*}[htp!]
\subfloat[N=200]{%
  \includegraphics[width=0.33\textwidth]{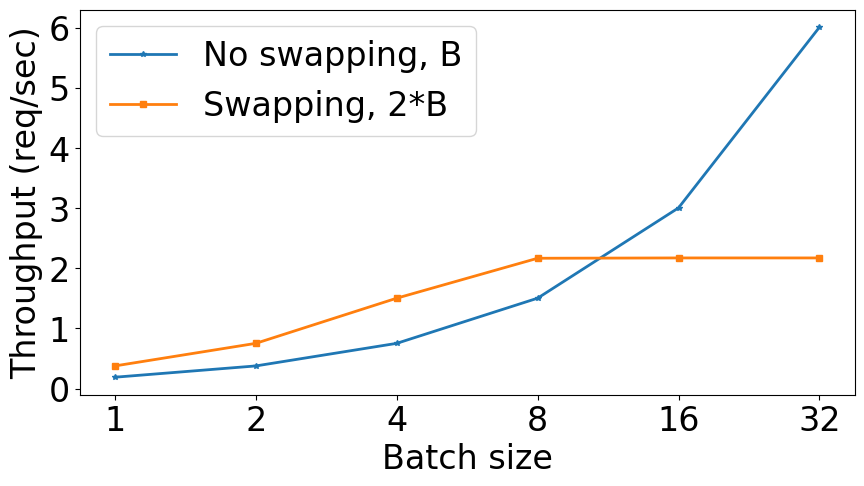}%
}
\subfloat[N=1200]{%
  \includegraphics[width=0.33\textwidth]{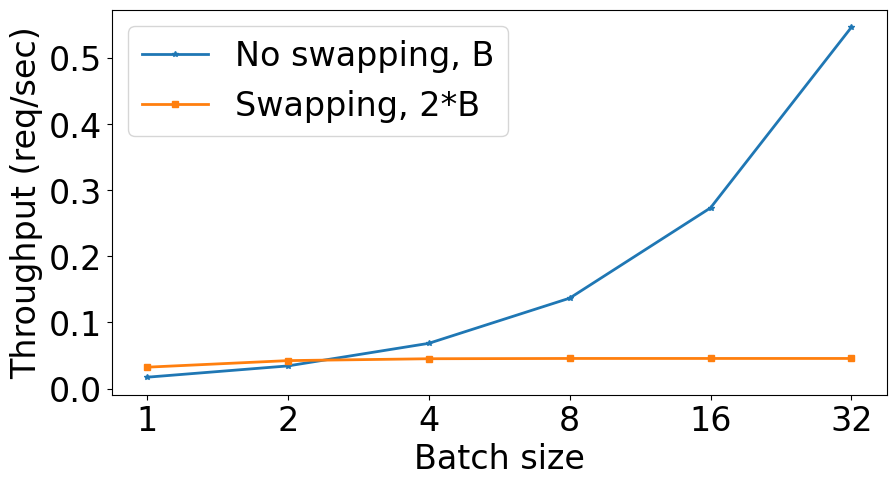}%
}
\subfloat[N=2000]{%
  \includegraphics[width=0.33\textwidth]{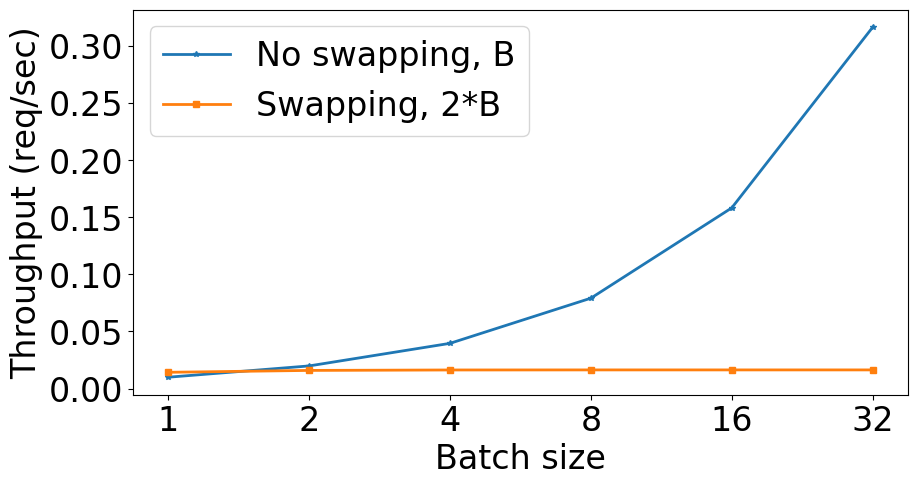}%
}
\caption{Throughput for BLOOM-176B, 5 A100-80GB}
\label{fig:swapping-BLOOM-Bsize}
\end{figure*}

Figures \ref{fig:swapping-OPT-30-Bsize}, \ref{fig:swapping-OPT-66-Bsize}, and \ref{fig:swapping-BLOOM-Bsize} show results in simulation, where we keep the total number of generated tokens constant, and vary the batch size. We measured the PCIe bandwidth in each of the examined machines and used the respective PCIe bandwidths for our simulations. The V100 machines are equipped with PCIe3x16, and A100 machines with PCIe4x16~\cite{pcie}.

%% file: main.bbl
\begin{thebibliography}{36}
\providecommand{\natexlab}[1]{#1}
\providecommand{\url}[1]{\texttt{#1}}
\expandafter\ifx\csname urlstyle\endcsname\relax
  \providecommand{\doi}[1]{doi: #1}\else
  \providecommand{\doi}{doi: \begingroup \urlstyle{rm}\Url}\fi

\bibitem[Agrawal et~al.(2023)Agrawal, Panwar, Mohan, Kwatra, Gulavani, and Ramjee]{agrawal2023sarathi}
Amey Agrawal, Ashish Panwar, Jayashree Mohan, Nipun Kwatra, Bhargav~S. Gulavani, and Ramachandran Ramjee.
\newblock Sarathi: Efficient llm inference by piggybacking decodes with chunked prefills, 2023.

\bibitem[Aminabadi et~al.(2022)Aminabadi, Rajbhandari, Zhang, Awan, Li, Li, Zheng, Rasley, Smith, Ruwase, and He]{aminabadi2022deepspeed}
Reza~Yazdani Aminabadi, Samyam Rajbhandari, Minjia Zhang, Ammar~Ahmad Awan, Cheng Li, Du~Li, Elton Zheng, Jeff Rasley, Shaden Smith, Olatunji Ruwase, and Yuxiong He.
\newblock Deepspeed inference: Enabling efficient inference of transformer models at unprecedented scale, 2022.

\bibitem[Boost(2021)]{Boost2021Asio}
Boost.
\newblock Boost.asio.
\newblock \url{https://www.boost.org/doc/libs/1_78_0/doc/html/boost_asio.html}, 2021.

\bibitem[Brown et~al.(2020)Brown, Mann, Ryder, Subbiah, Kaplan, Dhariwal, Neelakantan, Shyam, Sastry, Askell, Agarwal, Herbert-Voss, Krueger, Henighan, Child, Ramesh, Ziegler, Wu, Winter, Hesse, Chen, Sigler, Litwin, Gray, Chess, Clark, Berner, McCandlish, Radford, Sutskever, and Amodei]{Brown2020GPT}
Tom Brown, Benjamin Mann, Nick Ryder, Melanie Subbiah, Jared~D Kaplan, Prafulla Dhariwal, Arvind Neelakantan, Pranav Shyam, Girish Sastry, Amanda Askell, Sandhini Agarwal, Ariel Herbert-Voss, Gretchen Krueger, Tom Henighan, Rewon Child, Aditya Ramesh, Daniel Ziegler, Jeffrey Wu, Clemens Winter, Chris Hesse, Mark Chen, Eric Sigler, Mateusz Litwin, Scott Gray, Benjamin Chess, Jack Clark, Christopher Berner, Sam McCandlish, Alec Radford, Ilya Sutskever, and Dario Amodei.
\newblock Language models are few-shot learners.
\newblock In H.~Larochelle, M.~Ranzato, R.~Hadsell, M.F. Balcan, and H.~Lin, editors, \emph{Advances in Neural Information Processing Systems}, volume~33, pages 1877--1901. Curran Associates, Inc., 2020.
\newblock URL \url{https://proceedings.neurips.cc/paper_files/paper/2020/file/1457c0d6bfcb4967418bfb8ac142f64a-Paper.pdf}.

\bibitem[Chen et~al.(2023)Chen, Ye, Wu, Zhuo, Ceze, and Krishnamurthy]{chen2023punica}
Lequn Chen, Zihao Ye, Yongji Wu, Danyang Zhuo, Luis Ceze, and Arvind Krishnamurthy.
\newblock Punica: Multi-tenant lora serving, 2023.

\bibitem[Ding et~al.(2023)Ding, Ma, Dong, Zhang, Huang, Wang, Zheng, and Wei]{ding2023longnet}
Jiayu Ding, Shuming Ma, Li~Dong, Xingxing Zhang, Shaohan Huang, Wenhui Wang, Nanning Zheng, and Furu Wei.
\newblock Longnet: Scaling transformers to 1,000,000,000 tokens, 2023.

\bibitem[Dong et~al.(2024)Dong, Yang, Zhang, Wang, Chi, and Chen]{dong2024less}
Harry Dong, Xinyu Yang, Zhenyu Zhang, Zhangyang Wang, Yuejie Chi, and Beidi Chen.
\newblock Get more with less: Synthesizing recurrence with kv cache compression for efficient llm inference, 2024.

\bibitem[Eisenman et~al.(2022)Eisenman, Matam, Ingram, Mudigere, Krishnamoorthi, Nair, Smelyanskiy, and Annavaram]{Eisenman2022Checknrun}
Assaf Eisenman, Kiran~Kumar Matam, Steven Ingram, Dheevatsa Mudigere, Raghuraman Krishnamoorthi, Krishnakumar Nair, Misha Smelyanskiy, and Murali Annavaram.
\newblock {Check-N-Run}: a checkpointing system for training deep learning recommendation models.
\newblock In \emph{19th USENIX Symposium on Networked Systems Design and Implementation (NSDI 22)}, pages 929--943, Renton, WA, April 2022. USENIX Association.
\newblock ISBN 978-1-939133-27-4.
\newblock URL \url{https://www.usenix.org/conference/nsdi22/presentation/eisenman}.

\bibitem[Github(2023)]{Github-copilot}
Github.
\newblock Github copilot.
\newblock \url{https://github.com/features/copilot}, 2023.

\bibitem[Jeon et~al.(2019)Jeon, Venkataraman, Phanishayee, Qian, Xiao, and Yang]{Myeongjae2019Philly}
Myeongjae Jeon, Shivaram Venkataraman, Amar Phanishayee, Junjie Qian, Wencong Xiao, and Fan Yang.
\newblock Analysis of {Large-Scale} {Multi-Tenant} {GPU} clusters for {DNN} training workloads.
\newblock In \emph{2019 USENIX Annual Technical Conference (USENIX ATC 19)}, pages 947--960, Renton, WA, July 2019. USENIX Association.
\newblock ISBN 978-1-939133-03-8.
\newblock URL \url{https://www.usenix.org/conference/atc19/presentation/jeon}.

\bibitem[Jiang et~al.(2024)Jiang, Yan, Yao, Zhou, Chen, and Yuan]{jiang2024hexgen}
Youhe Jiang, Ran Yan, Xiaozhe Yao, Yang Zhou, Beidi Chen, and Binhang Yuan.
\newblock Hexgen: Generative inference of large-scale foundation model over heterogeneous decentralized environment, 2024.

\bibitem[Jin et~al.(2023)Jin, Wu, Brooks, and Wei]{jin2023s3}
Yunho Jin, Chun-Feng Wu, David Brooks, and Gu-Yeon Wei.
\newblock S$^{3}$: Increasing gpu utilization during generative inference for higher throughput, 2023.

\bibitem[Kwon et~al.(2023)Kwon, Li, Zhuang, Sheng, Zheng, Yu, Gonzalez, Zhang, and Stoica]{Kwon2023VLLM}
Woosuk Kwon, Zhuohan Li, Siyuan Zhuang, Ying Sheng, Lianmin Zheng, Cody~Hao Yu, Joseph Gonzalez, Hao Zhang, and Ion Stoica.
\newblock Efficient memory management for large language model serving with pagedattention.
\newblock In \emph{Proceedings of the 29th Symposium on Operating Systems Principles}, SOSP '23, page 611–626, New York, NY, USA, 2023. Association for Computing Machinery.
\newblock ISBN 9798400702297.
\newblock \doi{10.1145/3600006.3613165}.
\newblock URL \url{https://doi.org/10.1145/3600006.3613165}.

\bibitem[Li et~al.(2022)Li, Phanishayee, Murray, Tarnawski, and Kim]{Li2022Harmony}
Youjie Li, Amar Phanishayee, Derek Murray, Jakub Tarnawski, and Nam~Sung Kim.
\newblock Harmony: overcoming the hurdles of gpu memory capacity to train massive dnn models on commodity servers.
\newblock \emph{Proc. VLDB Endow.}, 15\penalty0 (11):\penalty0 2747–2760, jul 2022.
\newblock ISSN 2150-8097.
\newblock \doi{10.14778/3551793.3551828}.
\newblock URL \url{https://doi.org/10.14778/3551793.3551828}.

\bibitem[Miao et~al.(2023)Miao, Shi, Duan, Xi, Lin, Cui, and Jia]{miao2023spotserve}
Xupeng Miao, Chunan Shi, Jiangfei Duan, Xiaoli Xi, Dahua Lin, Bin Cui, and Zhihao Jia.
\newblock Spotserve: Serving generative large language models on preemptible instances, 2023.

\bibitem[Narayanan et~al.(2019)Narayanan, Harlap, Phanishayee, Seshadri, Devanur, Ganger, Gibbons, and Zaharia]{Narayanan2019Pipedream}
Deepak Narayanan, Aaron Harlap, Amar Phanishayee, Vivek Seshadri, Nikhil~R. Devanur, Gregory~R. Ganger, Phillip~B. Gibbons, and Matei Zaharia.
\newblock Pipedream: Generalized pipeline parallelism for dnn training.
\newblock In \emph{Proceedings of the 27th ACM Symposium on Operating Systems Principles}, SOSP '19, page 1–15, New York, NY, USA, 2019. Association for Computing Machinery.
\newblock ISBN 9781450368735.
\newblock \doi{10.1145/3341301.3359646}.
\newblock URL \url{https://doi.org/10.1145/3341301.3359646}.

\bibitem[Narayanan et~al.(2021)Narayanan, Shoeybi, Casper, LeGresley, Patwary, Korthikanti, Vainbrand, Kashinkunti, Bernauer, Catanzaro, Phanishayee, and Zaharia]{Narayanan2021Megatron}
Deepak Narayanan, Mohammad Shoeybi, Jared Casper, Patrick LeGresley, Mostofa Patwary, Vijay~Anand Korthikanti, Dmitri Vainbrand, Prethvi Kashinkunti, Julie Bernauer, Bryan Catanzaro, Amar Phanishayee, and Matei Zaharia.
\newblock Efficient large-scale language model training on gpu clusters using megatron-lm, 2021.

\bibitem[NVIDIA(2015)]{NVIDIA2015Streams}
NVIDIA.
\newblock Cuda c/c++ streams and concurrency.
\newblock \url{https://developer.download.nvidia.com/CUDA/training/StreamsAndConcurrencyWebinar.pdf}, 2015.

\bibitem[NVIDIA(2023{\natexlab{a}})]{NVIDIA2023NCCL}
NVIDIA.
\newblock Nvidia collective communications library (nccl).
\newblock \url{https://developer.nvidia.com/nccl}, 2023{\natexlab{a}}.

\bibitem[NVIDIA(2023{\natexlab{b}})]{NVIDIA21FasterTransformer}
NVIDIA.
\newblock Nvidia fastertransformer.
\newblock \url{https://github.com/NVIDIA/FasterTransformer}, 2023{\natexlab{b}}.

\bibitem[NVIDIA(2023{\natexlab{c}})]{NVIDIA21TensorRT}
NVIDIA.
\newblock Tensorrt-llm.
\newblock \url{https://github.com/NVIDIA/TensorRT-LLM}, 2023{\natexlab{c}}.

\bibitem[OpenAI(2023)]{openai}
OpenAI.
\newblock Openai developer platform.
\newblock \url{https://platform.openai.com/overview}, 2023.

\bibitem[OpenMPI(2023)]{OpenMPI2023MPI}
OpenMPI.
\newblock Open mpi: Open source high performance computing.
\newblock \url{https://www.open-mpi.org/}, 2023.

\bibitem[Ott et~al.(2019)Ott, Edunov, Baevski, Fan, Gross, Ng, Grangier, and Auli]{Ott2019fairseq}
Myle Ott, Sergey Edunov, Alexei Baevski, Angela Fan, Sam Gross, Nathan Ng, David Grangier, and Michael Auli.
\newblock fairseq: A fast, extensible toolkit for sequence modeling.
\newblock In Waleed Ammar, Annie Louis, and Nasrin Mostafazadeh, editors, \emph{Proceedings of the 2019 Conference of the North {A}merican Chapter of the Association for Computational Linguistics (Demonstrations)}, pages 48--53, Minneapolis, Minnesota, June 2019. Association for Computational Linguistics.
\newblock \doi{10.18653/v1/N19-4009}.
\newblock URL \url{https://aclanthology.org/N19-4009}.

\bibitem[Patel et~al.(2023)Patel, Choukse, Zhang, Íñigo Goiri, Shah, Maleki, and Bianchini]{patel2023splitwise}
Pratyush Patel, Esha Choukse, Chaojie Zhang, Íñigo Goiri, Aashaka Shah, Saeed Maleki, and Ricardo Bianchini.
\newblock Splitwise: Efficient generative llm inference using phase splitting, 2023.

\bibitem[Rajbhandari et~al.(2021)Rajbhandari, Ruwase, Rasley, Smith, and He]{Rajbhandari21Zero}
Samyam Rajbhandari, Olatunji Ruwase, Jeff Rasley, Shaden Smith, and Yuxiong He.
\newblock Zero-infinity: breaking the gpu memory wall for extreme scale deep learning.
\newblock In \emph{Proceedings of the International Conference for High Performance Computing, Networking, Storage and Analysis}, SC '21, New York, NY, USA, 2021. Association for Computing Machinery.
\newblock ISBN 9781450384421.
\newblock \doi{10.1145/3458817.3476205}.
\newblock URL \url{https://doi.org/10.1145/3458817.3476205}.

\bibitem[Sheng et~al.(2023{\natexlab{a}})Sheng, Cao, Li, Hooper, Lee, Yang, Chou, Zhu, Zheng, Keutzer, Gonzalez, and Stoica]{sheng2023slora}
Ying Sheng, Shiyi Cao, Dacheng Li, Coleman Hooper, Nicholas Lee, Shuo Yang, Christopher Chou, Banghua Zhu, Lianmin Zheng, Kurt Keutzer, Joseph~E. Gonzalez, and Ion Stoica.
\newblock S-lora: Serving thousands of concurrent lora adapters, 2023{\natexlab{a}}.

\bibitem[Sheng et~al.(2023{\natexlab{b}})Sheng, Zheng, Yuan, Li, Ryabinin, Fu, Xie, Chen, Barrett, Gonzalez, Liang, Ré, Stoica, and Zhang]{sheng2023flexgen}
Ying Sheng, Lianmin Zheng, Binhang Yuan, Zhuohan Li, Max Ryabinin, Daniel~Y. Fu, Zhiqiang Xie, Beidi Chen, Clark Barrett, Joseph~E. Gonzalez, Percy Liang, Christopher Ré, Ion Stoica, and Ce~Zhang.
\newblock Flexgen: High-throughput generative inference of large language models with a single gpu, 2023{\natexlab{b}}.

\bibitem[Wikipedia(2023)]{pcie}
Wikipedia.
\newblock Pci express.
\newblock \url{https://en.wikipedia.org/wiki/PCI_Express}, 2023.

\bibitem[Workshop(2023)]{workshop2023bloom}
BigScience Workshop.
\newblock Bloom: A 176b-parameter open-access multilingual language model, 2023.

\bibitem[Yu et~al.(2022)Yu, Jeong, Kim, Kim, and Chun]{GyeongIn2022Orca}
Gyeong-In Yu, Joo~Seong Jeong, Geon-Woo Kim, Soojeong Kim, and Byung-Gon Chun.
\newblock Orca: A distributed serving system for {Transformer-Based} generative models.
\newblock In \emph{16th USENIX Symposium on Operating Systems Design and Implementation (OSDI 22)}, pages 521--538, Carlsbad, CA, July 2022. USENIX Association.
\newblock ISBN 978-1-939133-28-1.
\newblock URL \url{https://www.usenix.org/conference/osdi22/presentation/yu}.

\bibitem[Zhang et~al.(2017)Zhang, Zheng, Xu, Dai, Ho, Liang, Hu, Wei, Xie, and Xing]{Zhang17Poseidon}
Hao Zhang, Zeyu Zheng, Shizhen Xu, Wei Dai, Qirong Ho, Xiaodan Liang, Zhiting Hu, Jinliang Wei, Pengtao Xie, and Eric~P. Xing.
\newblock Poseidon: An efficient communication architecture for distributed deep learning on {GPU} clusters.
\newblock In \emph{2017 USENIX Annual Technical Conference (USENIX ATC 17)}, pages 181--193, Santa Clara, CA, July 2017. USENIX Association.
\newblock ISBN 978-1-931971-38-6.
\newblock URL \url{https://www.usenix.org/conference/atc17/technical-sessions/presentation/zhang}.

\bibitem[Zhang et~al.(2022)Zhang, Roller, Goyal, Artetxe, Chen, Chen, Dewan, Diab, Li, Lin, Mihaylov, Ott, Shleifer, Shuster, Simig, Koura, Sridhar, Wang, and Zettlemoyer]{zhang2022opt}
Susan Zhang, Stephen Roller, Naman Goyal, Mikel Artetxe, Moya Chen, Shuohui Chen, Christopher Dewan, Mona Diab, Xian Li, Xi~Victoria Lin, Todor Mihaylov, Myle Ott, Sam Shleifer, Kurt Shuster, Daniel Simig, Punit~Singh Koura, Anjali Sridhar, Tianlu Wang, and Luke Zettlemoyer.
\newblock Opt: Open pre-trained transformer language models, 2022.

\bibitem[Zhang et~al.(2023)Zhang, Sheng, Zhou, Chen, Zheng, Cai, Song, Tian, Ré, Barrett, Wang, and Chen]{zhang2023h2o}
Zhenyu Zhang, Ying Sheng, Tianyi Zhou, Tianlong Chen, Lianmin Zheng, Ruisi Cai, Zhao Song, Yuandong Tian, Christopher Ré, Clark Barrett, Zhangyang Wang, and Beidi Chen.
\newblock H$_2$o: Heavy-hitter oracle for efficient generative inference of large language models, 2023.

\bibitem[Zheng et~al.(2023)Zheng, Chiang, Sheng, Li, Zhuang, Wu, Zhuang, Li, Lin, Xing, Gonzalez, Stoica, and Zhang]{zheng2023lmsyschat1m}
Lianmin Zheng, Wei-Lin Chiang, Ying Sheng, Tianle Li, Siyuan Zhuang, Zhanghao Wu, Yonghao Zhuang, Zhuohan Li, Zi~Lin, Eric.~P Xing, Joseph~E. Gonzalez, Ion Stoica, and Hao Zhang.
\newblock Lmsys-chat-1m: A large-scale real-world llm conversation dataset, 2023.

\bibitem[Zhong et~al.(2024)Zhong, Liu, Chen, Hu, Zhu, Liu, Jin, and Zhang]{zhong2024distserve}
Yinmin Zhong, Shengyu Liu, Junda Chen, Jianbo Hu, Yibo Zhu, Xuanzhe Liu, Xin Jin, and Hao Zhang.
\newblock Distserve: Disaggregating prefill and decoding for goodput-optimized large language model serving, 2024.

\end{thebibliography}
